\documentclass[journal,twocolumn,10pt]{IEEEtran}

\usepackage{bm}
\usepackage{enumitem}
\usepackage{numprint}
\usepackage{lipsum}
\usepackage{float}
\usepackage{comment}

\usepackage{nicefrac,xfrac}
\usepackage{amsmath,amssymb,amsfonts,accents,mathrsfs,nccmath}
\usepackage{mathtools,array}
\usepackage{cuted}
\usepackage{scalerel}

\usepackage{bigints} 
\usepackage{multicol}
\usepackage{stfloats}
\usepackage[linesnumberedhidden,lined,ruled]{algorithm2e}

\usepackage{multirow}
\usepackage{tabularx}
\usepackage{booktabs}
\usepackage{makecell}
\usepackage{colortbl} 
\usepackage{multicol}
\newcolumntype{C}[1]{>{\centering\arraybackslash}m{#1}} 

\usepackage{float}
\usepackage{graphicx}
\usepackage{textcomp}
\usepackage{rotating} 

\usepackage{pgfplots}
\usepackage{tikz}
\pgfplotsset{compat=newest}
\newlength\figureheight
\newlength\figurewidth

\def\bphi{{\boldsymbol{\phi}}}

\def\bGamma{{\boldsymbol{\Gamma}}}

\def\bc{{\mathbf{c}}}
\def\bd{{\mathbf{d}}}
\def\bX{{\mathbf{X}}}
\def\bx{{\mathbf{x}}}
\def\bH{{\mathbf{H}}}
\def\bh{{\mathbf{h}}}
\def\bY{{\mathbf{Y}}}
\def\by{{\mathbf{y}}}

\def\bs{{\mathbf{s}}}

\def\bW{{\mathbf{W}}}
\def\bZ{{\mathbf{Z}}}

\def\brho{{\boldsymbol{\rho}}}

\def\bphi{{\boldsymbol{\phi}}}

\def\bsfx{\boldsymbol{\mathsf{x}}}
\def\bsfh{\boldsymbol{\mathsf{h}}}
\def\bsfy{\boldsymbol{\mathsf{y}}}
\def\bsfz{\boldsymbol{\mathsf{z}}}
\def\bsfp{\boldsymbol{\mathsf{p}}}
\def\bsfq{\boldsymbol{\mathsf{q}}}
\def\bsfr{\boldsymbol{\mathsf{r}}}
\def\bsfX{\boldsymbol{\mathsf{X}}}
\def\bsfH{\boldsymbol{\mathsf{H}}}

\def\bsfZ{\boldsymbol{\mathsf{Z}}}

\def\bsfgamma{\boldsymbol{\mathsf{\gamma}}}

\def\onu{\overline{\nu}}

\definecolor{amethyst}{rgb}{0.6, 0.4, 0.8}
\definecolor{purple}{rgb}{0.49020,0.18039,0.56078}%
\definecolor{mustard}{rgb}{0.92941,0.69020,0.12941}%
\definecolor{gold}{rgb}{0.7607843137,0.6588235294,0}%
\definecolor{orange}{rgb}{0.82353,0.41176,0.11765}%
\definecolor{dark_green}{rgb}{0.00000,0.50196,0.00000}%
\definecolor{lgreen}{rgb}{0.00000,0.7411764706,0.00000}%
\definecolor{wine}{rgb}{0.3725490196,0,0}%
\definecolor{gray}{rgb}{0.50196,0.50196,0.50196}%
\definecolor{mblue}{rgb}{0.00000,0.45098,0.74118}%
\definecolor{lblue}{rgb}{0.5,0.8,0.9}%
\definecolor{lgray}{rgb}{0.65098,0.65098,0.65098}%
\definecolor{llgray}{rgb}{0.85098,0.85098,0.85098}%
\definecolor{lllgray}{rgb}{0.97,0.97,0.97}%
\definecolor{lpink}{rgb}{0.9843137255,0.7333333,0.8862745098}%
\definecolor{mpink}{rgb}{0.933333,0.1490196078,0.6666666667}

\definecolor{modred_VR}{RGB}{192,51,77}
\definecolor{modpink_VR}{RGB}{214, 97, 143}
\definecolor{vsoftorange_VR}{RGB}{243,212,160}
\definecolor{vividorange_VR}{RGB}{241,147,29}
\definecolor{mddarkorange_VR}{RGB}{143,113,91}
\definecolor{modcyan_VR}{RGB}{97, 202, 214}
\definecolor{modlimegreen_VR}{RGB}{97, 214, 110}

\definecolor{vdarkblue_CA}{RGB}{20,50,92}
\definecolor{softblue_CA}{RGB}{83,152,217}
\definecolor{vsoftorange_CA}{RGB}{244,227,177}
\definecolor{vividorange_CA}{RGB}{217,107,12}
\definecolor{darkmodred_CA}{RGB}{165,58,59}
\definecolor{darkmodlimegreen_CA}{RGB}{59,165,58}
\definecolor{darkmodviolet_CA}{RGB}{111,58,165}

\definecolor{Maroon_ST}{RGB}{128,0,0}
\definecolor{Red_ST}{RGB}{230,25,75}
\definecolor{MedRed_ST}{RGB}{240,65,105}
\definecolor{Pink_ST}{RGB}{250,190,212}
\definecolor{Pink_ST2}{rgb}{0.83922,0.38039,0.56078}
\definecolor{Brown_ST}{RGB}{170,110,40}
\definecolor{Orange_ST}{RGB}{245,130,48}
\definecolor{Olive_ST}{RGB}{128,128,0}
\definecolor{Yellow_ST}{RGB}{255,255,25}
\definecolor{Apricot_ST}{RGB}{255,215,180}
\definecolor{Beige_ST}{RGB}{255,250,200}
\definecolor{Green_ST}{RGB}{60,180,75}
\definecolor{Lime_ST}{RGB}{210,245,60}
\definecolor{Mint_ST}{RGB}{170,255,195}
\definecolor{Teal_ST}{RGB}{0,128,128}
\definecolor{Navy_ST}{RGB}{0,0,128}
\definecolor{Blue_ST}{RGB}{0,130,200}
\definecolor{MedBlue_ST}{RGB}{0,150,220}
\definecolor{Cyan_ST}{RGB}{70,240,240}
\definecolor{Purple_ST}{RGB}{145,30,180}
\definecolor{Lavender_ST}{RGB}{220,190,255}
\definecolor{DarkLavender_ST}{RGB}{190,160,225}
\definecolor{Magenta_ST}{RGB}{240,50,230}
\definecolor{Grey_ST}{RGB}{128,128,128}

\newcommand{\AMPcolor}{Brown_ST}
\newcommand{\BIGAMPcolor}{MedBlue_ST}
\newcommand{\JEMAMPcolor}{DarkLavender_ST}
\newcommand{\HyGAMPcolor}{Grey_ST}
\newcommand{\RBPcolor}{Green_ST}
\newcommand{\AUDcolor}{Orange_ST}
\newcommand{\TBiGAMPcolor}{Pink_ST2}

\makeatletter%
    \if@twocolumn%
    \else
    \fi
\makeatother

\begin{document}
%
\title{\huge Joint Channel Estimation, Activity Detection and Decoding using Dynamic Message-Scheduling for Machine-Type Communications}

\author{Roberto~B.~Di~Renna, \textit{Graduate Student Member, IEEE}~and~Rodrigo~C.~de~Lamare, \textit{Senior Member, IEEE}  
\thanks{The authors are with the Centre for Telecommunications Studies (CETUC),
Pontifical Catholic University of Rio de Janeiro (PUC-Rio), Rio de Janeiro 22453-900,
Brazil (e-mail: \{robertobrauer, delamare\}@cetuc.puc-rio.br).
This work was supported by the Conselho Nacional de Desenvolvimento Cient\'{i}fico e Tecnol\'{o}gico (CNPq).}}


\maketitle

\begin{abstract}
In this work, we present a joint channel estimation, activity
detection and data decoding scheme for massive machine-type
communications. By including the channel and the a priori activity
factor in the factor graph, we present the bilinear
message-scheduling GAMP (BiMSGAMP), a message-passing solution that
uses the channel decoder beliefs to refine the activity detection
and data decoding. We include two  message-scheduling strategies
based on the residual belief propagation and the activity user
detection in which messages are evaluated and scheduled in every new
iteration. An analysis of the convergence of BiMSGAMP along with a
study of its computational complexity is carried out. Numerical
results show that BiMSGAMP outperforms state-of-the-art algorithms,
highlighting the gains achieved by using the dynamic scheduling
strategies and the effects of the channel decoding part in the
system.

\end{abstract}

\begin{IEEEkeywords}
    mMTC, message-passing, joint activity detection, channel estimation and data decoding, grant-free massive MIMO.
\end{IEEEkeywords}

\IEEEpeerreviewmaketitle

\section{Introduction}
\label{sec:intro}
\IEEEPARstart{I}n 5th generation (5G) of wireless systems, massive
machine-type communications (mMTC) covers emerging smart service
such as industrial automation, environmental sensing and remote
manufacturing ~\cite{DiRennaAccess2020}. mMTC focuses on the uplink
and aims to provide massive connectivity to different types of
devices that behave differently from the well-established human-type
communications (HTC). Designed for specific applications, mMTC
devices (MTCDs) exhibit a sporadic data traffic, where small packets
are transmitted at low rates. Since most MTCDs are battery
operated~\cite{PopovskiAccess2018}, they are energy-constrained.
These unique aspects of mMTC impose new demands and challenges to
random access (RA) design.

Although solutions based on physical random access channel
(PRACH)~\cite{3GPPTR36888, 3GPP2016} have recently been proposed to
fit the mMTC traffic in the Long Term Evolution (LTE) standards,
they are still not suitable to fulfill the mMTC requirements. Due to
the aforementioned mMTC characteristics, the limited number of
available preambles for the access reservation procedure, the
massive number of concurrent transmissions of the same preambles
would cause the overload of the RA procedure. This issue results in
high collision probability, access failure rate and delay. Moreover,
the signalling overhead degrades the overall system efficiency since
the size of the upload data payload from MTCDs is signiﬁcantly
smaller than the traditional HTC~\cite{CentenaroTCom2017}. In this
way, a new approach with reduced signalling is required which does
not demand orthogonal preambles. A promising approach is the
Grant-Free Random Access
(GFRA)~\cite{BockelmannAccess2018,TSalamAccess2019}, which allows
MTCDs to transmit their packages to the base station (BS) directly,
without the need to wait for a specific uplink grant from the BS.
The main advantages of GFRA are the reduced transmission latency,
smaller signalling overhead due to the simplification of the
scheduling procedure and improved energy efficiency (battery life)
of MTCDs. With a massive number of MTCDs requiring access without
coordination, even the use of non-orthogonal preambles with a
time-slotted transmission would cause significant overhead. In
scenarios where MTCDs can transmit their packets only at the
beginning of each time-slot, any device that fails to align its time
slots properly may degrade its detection and estimation performance.
Hence, a non-time-slotted (or asynchronous) transmission would
further simplify scheduling, resulting in smaller signalling
overhead, reduced transmission latency and improved energy
efficiency~\cite{F5G1}. Despite the fact that in asynchronous
scenarios the preamble and data signals are superposed in a
non-orthogonal manner and interfere with each other, due to the
asymptotic favourable propagation in massive multiple-input
multiple-output (mMIMO), their spatial subspaces are approximately
mutually orthogonal~\cite{LLuJSTSP2014,Tse2005}. Thus, the BS can
decode the data of MTCDs that transmitted first and then employ
successive interference cancellation (SIC) or other interference
cancellation techniques
\cite{itic,spa,jidf,mmimo,wence,rrser,mfsic,mbdf,tds,bfidd,aaidd,listmtc,1bitidd,detmtc,dynovs}
to decode data for the received packets~\cite{JDingTCom2020}.

In mMTC, the BS load is increased due to random transmissions of
many MTCDs, which calls for the reception of many simultaneous
packets and mitigation of multiuser interference. Furthermore, the
BS has no knowledge which MTCD is active at a given time instant
such that the physical layer task is to jointly estimate the
channels and detect the activity and the data of the
devices~\cite{DiRennaAccess2020}. From a physical layer perspective,
the mMTC scenario with intermittent transmissions can be seen as a
sparse recovery problem. Considering perfect channel estimation at
the BS, several joint activity and data detection techniques based
on compressed sensing (CS)~\cite{JWChoiCST2017} have been proposed.
An approach that adapts classic algorithms as maximum likelihood
(ML)~\cite{Zhu2011}, sphere decoder (SD)\cite{Knoop2014} and minimum
mean squared error (MMSE)~\cite{Ahn2018, DiRennaSCC2019} to the
sparse scenario is the addition of a regularization parameter into
the cost function. The sparsity scenario also admits greedy
solutions in which variations of orthogonal matching pursuit (OMP)
and orthogonal least squares
(OLS)~\cite{Wang2015,SchepkerTCOM2015,LiuICC2017} have been devised.
Employing channel coding, the schemes
in~\cite{BockelmannIEEECommLet2015,BKJeongICASSP2018,
DiRennaWCL2019, DiRennaTCom2020, DiRennaISWCS21} propose adaptive
and iterative solutions that exchange extrinsic information between
activity and symbol detectors. In order to reduce complexity,
approximate message passing (AMP) techniques
~\cite{Donoho2009,ZChenTWirCom2019, KSenelGlobecom2017,
CWeiCommLet2017, LLiuTWC2019} have been reported, where tools like
expectation maximization (EM)~\cite{JVilaTSP2013} and expectation
propagation (EP)~\cite{TMinkaPhD2001} are employed. With the
message-passing approach, there are plenty of solutions that address
the activity detection and channel estimation problems as in
~\cite{LLiuTSigPr2018,ZTangAccess2020,YZhangTVT2018,AhnTCom2019,QZouSPL2020,DiRennaWCL2021}.
There are also works
~\cite{YBaiVTC2019,ZZhangTVT2019,GGuiTVT2018,WZhuTWC2021,YCuiJSAC2021}
that use machine-learning to estimate the channels. Furthermore,
variational inference techniques combined with AMP that use
Kullback-Leibler divergence to transform an intractable inference
problem into a tractable optimization problem have been reported
~\cite{AhnTCom2019,XMengJSAC2021,LBaiTVT2020,FWeiTWC2019}. Recently,
approaches that jointly perform the activity and data detection and
channel estimation using message passing have been
studied~\cite{FWeiTWC2019,ZHanWCL2021,XKuaiTCom2020,YZhangTVT2020,QZouSPLComp2020,TDingTWC2019,SJiangTWC2020}.
The works in~\cite{FWeiTWC2019,ZHanWCL2021} design SCMA receivers,
where~\cite{ZHanWCL2021} proposes a Kronecker-product coding scheme
to the data detection part. The work in~\cite{XKuaiTCom2020} studies
a scenario considering the angular-domain sparsity and spatial
correlation in a large-scale antenna array. In a message-passing
framework, the work in~\cite{YZhangTVT2020} studies the overhead
reduction in a low density signature OFDM scenario. The work
in~\cite{QZouSPLComp2020} studied a low-complexity joint user
activity, channel and data estimation scheme based on the BiG-AMP
approach, while~\cite{TDingTWC2019} considers a coherent detection
scenario. On the other hand, the approach of~\cite{SJiangTWC2020}
focuses on the mitigation of phase ambiguity issues.

In this work, based on the generalized approximate message passing
(GAMP)~\cite{Rangan2012} algorithm, we propose a novel bilinear
message-scheduling GAMP (BiMSGAMP), that jointly performs device
activity detection, channel estimation and data decoding in a
grant-free massive MIMO scenario. Unlike existing works, based on 5G
channel coding techniques, we exploit the decoding of Low-Density
Parity-Check (LDPC) codes that is also based on message-passing and
devise a solution that uses the channel decoder beliefs to refine
the activity detection and data decoding. To the best of our
knowledge, it is the first work that fully integrates the joint
channel estimation, activity and data detection to the decoding
part. Unlike most of the message-passing works in the literature,
that consider a completely parallel update of the messages, we
introduce the dynamic message-scheduling concept. Dynamic
message-scheduling schemes dramatically reduce the computational
cost since there is no need to update every node of the factor
graph, differently from existing schemes. In particular, BiMSGAMP
updates messages according to the activity user detection (AUD) and
the residual belief propagation (RBP), metrics already available in
the factor graph. Additionally, we examine the mMTC overhead issues
described before by considering in our framework non-orthogonal
pilots and investigate the asynchronous mMTC
scenario~\cite{TDingTWC2019, XMaAccess2020, JZhangOJVT2020,
WZhuTWC2021, SKimTWC2021}. Therefore, BiMSGAMP departs from the
common synchronous transmissions and addresses the problem without
requiring frame-level synchronization. We also carry out an analysis
of the convergence of BiMSGAMP along with a study of its
computational complexity. In order to verify the BiMSGAMP
performance, we compare its efficiency against other approaches in
terms of normalized MSE (NMSE), false alarm rate (FAR), missed
detection rate (MDR) and frame error rate (FER).

    Therefore, the main contributions of this paper include the following four aspects:
    \begin{itemize}
        \item The development of the BiMSGAMP that introduces channel decoder beliefs into the framework of Bayesian inference wherein the resulting factor graph is a fully connected structure, where the messages are exchanged between the joint channel estimation, activity and data detection parts, and the LDPC decoder.

        \item Novel dynamic message-scheduling techniques that accelerate the convergence and dramatically reduces the computational cost of the algorithm which is key for 5G and beyond systems, where the mMTC network must support a massive number of devices. A complexity study based on the required floating-point operations (FLOPs) of the proposed and existing techniques is also presented.

        \item An analysis of BiMSGAMP based on state-evolution (SE) is developed, which shows that the SE method predicts the performance accurately and may provide useful insights for system design.

        \item Comparisons in terms of NMSE, FAR, MDR and FER for synchronous and asynchronous grant-free uplink mMTC scenarios that assess the efficiency of BiMSGAMP and other algorithms, the gains achieved by the proposed dynamic scheduling strategies and the effects of channel decoding.
    \end{itemize}

    The remainder of this work is structured as follows. In Section~\ref{sec:SysMod} we describe the system model, divided in synchronous and asynchronous grant-free random access and in Section~\ref{sec:prob_form}, the problem is formulated. The proposed joint activity detection, channel estimation and data decoding structure is presented in Section~\ref{sec:jointAUDCEDATA}, where the messages are derived and shown, as the LLR conversion and the integration with the sum-product algorithm for LDPC decoding. Section~\ref{sec:DynSchStr} explains the dynamic scheduling strategies used in different parts of the factor graph while Section~\ref{sec:analysis} analyzes their computational cost and convergence. Numerical results in terms of frame error rates, normalized mean squared errors and activity error rates are shown in Section~\ref{sec:Num_res} as long as Section~\ref{sec:conc} draws the conclusions. \textit{Notations:} Matrices and vectors are denoted by boldfaced capital letters  and lowercase letters, respectively. The space of complex (real) $N$-dimensional vectors is denoted by $\mathbb{C}^N\left(\mathbb{R}^N\right)$. The $i$-th column of a matrix $\mathbf{A} \in \mathbb{C}^{M\times N}$ is denoted by $\mathbf{a}_i \in \mathbb{C}^M$.
    For a vector $\mathbf{x} \in \mathbb{C}^N, ||\mathbf{x}||$ denotes its Euclidean norm, $||\mathbf{x}||_\text{F}$ the Frobenius norm and $\mathcal{P}\left(\cdot\right)$ the probability density/mass distributions. A summary of key notations in this paper is given in Table~\ref{tab:notations}.

    \makeatletter%
    \if@twocolumn%
\begin{table}[t]
    \caption{Summary of key notations.}
    \label{tab:notations}
    \centering
\begin{tabular}{llp{4.8cm}}
\hline
    \multicolumn{3}{c}{System model}                         \\ \hline
    $N$ & \multicolumn{2}{p{6.4cm}}{Number of single-antenna devices}    \\
    $M$ & \multicolumn{2}{p{6.4cm}}{Number of  BS antennas}\\
    $L$ & \multicolumn{2}{p{6.4cm}}{Number symbols per frame, divided in pilots $\left(L_p\right)$ and data $\left(L_d\right)$} \\
    $\gamma_{nt}$   &   \multicolumn{2}{p{6.4cm}}{Activity ind. of the $n$-th device at the $t$-th symb.interval}                \\
$T$                                         & \multicolumn{2}{p{6.4cm}}{Sliding window size} \\
$\Delta t$                                   & \multicolumn{2}{p{6.4cm}}{Step size of the sliding window} \\
$\rho_n$    & \multicolumn{2}{p{6.4cm}}{Probability of being active of the $n$-th device} \\
$|\mathcal{S}^{(i)}|$                        & \multicolumn{2}{p{6.4cm}}{Group of nodes to be updated in the $i$-th iteration of a message scheduling technique} \\[0.2cm] \hline
%
\multicolumn{3}{c}{SPA message definitions at $i$-th iteration, $\forall i \in \mathcal{Z}$}                                                                  \\ \hline
$\Delta^i_{h_{mn}}$                                             &  & SPA-approx. log posterior pdf of $\bsfh_{mn}$                                         \\
$\Delta^{i}_{x_{nt}}$                       &  & SPA-approx. log posterior pdf of $\bsfx_{nt}$  \\[0.1cm]
 \arrayrulecolor{llgray}\hline \\[-0.2cm]
$\Delta^{i+1}_{k_{mn} \rightarrow \gamma_n}$ &
  \multicolumn{1}{p{1.25cm}}{\multirow{2}{*}{\begin{tabular}[c]{@{}l@{}}Activity\\ prior\end{tabular}}} &
  SPA message from factor node $k_{mn}$ to variable node $\gamma_{n}$ \\
$\Delta^{i+1}_{\gamma_n \rightarrow k_{mn}}$ & &
  SPA message from variable node $\gamma_{n}$ to factor node $k_{mn}$\\
 \arrayrulecolor{llgray}\hline \\[-0.2cm]
$\Delta^{i+1}_{g_{ml} \rightarrow h_{mn}} $ &\multicolumn{1}{p{1.25cm}}{\multirow{2}{*}{\begin{tabular}[c]{@{}l@{}}Channel\\ estimation\end{tabular}}}  & SPA message from factor node $g_{ml}$ to variable node $h_{mn}$ \\
$\Delta^{i+1}_{h_{mn} \rightarrow g_{ml}} $ & & SPA message from variable node $h_{mn}$ to factor node $g_{ml}$\\
\arrayrulecolor{llgray}\hline \\[-0.2cm]
$\Delta^{i+1}_{g_{ml} \rightarrow x_{nl}} $ &\multicolumn{1}{p{1.25cm}}{\multirow{2}{*}{\begin{tabular}[c]{@{}l@{}}Data\\ detection\end{tabular}}}  & SPA message from factor node $g_{nl}$ to variable node $x_{nl}$ \\
$\Delta^{i+1}_{x_{nl} \rightarrow g_{ml}} $ & & SPA message from variable node $x_{nl}$ to factor node $g_{ml}$\\
\arrayrulecolor{llgray}\hline \\[-0.2cm]
$\Delta^{i+1}_{f_{nl} \rightarrow d_{nl}} $ &\multicolumn{1}{p{1.25cm}}{\multirow{4}{*}{\begin{tabular}[c]{@{}l@{}}LLR\\ conversion\end{tabular}}}  & SPA message from factor node $f_{nl}$ to variable node $d_{nl}$ \\
$\Delta^{i+1}_{d_{nl} \rightarrow f_{nl}} $ & & SPA message from variable node $d_{nl}$ to factor node $f_{nl}$ \\
$\Delta^{i+1}_{f_{nl} \rightarrow \xi_{nl}} $ & & SPA message from factor node $f_{nl}$ to variable node $\xi_{nl}$ \\
$\Delta^{i+1}_{\xi_{nl} \rightarrow f_{nl}} $ & & SPA message from variable node $\xi_{nl}$ to factor node $f_{nl}$ \\[-0.2cm]
\\ \arrayrulecolor{black}\hline
\end{tabular}
\end{table}

    \else
\begin{table}[t]
    \caption{Summary of key notations.}
    \vspace{-0.25cm}
    \label{tab:notations}
    \centering
\begin{tabular}{lll}
\hline
\multicolumn{3}{c}{System model}                         \\ \hline
$N$   & \multicolumn{2}{l}{Number of single-antenna devices} \\[-0.1cm]
$M$ & \multicolumn{2}{l}{Number of  BS antennas}  \\[-0.1cm]
$L$ & \multicolumn{2}{l}{Number symbols per frame, divided in pilots $\left(L_p\right)$ and data $\left(L_d\right)$} \\[-0.1cm]
$\gamma_{nt}$   & \multicolumn{2}{l}{Activity indicator of the $n$-th device at the $t$-th symbol interval} \\[-0.1cm]
$T$     & \multicolumn{2}{l}{Sliding window size} \\[-0.1cm]
$\Delta t$  & \multicolumn{2}{l}{Step size of the sliding window}
\\[-0.1cm]
$\rho_n$ & \multicolumn{2}{l}{Probability of being active of the $n$-th device}                     \\[-0.1cm]
$|\mathcal{S}^{(i)}|$    & \multicolumn{2}{l}{\begin{tabular}[c]{@{}l@{}}Group of nodes to be updated in the $i$-th iteration of a message\\[-0.2cm] scheduling technique\end{tabular}} \\ \hline
%
\multicolumn{3}{c}{SPA message definitions at $i$-th iteration, $\forall i \in \mathcal{Z}$}                                                                  \\ \hline
$\Delta^i_{h_{mn}}$                                             &  & SPA-approximated log posterior pdf of $\bsfh_{mn}$                                         \\[-0.1cm]
$\Delta^{i}_{x_{nt}}$                       &  & SPA-approximated log posterior pdf of $\bsfx_{nt}$  \\
 \arrayrulecolor{llgray}\hline \\[-0.65cm] 
$\Delta^{i+1}_{k_{mn} \rightarrow \gamma_n}$ &
  \multicolumn{1}{p{1.25cm}}{\multirow{2}{*}{\begin{tabular}[c]{@{}l@{}}Activity\\[-0.25cm] prior\end{tabular}}} &
  SPA message from factor node $k_{mn}$ to variable node $\gamma_{n}$ \\[-0.1cm]
$\Delta^{i+1}_{\gamma_n \rightarrow k_{mn}}$ & &
  SPA message from variable node $\gamma_{n}$ to factor node $k_{mn}$\\
 \arrayrulecolor{llgray}\hline \\[-0.65cm] 
$\Delta^{i+1}_{g_{ml} \rightarrow h_{mn}} $ &\multicolumn{1}{p{1.25cm}}{\multirow{2}{*}{\begin{tabular}[c]{@{}l@{}}Channel\\[-0.25cm] estimation\end{tabular}}}  & SPA message from factor node $g_{ml}$ to variable node $h_{mn}$ \\[-0.1cm]
$\Delta^{i+1}_{h_{mn} \rightarrow g_{ml}} $ & & SPA message from variable node $h_{mn}$ to factor node $g_{ml}$\\
\arrayrulecolor{llgray}\hline \\[-0.65cm]
$\Delta^{i+1}_{g_{ml} \rightarrow x_{nl}} $ &\multicolumn{1}{p{1.25cm}}{\multirow{2}{*}{\begin{tabular}[c]{@{}l@{}}Data\\[-0.25cm] detection\end{tabular}}}  & SPA message from factor node $g_{nl}$ to variable node $x_{nl}$ \\[-0.1cm]
$\Delta^{i+1}_{x_{nl} \rightarrow g_{ml}} $ & & SPA message from variable node $x_{nl}$ to factor node $g_{ml}$\\
\arrayrulecolor{llgray}\hline \\[-0.65cm]
$\Delta^{i+1}_{f_{nl} \rightarrow d_{nl}} $ &\multicolumn{1}{p{1.25cm}}{\multirow{4}{*}{\begin{tabular}[c]{@{}l@{}}LLR\\[-0.25cm] conversion\end{tabular}}}  & SPA message from factor node $f_{nl}$ to variable node $d_{nl}$ \\[-0.1cm]
$\Delta^{i+1}_{d_{nl} \rightarrow f_{nl}} $ & & SPA message from variable node $d_{nl}$ to factor node $f_{nl}$ \\[-0.1cm]
$\Delta^{i+1}_{f_{nl} \rightarrow \xi_{nl}} $ & & SPA message from factor node $f_{nl}$ to variable node $\xi_{nl}$ \\[-0.1cm]
$\Delta^{i+1}_{\xi_{nl} \rightarrow f_{nl}} $ & & SPA message from variable node $\xi_{nl}$ to factor node $f_{nl}$ \\
\\[-0.55cm] \arrayrulecolor{black}\hline
\end{tabular}
\end{table}

    \fi
    \makeatother

\section{System Model}\label{sec:SysMod}
    In this section, we describe asynchronous and synchronous grant-free uplink massive MIMO scenarios. We consider a synchronous scenario, where each observation window employs frame-level synchronization, and an asynchronous scenario, where in each observation window symbol-level but not frame-level synchronization is assumed. Thus, the synchronous scenario can be seen as a special case of the asynchronous one. In the uplink, we have $N$ single-antenna MTDs communicating with a BS equipped with $M'$ antennas. In the grant-free system model, each frame consists of pilot and data symbols~\cite{DiRennaAccess2020}.

    \begin{figure*}[t]
        \begin{minipage}[b]{0.5\linewidth}
            \centering
            \includegraphics{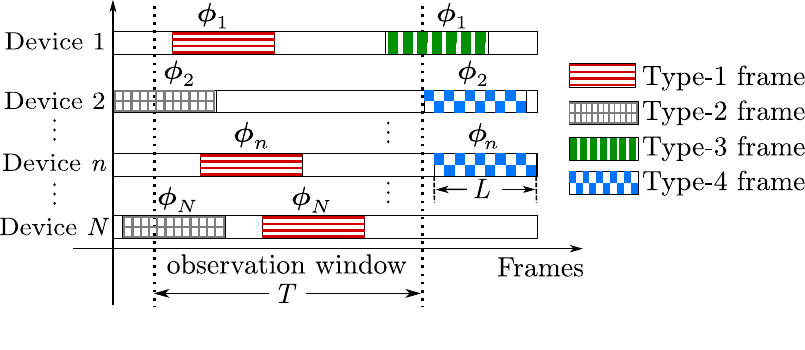}
            \centerline{\footnotesize (a)}\medskip
        \end{minipage}
        \begin{minipage}[b]{0.5\linewidth}
            \centering
            \includegraphics{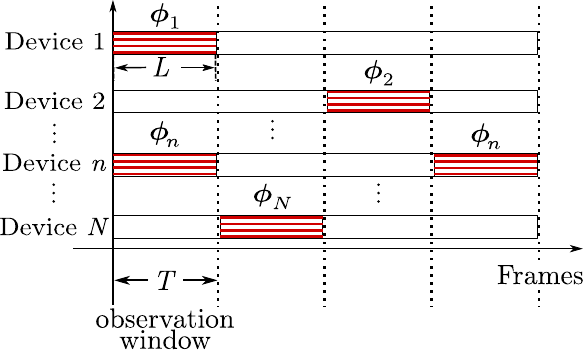}
            \centerline{\footnotesize (b)}\medskip
        \end{minipage}
        \vspace{-1cm}
        \caption{(a) Asynchronous and (b) synchronous frames of the grant-free mMTC scenario.}
        \label{fig:asy_and_syn_mmtc}%
    \end{figure*}

    \subsection{Asynchronous Grant-Free Random Access}

    Following asynchronous schemes of the literature~\cite{TDingTWC2019,WZhuTWC2021}, in this case each device is allowed to transmit $L$ symbols, which we denote here as a frame, at the beginning of any symbol interval $t$. The sparsity of the mMTC scenario is represented by the Boolean variable $\gamma_{nt} = 1$ that indicates that the $n$-th device is active in the $t$-th symbol interval and $\gamma_{nt} = 0$, otherwise. Thus, considering  $\rho_n$ as the probability of being active of the $n$-th device, $P\left(\gamma_{nt} = 1\right) = 1 - P\left(\gamma_{nt} = 0\right) = \rho_n$, where all activity indicators $\gamma_{nt}$ are considered i.i.d. with respect to $n$ and each device has its own activity probability.

    As depicted in Fig.~\ref{fig:asy_and_syn_mmtc}(a), it is possible that just part of the transmitted frame falls within the observation window. Since the problem of interest here is to jointly estimate the channels, decode the signals and detect the activity of devices, the BS is only able to deal with the type-1 frames. Thus, type-2 and type-3 frames are seen as noise in this window. In order to be treated, those frames should be re-transmitted until they fully occupy another observation window. Despite the intermittent pattern of transmissions, each device should wait, at least, for the guard period interval to transmit again. Accordingly, the BS generates a sequence of observation windows $\left\{t_v, t_v + T\right\}_{v \in \mathbb{Z}_+}$ where $t_v =0$, if $v=1$ and $t_v = t_{v-1} + \Delta t$, otherwise. This sequence can be seen as a sliding window with window size $T$ and step size $\Delta t$. Since in the asynchronous GFRA $T > L$, consecutive observation windows have an intersection of $T - \Delta t$ symbol intervals, this enables BS to jointly estimate the channels and detect the activity and data of all frames.

    Considering $t_n$ as the symbol interval in which the $n$-th device initiates its transmission, the channel matrix is modeled as given by

    \makeatletter%
    \if@twocolumn%
        \begin{align}\label{eq:Hasy}
            &\hspace{2cm} H = \left[
            \begin{array}{c}
                \tilde{H}^{1}_{\left(M'\times N\right)} \\
                \tilde{H}^{2}_{\left(M'\times N\right)} \\
                \vdots \\
                \tilde{H}^{T}_{\left(M'\times N\right)} \\
            \end{array}\right], \\ \nonumber
           & \tilde{h}_{m'n}^t = \left\{
            \begin{array}{rl}
                \hspace{-5pt}\sqrt{\beta_n}\,a_{m'n}^{\left(t-t_n+1\right)}, &\hspace{-2.5pt}\forall\, \left(t_{n} \leq t < t_{n} + L\right), \\
                0, &\hspace{-2.5pt} \text{otherwise.}
            \end{array}\right.
        \end{align}
    \else
        \begin{equation}\label{eq:Hasy}
            \hspace{-2.6pt}H = \left[
            \begin{array}{c}
                \tilde{H}^{1}_{\left(M'\times N\right)} \\
                \tilde{H}^{2}_{\left(M'\times N\right)} \\
                \vdots \\
                \tilde{H}^{T}_{\left(M'\times N\right)} \\
            \end{array}\right], \text{ where }
            \tilde{h}_{m'n}^t = \left\{
            \begin{array}{rl}
                \hspace{-5pt}\sqrt{\beta_n}\,a_{m'n}^{\left(t-t_n+1\right)}, &\hspace{-2.5pt}\forall\, \left(t_{n} \leq t < t_{n} + L\right), \\
                0, &\hspace{-2.5pt} \text{otherwise.}
            \end{array}\right.
        \end{equation}
    \fi
    \makeatother

    Therefore, the channel matrix that gathers the realizations of the whole observation window has dimensions $M \times N$, where $M = M' \times T$.

    In our work, we consider a block fading model, where a channel realization is constant over a transmission frame duration and changes independently from frame to frame.
    The channel matrix gathers independent fast fading, geometric attenuation and log-normal shadow fading at the observation window. Referring to the $t$-th symbol interval, matrix $\mathbf{A}^t$ represents the fast fading coefficients circularly symmetric complex Gaussian distributed with zero mean and unit variance. The channel variance of each device is given by $\beta_n$ and represents the path-loss and shadowing component, which depends on the device's location and remains the same for all frames transmitted by the $n$-th device. Thus, for an arbitrary observation window, the received matrix $\bY \in \mathbb{C}^{M\times T}$ that gathers the received signals is given by

    \begin{align}\label{eq:sig_model_asyn}
        \bY = \bH \bX + \bW
    \end{align}

    \noindent where $\bW \in \mathbb{C}^{M\times T}$ is a complex Gaussian noise matrix with $\mathcal{N}_c\left(0,\sigma_w^2\right)$ and $\bX \in \mathbb{C}^{N\times T}$ is the transmission matrix. With the massive number of MTDs requiring access, assigning orthogonal sequences to the MTDs would be impractical. Thus, in order to eliminate the need for round-trip signaling, firstly the BS broadcasts a set of non-orthogonal pilot sequences and then each active device directly transmits frames without previous scheduling~\cite{BockelmannAccess2018}. Thus, considering $L_p$ as the size of the pilot sequence, the pilot part of the frame of the $n$-th active device is composed by $\bphi^{n} = \nicefrac{\exp{\left(j \pi \boldsymbol{\alpha}\right)}}{\|\exp{\left(j \pi \boldsymbol{\alpha}\right)}\|}$, where each element of vector $\boldsymbol{\alpha} \in \mathbb{R}^{1\times L_p}$ is drawn according to a uniform distribution in the interval $\left[-1,1\right]$. After the encoding of the information bits, the data symbols of the $n$-th device are mapped into a modulation alphabet $\mathcal{A}$, such as quadrature phase shift keying (QPSK), resulting in a vector $\bd^n$ of $L_d$ symbols. Thus, the frame size of an active device is given by $L = L_p + L_d$. Therefore, for any observation window, the
    transmitted symbols that belongs to the $N \times T$ transmission matrix $\bX$ are given by
    \begin{equation}
        x_{nt} = \left\{
            \begin{array}{rl}
                \hspace{-5pt}\phi^n_{t-t_n+1}, &\hspace{-2.5pt}\forall\, \left(t_{n} \leq t < t_{n} + L_p\right), \\
                \hspace{-5pt}d^{n}_{t-(t_n+L_p)+1}, &\hspace{-2.5pt}\forall\, \left(t_{n}+L_p \leq t < t_{n} + L\right), \\
                0, &\hspace{-2.5pt} \text{otherwise.}
            \end{array}\right.
    \end{equation}

    Despite the throughput advantages commented in the previous section, this scenario is even more challenging, since it requires an even lower activity detection rate. The BS has the additional work to identify the frames that are not fully within the observation window, which harms the activity detection. Since the BS does not have the knowledge of which device is active or not, it should deal with each symbol interval independently. That is, the BS can consider a received frame only if a set of $L$ symbols in sequence, in the same observation window, are detected as active.

    \subsection{Synchronous Grant-Free Random Access}

    Common in the literature, the synchronous GFRA scenario, as depicted in Fig.~\ref{fig:asy_and_syn_mmtc}(b), allows each active device to transmit only at the beginning of a new observation window. Since in this case $L=T$, following the signal model in~(\ref{eq:sig_model_asyn}), the main difference is that the Boolean variable $\gamma$ is the same for the whole observation window, which facilitates detection. Thus, the received signals can be written as in~(\ref{eq:sig_model_asyn}), but the matrices depend only on the frame size, as $\mathbf{W} \in \mathbb{C}^{M\times L}$, $\bY \in \mathbb{C}^{M\times L}$, $\bX\in \mathbb{C}^{N\times L}$ and $\mathbf{H} \in \mathbb{C}^{M\times N}$. As in both scenarios we have a massive number of devices, and the size of the window $T$ and the frame size $L$ are smaller than $N$, which characterizes the system as overloaded. However, as seen before, $\bH$ and $\bX$ are sparse, which makes their recovery possible through the theory of compressed sensing (CS)~\cite{JWChoiCST2017}.

    \section{Problem formulation}\label{sec:prob_form}

    We formulate the problem for the asynchronous scenario since the use in the synchronous form is straightforward. To perform the joint activity, data and channel estimation, we treat the problem under the framework of Bayesian inference, which provides optimal estimation in the MSE sense via the minimum mean square error (MMSE) estimator. Following the literature, we start the formulation by marginalizing the joint distribution $\mathcal{P}(\bH,\bX,\bGamma,\bY)$, so that we can take over the expectations of $\mathcal{P}\!\left(x_{nl}|\by\right)$ and $\mathcal{P}\!\left(h_{mn}|\by\right)$. Considering $\mathcal{L}_d = \left[t_{n}+L_p, t_{n} + L\right)$ and $\mathcal{L}_p = \left[t_{n}, t_{n} + L_p\right)$, the MMSE estimates of $\bX_d$ and $\bH$ are respectively given by
    \begin{equation}\label{eq:MMSE}
        \begin{array}{rll}
        \forall\, n, t \in \mathcal{L}_d:& \hat{x}_{nt}\hspace{4pt} = \mathbb{E}\left[x_{nt}|\by\right] \\
        \forall\, m, n:& \hat{h}_{mn} = \mathbb{E}\left[h_{mn}|\by\right]
        \end{array}
    \end{equation}

    \noindent where the expectations are taken over $\mathcal{P}\!\left(x_{nt}|\by\right)$ and $\mathcal{P}\!\left(h_{mn}|\by\right)$ both of which are marginalization of $\mathcal{P}(\bH,\bX,\bGamma|\bY)$ from the joint distribution $\mathcal{P}(\bH,\bX,\bGamma,\bY)$ given by

    \makeatletter%
    \if@twocolumn%
        \begin{equation}\label{eq:ch6_jointdist}
            \begin{split}
                &\mathcal{P} \left(\bH,\bX,\bGamma|\bY\right)\\
                &\hspace{0.25cm} = \mathcal{P}\!\left(\bY|\bH,\bX,\bGamma\right) \mathcal{P}\!\left(\bX\right) \mathcal{P}\!\left(\bH|\bGamma\right) \mathcal{P}\!\left(\bGamma\right)/\mathcal{P}\!\left(\bY\right)\\
                &\hspace{0.25cm} \propto \mathcal{P}\!\left(\bY|\bH\bX\right) \mathcal{P}\!\left(\bX\right) \mathcal{P}\!\left(\bH|\bGamma\right) \mathcal{P}\!\left(\bGamma\right)\\
                &\hspace{0.25cm} = \prod^M_{m=1} \prod^T_{t=1} \mathcal{P}\!\left(y_{mt}\Big|\sum_{n=1}^{N} h_{mn}x_{nt}\right) \prod^N_{n=1} \prod^T_{t=1} \mathcal{P}\!\left(x_{nt}\right) \times\\
                &\hspace{0.40cm} \prod^M_{m=1} \prod^N_{n=1} \prod^T_{t=1} \mathcal{P}\!\left(h_{mn}|\gamma_{nt}\right)  \prod^N_{n=1}\prod^T_{t=1} \mathcal{P}\!\left(\gamma_{nt}\right),
            \end{split}
        \end{equation}
    \else
        \begin{equation}\label{eq:ch6_jointdist}
            \begin{split}
                &\mathcal{P} \left(\bH,\bX,\bGamma|\bY\right)\\
                &\hspace{0.5cm} = \mathcal{P}\!\left(\bY|\bH,\bX,\bGamma\right) \mathcal{P}\!\left(\bX\right) \mathcal{P}\!\left(\bH|\bGamma\right) \mathcal{P}\!\left(\bGamma\right)/\mathcal{P}\!\left(\bY\right)\\
                &\hspace{0.5cm} \propto \mathcal{P}\!\left(\bY|\bH\bX\right) \mathcal{P}\!\left(\bX\right) \mathcal{P}\!\left(\bH|\bGamma\right) \mathcal{P}\!\left(\bGamma\right)\\
                &\hspace{0.5cm} = \prod^M_{m=1} \prod^T_{t=1} \mathcal{P}\!\left(y_{mt}\Big|\sum_{n=1}^{N} h_{mn}x_{nt}\right) \prod^N_{n=1} \prod^T_{t=1} \mathcal{P}\!\left(x_{nt}\right) \prod^M_{m=1} \prod^N_{n=1} \prod^T_{t=1} \mathcal{P}\!\left(h_{mn}|\gamma_{nt}\right)  \prod^N_{n=1}\prod^T_{t=1} \mathcal{P}\!\left(\gamma_{nt}\right),             %
            \end{split}
        \end{equation}
    \fi
    \makeatother

    \noindent where the normalization to unit area is omitted. Moreover, the transition distribution are separable as $\bsfz_{mt} = \sum_{n=1}^N \bsfh_{mn}\bsfx_{nt}$ with $\bsfZ= \bsfH \bsfX$. Since one of the goals is to decode the data symbols, the transmitted signal also depends on code symbols $\bc_t$ and $\bs_t$ activity variables, as given by

    \makeatletter%
    \if@twocolumn
        \begin{equation}\label{eq:px_all}
            \begin{split}
                & \mathcal{P}\left(\bX\right) =\\
                & \prod_{n=1}^N\prod_{t\in \mathcal{L}_p} \mathcal{P}_{\bsfx_p} \left(x_{nt}\right) \sum_{s_{nt}} \sum_{c_{nt}} \prod_{n=1}^{N}\prod_{t \in \mathcal{L}_d} \mathcal{P}_{\bsfx_d}\left(x_{nt},c_{nt},s_{nt}\right),
            \end{split}
        \end{equation}
    \else
        \begin{equation}\label{eq:px_all}
            \mathcal{P}\left(\bX\right) = \prod_{n=1}^N\prod_{t\in \mathcal{L}_p} \mathcal{P}_{\bsfx_p} \left(x_{nt}\right) \sum_{s_{nt}} \sum_{c_{nt}} \prod_{n=1}^{N}\prod_{t \in \mathcal{L}_d} \mathcal{P}_{\bsfx_d}\left(x_{nt},c_{nt},s_{nt}\right),
        \end{equation}
    \fi
    \makeatother

    \noindent where $c_{nt}$ and $s_{nt}$  $\in \left\{0,1\right\}$. Accordingly, the MSE of those MMSE estimators are presented by $\textsc{mse}\left(\bX_d\right) = \left(\nicefrac{1}{N L_d}\right) \mathbb{E}\left[\|\hat{\bX}_d - \bX_d\|^2_\text{F}\right]$ and $\textsc{mse}\left(\bH\right) = \left(\nicefrac{1}{M N}\right) \mathbb{E}\left[\|\hat{\bH} - \bH\|^2_\text{F}\right]$. Besides, the activity of the $n$-th device is decided by the log-likelihood ratio (LLR) as described by
    \def\useanchorwidth{T}
    \begin{equation}\label{eq:hyp}
        \text{LLR}(s_{nt}) = \log \frac{\mathcal{P}\left(\gamma_{nt} = 1 | \bY\right)}{\mathcal{P}\left(\gamma_{nt} = 0 | \bY\right)}\, \mathop{\lessgtr}^{H_0}_{H_1}\, 0,
    \end{equation}

    \noindent where $\mathcal{P}\left(\gamma_{nt}|\bY\right)$ is marginalization of $\mathcal{P}\!\left(\bH,\bX,\bGamma|\bY\right)$ and the hypothesis $H_0$ and $H_1$ are about the $n$-th device activity.
        \begin{figure*}[t]
            \centering
            \centerline{\includegraphics[scale=0.5]{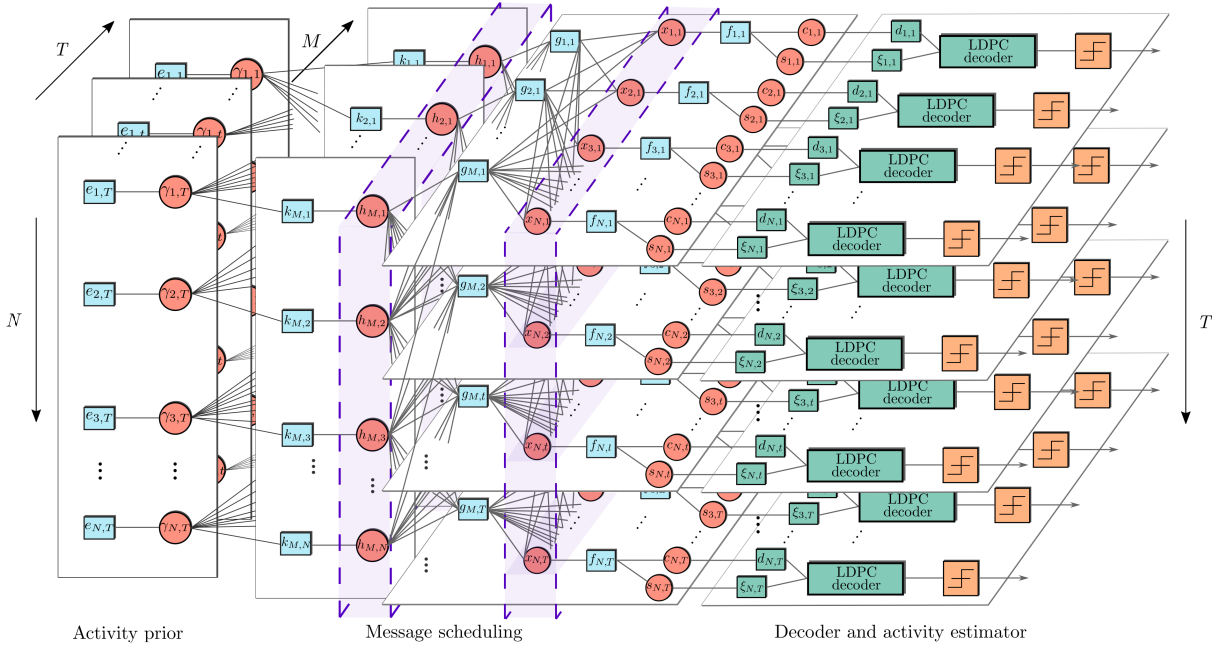}}
            \caption{Factor graph of the asynchronous problem. Rectangles represents factor nodes while spheres are the variable nodes.}
            \label{fig:FG}%
        \end{figure*}

    The MMSE estimators in~(\ref{eq:MMSE}) and the hypothesis test in~(\ref{eq:hyp}) involve multi-dimensional integrals due to the marginalization of the posterior distribution. Thus, exact message passing based on the sum-product rule is too complicated to implement, while the computational cost is impractical for the mMTC scenario. Thus, an alternative applied in recent years~\cite{TDingTWC2019,SRanganTSP2017,HIimoriTWC2021} is to approximate these quantities using loopy belief propagation (LBP)~\cite{FreyMIT1997}. In LBP, beliefs about the random variables are exchanged between the nodes of the factor graph until they converge. Those beliefs are in the form of probability density functions (pdfs) or logarithm pdfs (log-pdfs) and are computed using the sum-product algorithm (SPA)~\cite{Kschischang2001}. The procedure of SPA is that the belief sent by a variable node along a edge of the graph is computed as the integral of the product of the factor associated with that node and the incoming beliefs on all other edges. The product of all beliefs impinging on a given variable node yields the posterior pdf for that variable. To this end, we propose an efficient algorithm that incorporates the channel coding into the message-passing approach and uses specific message-scheduling schemes that dramatically reduce the computational cost.

    \section{Proposed BiMSGAMP Structure} \label{sec:jointAUDCEDATA}

    In this section a novel and low-complexity joint estimation, detection and decoding algorithm is proposed. Firstly presented in~\cite{JTParkerTSP2014}, we use the bilinear FG model to exchange messages between function and variable nodes and then compute the \textit{a posteriori} densities. Drawing inspiration from~\cite{Rangan2012}     and~\cite{FMonseesPhD2017}, the problem is divided in three parts, the activity prior, where a loopy belief propagation (LBP) part is considered, the channel estimation and symbol detector block, where the message scheduling is applied, and the decoder and activity detector block, as depicted in Fig.~\ref{fig:FG}. Unlike existing works, based on 5G channel coding techniques, we propose a low-complexity complete solution, that exploits the fact that the decoding of Low-Density Parity-Check (LDPC) codes is also based on message-passing and devise a solution that uses the channel decoder beliefs to refine the activity detection and data decoding.
    In the factor graph (FG), rectangles denote factor nodes $\{e_{nt}\}$, $\{k_{mn}\}$, $\{g_{mt}\}$, $\{f_{nt}\}$ corresponding to the marginal \textit{a priori} distributions and equality constraints while spheres $\{\gamma_{nt}\}$, $\{h_{mn}\}$, $\{x_{nt}\}$, $\{c_{nt}\}$, $\{s_{nt}\}$ are the variable nodes. The belief propagation (BP) framework consists of $T$ multiuser detectors computing probabilistic information about the symbols $x_{nt}$ in the vectors $\bx_t$. This information is exchanged between the $T$ detectors and processed by the detector and decoder blocks. The function nodes $f_{nt}$ with the variables $c_{nt}$ and $s_{nt}$\footnote{Despite the fact that the joint distribution in~(\ref{eq:ch6_jointdist}) explicitly shows the activity indicators $\gamma_{nt}$ since the algorithm has previous knowledge of it, this quantity is estimated by the variable $s_{nt}$.} are the connecting points for the channel decoder and the symbol detector block.

    \subsection{Activity prior}\label{subsection:actprior}

    Originally from~\cite{DiRennaWCL2021} and~\cite{SRanganTSP2017}, the LBP part is included to provide an initial activity detection, which is further estimated by the $s_{nt}$ variables. During the channel estimation phase, BiMSGAMP uses this activity detection to refine the means and variances of the channels, beyond defining the message scheduling. Computed using Gaussian approximations of likelihood functions, these estimates are then used to define the message scheduling strategies proposed in this work. Let $k_{mn}\left(h_{mn}|\gamma_{nt}\right) = \mathcal{P}\left(h_{mn}| \gamma_{nt}\right)$ to $\gamma_{nt}$, the messages in the activity prior stage are given by

    \makeatletter%
    \if@twocolumn
        \begin{equation}\label{eq:LBP1}
        \resizebox{.9\hsize}{!}{$
            \Delta^{i+1}_{k_{mn} \rightarrow \gamma_{nt}}\left(\gamma_{nt}\right) \propto \int k_{mn}\left(h_{mn}|\gamma_{nt}\right)\! \Delta^{i}_{k_{mn} \rightarrow \gamma_{nt}}\left(h\right)\, \text{d}h_{mn},
        $}
        \end{equation}

        \vspace{-0.5cm}
        \begin{align} \label{eq:LBP2}
            \Delta^{i+1}_{\gamma_{nt} \rightarrow k_{mn}}\left(\gamma_{nt}\right) \propto&\, \mathcal{P}\!\left(\gamma_{nt}\right)\! \prod_{\underset{l=M'(t-1)+1}{l \neq m}}^{M't}\! \Delta^{i+1}_{k_{ln} \rightarrow \gamma_{nt}}\left(\gamma_{nt}\right),
        \end{align}
    \else
        \begin{equation}\label{eq:LBP1}
            \Delta^{i+1}_{k_{mn} \rightarrow \gamma_{nt}}\left(\gamma_{nt}\right) \propto \int k_{mn}\left(h_{mn}|\gamma_{nt}\right)\! \Delta^{i}_{k_{mn} \rightarrow \gamma_{nt}}\left(h\right)\, \text{d}h_{mn},
        \end{equation}
        \vspace{-0.5cm}
        \begin{align} \label{eq:LBP2}
            \Delta^{i+1}_{\gamma_{nt} \rightarrow k_{mn}}\left(\gamma_{nt}\right) \propto&\, \mathcal{P}\!\left(\gamma_{nt}\right)\! \prod_{\underset{l=M'(t-1)+1}{l \neq m}}^{M't}\! \Delta^{i+1}_{k_{ln} \rightarrow \gamma_{nt}}\left(\gamma_{nt}\right),
        \end{align}
    \fi
    \makeatother

    \noindent where~(\ref{eq:LBP1}) refers to the message from $\mathcal{P}\!\left(h_{mn}|\gamma_{nt}\right)$ to $\gamma_{nt}$ while (\ref{eq:LBP2}) denotes the message in opposite direction and each belief at $\gamma_{nt}$ is given by
    \begin{equation}
        \Delta^{i}_{\gamma_{nt}}\left(\gamma_{nt}\right) \propto \mathcal{P}\!\left(\gamma_{nt}\right)\!\prod_{m=M'(t-1)+1}^{M't}\! \Delta^{i}_{k_{mn} \rightarrow \gamma_{nt}}(\gamma_{nt}).
    \end{equation}

    In order to perform the activity detection in the channel estimation phase, similarly as in~\cite{SRanganTSP2017}, the likelihood functions are given by
    \begin{equation}\label{eq:LLR_prior}
        L_{k_{mn}\rightarrow \gamma_{nt}}^{i+1} := \log \frac{\Delta^i_{k_{mn}\rightarrow \gamma_{nt}}\left(\gamma_{nt}=1\right)}{\Delta^i_{k_{mn}\rightarrow \gamma_{nt}}\left(\gamma_{nt}=0\right)},
    \end{equation}

    \noindent similarly, $L_{\gamma_{nt}\rightarrow k_{mn}}^{i+1} := \log \frac{\gamma_{nt}\rightarrow \Delta^i_{k_{mn}}\left(\gamma_{nt}=1\right)}{\gamma_{nt}\rightarrow \Delta^i_{k_{mn}}\left(\gamma_{nt}=0\right)}$ and $L_{\gamma_{nt}}^{i+1} := \log \frac{\gamma_{nt}\left(\gamma_{nt}=1\right)}{\gamma_{nt}\left(\gamma_{nt}=0\right)}$. Substituting~(\ref{eq:LLR_prior}) in~(\ref{eq:LBP2}), we have
    \begin{align}\label{eq:LLRnleftnm2}
        L_{\gamma_{nt}\rightarrow k_{mn}}^{i+1} = \log \left(\frac{\rho_n}{1- \rho_n}\right) + \sum_{k\neq m} L_{k_{kn}\rightarrow \gamma_{nt}}^{i+1},
    \end{align}

    \noindent where one can see that we consider that the BS has the knowledge of the probability of being active of each device. The previous LLR is used to estimate the activity of devices, as given by
    \makeatletter%
    \if@twocolumn
        \begin{equation}\label{eq:est_rho}
        \resizebox{.88\hsize}{!}{$
            \hat{\rho}_{mnt}^i := \Delta^{i+1}_{\gamma_{nt} \rightarrow k_{mn}}\left(\gamma_{nt} = 1\right) = 1 - \left(1+\exp{\left(L^i_{\gamma_{nt}\rightarrow k_{mn}}\right)}\right)^{-1}
        $}.
        \end{equation}
    \else
        \begin{equation}\label{eq:est_rho}
            \hat{\rho}_{mnt}^i := \Delta^{i+1}_{\gamma_{nt} \rightarrow k_{mn}}\left(\gamma_{nt} = 1\right) = 1 - \frac{1}{1+\exp{\left(L^i_{\gamma_{nt}\rightarrow k_{mn}}\right)}}.
        \end{equation}
    \fi
    \makeatother

    Since $m$  represents the BS antennas, we can proceed with the mean value of $\hat{\rho}_{mnt}$, $\hat{\rho}_{nt}$ in relation to $m$. Thus, with the messages associated to the activity prior in the system, the next step is the channel estimation and data detection part.

    \subsection{Channel estimation, activity and data detection}
    To compute the a posteriori densities within such graph, messages between function and variable nodes have to be exchanged. Applying the sum product algorithm to the FG, the underlying iterative update equations at iteration number $i$ summarize the messages from functions to variable nodes
    \makeatletter%
    \if@twocolumn
        \begin{figure*}
            \begin{equation}\label{eq:gtoh}
                \Delta^{i+1}_{g_{mt}\rightarrow h_{mn}}\left(h_{mn}\right) =  \int_{\left\{h_{mr}\right\}_{r\neq n},\left\{x_{kt}\right\}_{k=1}^{N}} g_{mt}\left(y_{mt}|z_{mt}\right)\,
                \prod_{k=1}^N \Delta^i_{x_{kt}\rightarrow g_{mt}}\left(x_{kt}\right) \,
                \prod_{r \neq n} \Delta^i_{h_{mr}\rightarrow g_{mt}}\left(h_{mr}\right)
            \end{equation}
            \begin{equation}\label{eq:gtox}
                \Delta^{i+1}_{g_{mt}\rightarrow x_{nt}}\left(x_{nt}\right) = \int_{\left\{h_{mk}\right\}_{k=1}^{N},\left\{x_{rt}\right\}_{r\neq n}} g_{mt}\left(y_{mt}|z_{mt}\right)\,
                \prod_{r \neq n} \Delta^i_{x_{rt}\rightarrow g_{mt}}\left(x_{rl}\right)\,
                \prod_{k=1}^N \Delta^i_{h_{mk}\rightarrow g_{mt}}\left(h_{mk}\right)
            \end{equation}
            \hrulefill
            \vspace*{4pt}
        \end{figure*}
    \else
        \begin{equation}\label{eq:gtoh}
            \Delta^{i+1}_{g_{mt}\rightarrow h_{mn}}\left(h_{mn}\right) =  \int_{\left\{h_{mr}\right\}_{r\neq n},\left\{x_{kt}\right\}_{k=1}^{N}} g_{mt}\left(y_{mt}|z_{mt}\right)\,
            \prod_{k=1}^N \Delta^i_{x_{kt}\rightarrow g_{mt}}\left(x_{kt}\right) \,
            \prod_{r \neq n} \Delta^i_{h_{mr}\rightarrow g_{mt}}\left(h_{mr}\right)
        \end{equation}
        \begin{equation}\label{eq:gtox}
            \Delta^{i+1}_{g_{mt}\rightarrow x_{nt}}\left(x_{nt}\right) = \int_{\left\{h_{mk}\right\}_{k=1}^{N},\left\{x_{rt}\right\}_{r\neq n}} g_{mt}\left(y_{mt}|z_{mt}\right)\,
            \prod_{r \neq n} \Delta^i_{x_{rt}\rightarrow g_{mt}}\left(x_{rl}\right)\,
            \prod_{k=1}^N \Delta^i_{h_{mk}\rightarrow g_{mt}}\left(h_{mk}\right)
        \end{equation}
    \fi
    \makeatother

    \noindent where $g_{mt}\left(y_{mt}|z_{mt}\right) = \mathcal{P}\!\left(y_{mt}\big|\sum_{k=1}^{N}h_{mk}x_{kt}\right)$ and the messages from variable to factor nodes
    \makeatletter%
    \if@twocolumn
        \begin{equation}\label{eq:htog}
            \resizebox{.89\hsize}{!}{$
                \Delta^{i+1}_{h_{mn}\rightarrow g_{mt}}\left(h_{mn}\right) = \underbrace{k_{mn}\left(h_{mn}|\gamma_{nt}\right)}_{\Delta_{k_{mn}\rightarrow h_{mn}\left(h_{mn}\right)}} \prod_{p \neq l} \Delta^i_{g_{mp}\rightarrow h_{mn}}\left(h_{mn}\right)
            $}
        \end{equation}

        \begin{equation}\label{eq:xtoh}
            \resizebox{.89\hsize}{!}{$
            \Delta^{i+1}_{x_{nt}\rightarrow g_{mt}}\left(x_{nt}\right) =   \underbrace{f_{nt}\left(x_{nt},c_{nt},s_{nt}\right)}_{\Delta_{f_{nt}\rightarrow x_{nt}\left(x_{nt}\right)}} \prod_{p \neq m} \Delta^i_{g_{pl}\rightarrow x_{nt}}\left(x_{nt}\right).
            $}
        \end{equation}
    \else
        \begin{equation}\label{eq:htog}
            \Delta^{i+1}_{h_{mn}\rightarrow g_{mt}}\left(h_{mn}\right) =  \underbrace{k_{mn}\left(h_{mn}|\gamma_{nt}\right)}_{\Delta_{k_{mn}\rightarrow h_{mn}\left(h_{mn}\right)}} \prod_{p \neq l} \Delta^i_{g_{mp}\rightarrow h_{mn}}\left(h_{mn}\right)
        \end{equation}

        \begin{equation}\label{eq:xtoh}
            \Delta^{i+1}_{x_{nt}\rightarrow g_{mt}}\left(x_{nt}\right) =   \underbrace{f_{nt}\left(x_{nt},c_{nt},s_{nt}\right)}_{\Delta_{f_{nt}\rightarrow x_{nt}\left(x_{nt}\right)}} \prod_{p \neq m} \Delta^i_{g_{pl}\rightarrow x_{nt}}\left(x_{nt}\right).
        \end{equation}
    \fi
    \makeatother

    The resulting frame belief propagation (BP) consists of $T$ multiuser detectors yielding probabilistic information about the symbols $x_{nt}$ in the vectors $\bx_t$. The information is then processed within a joint estimation block and a decoder block. Both blocks exchange extrinsic information with $T$ multiuser detectors. The function nodes $f_{nt}$ with the variables $c_{nt}$ and $s_{nt}$ is the connecting point for the channel decoder and the joint estimation block. Therefore, in a BPSK scenario, we now use the description of the function node according to
    \makeatletter%
    \if@twocolumn
        \begin{align}
            &f_{nt}\left(x_{nt}, c_{nt}, s_{nt}\right) =\\ \nonumber
            &\resizebox{.98\hsize}{!}{$
            s_{nt} \left[\,c_{nt}\, \delta\left(x_{nt} - 1\right) + \left(1 - c_{nt}\right) \delta \left(x_{nt} +1\right)\right] + \left(1- s_{nt}\right) \delta\left(x_{nt}\right).
            $}
        \end{align}
    \else
        \begin{equation}
            f_{nt}\left(x_{nt}, c_{nt}, s_{nt}\right) = \, s_{nt} \left[\,c_{nt}\, \delta\left(x_{nt} - 1\right) + \left(1 - c_{nt}\right) \delta \left(x_{nt} +1\right)\right] + \left(1- s_{nt}\right) \delta\left(x_{nt}\right).
        \end{equation}
    \fi
    \makeatother

    The function $f_{nt}$ summarizes the connection between variables $\left\{x_{nt}, c_{nt}, s_{nt}\right\}$ accounts for their probabilistic dependencies and can be seen as a check node that is zero for any invalid combination of the variables involved. As an example, for the prior probability of $x_{nt}$ and considering $\rho_{nt}$ as the probability of being active of the $n$-th device, we have $s_{nt} = \rho_{nt}$ and $c_{nt} = 0.5$, then the function node is
    \makeatletter%
    \if@twocolumn
        \begin{align}
            &f_{nt}\left(x_{nt}\right) =\\ \nonumber
            &\rho_{nt} \left[\frac{1}{2}\, \delta\left(x_{nt} -1\right) + \frac{1}{2}\, \delta \left(x_{nt} +1\right)\right] + \left(1-\rho_{nt}\right) \delta(x_{nt}).
        \end{align}
    \else
        \begin{equation}
            f_{nt}\left(x_{nt}\right) = \rho_{nt} \left[\frac{1}{2}\, \delta\left(x_{nt} -1\right) + \frac{1}{2}\, \delta \left(x_{nt} +1\right)\right] + \left(1-\rho_{nt}\right) \delta(x_{nt}).
        \end{equation}
    \fi
    \makeatother

    To process the code bits $c_{nt}$ and the hidden activity variable $s_{nt}$, we require two new function nodes corresponding to the decoder and activity detector. These nodes are subsequently denoted as $d_{nt}$ and $\xi_{nt}$. Since that channel code and activity states are node specific, the branch corresponding to one node connects to one function $d_{nt}$ and $\xi_{nt}$ only. Specifically, we follow the GAMP algorithm~\cite{Rangan2012}. The message passed between the nodes are approximated by Gaussian distributions, so that only the means and variances of the messages are involved in message exchanges. We next outline each step by following the GAMP algorithms~\cite{Rangan2012, JTParkerTSP2014}. In order to make this work self-contained, the detailed derivations of the messages of interest are in Appendix~\ref{app:B}.

    Based on the linear model, $z_{mt} = \sum_{n=1}^{N} h_{mn}x_{nt}$, the messages of $y_{mt}$ are accumulated to obtain an estimate of $z_{mt}$. With the ``Onsager'' correction applied, the messages in~(\ref{eq:gtoh}) and~(\ref{eq:gtox}), in the form of means $\hat{p}^{i}_{mt}$ and variances $\nu^{pi}_{mt}$, for all $m, n$ and $t$, are computed as~\cite{JTParkerTSP2014}
    \begin{align}\label{eq:p_var}
        \nu^{pi}_{mt} \triangleq &\,\sum^N_{k=1}\, |\hat{h}^i_{mn}|^2\, \nu^{xi}_{nt} + \nu^{hi}_{nt}\, |\hat{x}^i_{nt}|^2 + \nu^{hi}_{nt}\,\nu^{xi}_{nt}, \\ \label{eq:p_hat}
        \hat{p}^i_{mt} \triangleq &\, \sum^N_{k=1}\, \hat{h}^i_{mn}\, \hat{x}^i_{nt} - \hat{s}_{mt}^{i-1} \left(|\hat{h}^i_{mn}|^2\, \nu^{xi}_{nt} + \nu^{hi}_{nt}\, |\hat{x}^i_{nt}|^2\right),
    \end{align}

    \noindent where we initially set $\hat{s}_{mt} = 0,\, \forall t$. Then, the means $\hat{z}^i_{mt}$ and variances $\nu^{zi}_{mt}$ are computed by using the observations $\hat{r}^i_{nt}$ and $\hat{q}^i_{mn}$ as
    \makeatletter%
    \if@twocolumn
        %
        \begin{equation}\label{eq:z_var}
            \nu^{zi}_{mt} = \, \textsc{var}\left\{\bsfz_{mt}| \bsfp_{mt} = \hat{p}^i_{mt}, \nu^{pi}_{mt}\right\},
        \end{equation}

        \vspace{-0.25cm}
        \begin{equation}\label{eq:z_hat}
               \text{and \hspace{0.1cm} }\, \hat{z}^{i}_{mt} = \, \mathbb{E}\left\{\bsfz_{mt}| \bsfp_{mt} = \hat{p}^i_{mt}, \nu^{pi}_{mt}\right\}.
        \end{equation}
    \else
    \vspace{0.3cm}
        \begin{minipage}{0.45\linewidth}
            \begin{equation}\label{eq:z_var}
            \nu^{zi}_{mt} = \, \textsc{var}\left\{\bsfz_{mt}| \bsfp_{mt} = \hat{p}^i_{mt}, \nu^{pi}_{mt}\right\},
            \end{equation}
        \end{minipage}
        \begin{minipage}{0.5\linewidth}
            \begin{equation}\label{eq:z_hat}
                \text{and \hspace{0.1cm} }\, \hat{z}^{i}_{mt} = \, \mathbb{E}\left\{\bsfz_{mt}| \bsfp_{mt} = \hat{p}^i_{mt}, \nu^{pi}_{mt}\right\}.
            \end{equation}
        \end{minipage}
    \fi
    \makeatother

    \noindent where the mean and variance operations are taken with respect to the \textit{a posteriori} distribution of $z_{mt}$ given the \textit{a priori} distribution $z_{mt} \sim \mathcal{N}_c\left(\hat{p}_{mt}^i,\nu^{pi}_{mt}\right)$ and the observation $\hat{r}^i_{nt} = z_{mt}+w_{mt}$. Lastly, the residual $\hat{s}^i_{mt}$ and the inverse-residual-variances $\nu^{si}_{mt}$ are computed by\\

    \begin{minipage}{0.45\linewidth}
        \begin{equation}\label{eq:s_var}
            \nu^{si}_{mt} = \, \left(\frac{1-\nu^{zi}_{mt}}{\nu^{pi}_{mt}}\right) \frac{1}{\nu^{pi}_{mt}},
        \end{equation}
    \end{minipage}
    \begin{minipage}{0.5\linewidth}
        \begin{equation}\label{eq:s_hat}
            \text{and \hspace{0.1cm}}\, \hat{s}^{i}_{mt} = \, \frac{\left(\hat{z}^i_{mt}-\hat{p}^i_{mt}\right)}{\nu^{pi}_{mt}}.
        \end{equation}
    \end{minipage}

    \vspace{0.35cm}
    With the quantities $\hat{s}^{i}_{mt}$ and $\nu^{si}_{mt}$ computed, the means and variances derived from the messages\\ $\Delta^{i+1}_{h_{mn}\rightarrow g_{mt}}\left(h_{mn}\right)$, used to estimate the channels are given by
    \makeatletter%
    \if@twocolumn
        %
        \begin{equation}\label{eq:hvar}
            \nu^{h\,i+1}_{mn} = \, \textsc{var}\left\{\bsfh_{mn}|\bsfq_{mn}=\hat{q}_{mn}^i, \nu^{qi}_{mn}\right\}
            \end{equation}

        \vspace{-0.25cm}
        \begin{equation}\label{eq:hhat}
                \text{and \hspace{0.1cm} }\, \hat{h}^{i+1}_{mn} =\, \mathbb{E}\left\{\bsfh_{mn}|\bsfq_{mn}=\hat{q}_{mn}^i, \nu^{qi}_{mn}; \hat{\rho}_{mnt}\right\},
            \end{equation}
    \else
        \vspace{0.3cm}
        \begin{minipage}{0.45\linewidth}
            \begin{equation}\label{eq:hvar}
            \nu^{h\,i+1}_{mn} = \, \textsc{var}\left\{\bsfh_{mn}|\bsfq_{mn}=\hat{q}_{mn}^i, \nu^{qi}_{mn}\right\}
            \end{equation}
        \end{minipage}
        \begin{minipage}{0.5\linewidth}
            \begin{equation}\label{eq:hhat}
                \text{and \hspace{0.1cm} }\, \hat{h}^{i+1}_{mn} =\, \mathbb{E}\left\{\bsfh_{mn}|\bsfq_{mn}=\hat{q}_{mn}^i, \nu^{qi}_{mn}; \hat{\rho}_{mnt}\right\},
            \end{equation}
        \end{minipage}
    \fi
    \makeatother

    \vspace{0.3cm}
    \noindent where $\nu^{qi}_{mn}$ and $\hat{q}^i_{mn}$ are

    \begin{equation} \label{eq:qvar}
        \nu^{qi}_{mn} =\left[\nu^{si}_{mt}\left(\sum_{t \in \mathcal{L}_p} |x_{nt}|^2 + \sum_{t \in \mathcal{L}_d} |\hat{x}^{i}_{nt}|^2 \right)\right]^{-1}
    \end{equation}

    \makeatletter%
    \if@twocolumn
        \begin{align} \label{eq:qhat}
            &\hat{q}^i_{mn} = \hat{h}^{i}_{mn} \left(\!1 - \nu^{qi}_{mn} \sum_{t \in \mathcal{L}_d} \nu^{xi}_{nt}\, \nu^{si}_{mt}\right) +\\ \nonumber
            &\hspace{2cm} \nu^{qi}_{mn} \left(\sum_{t \in \mathcal{L}_p} x^{i\ast}_{nt}\,\hat{s}^i_{mt} + \sum_{t \in \mathcal{L}_d} \hat{x}^{i\ast}_{nt}\,\hat{s}^i_{mt}\right).
        \end{align}
    \else
        \begin{equation} \label{eq:qhat}
            \hat{q}^i_{mn} = \hat{h}^{i}_{mn} \left(\!1 - \nu^{qi}_{mn} \sum_{t \in \mathcal{L}_d} \nu^{xi}_{nt}\, \nu^{si}_{mt}\right) + \nu^{qi}_{mn} \left(\sum_{t \in \mathcal{L}_p} x^{i\ast}_{nt}\,\hat{s}^i_{mt} + \sum_{t \in \mathcal{L}_d} \hat{x}^{i\ast}_{nt}\,\hat{s}^i_{mt}\right).
        \end{equation}
    \fi
    \makeatother

    Similarly, for the data we have
    \makeatletter%
    \if@twocolumn
        \begin{equation}\label{eq:xvar}
                \nu^{x\,i+1}_{nt} =\, \textsc{var}\left\{\bsfx_{nt}|\bsfr_{nt}=\hat{r}_{nt}^i,\, \nu^{ri}_{nt}\right\},
        \end{equation}

        \vspace{-0.3cm}
        \begin{equation}\label{eq:xhat}
            \text{and \hspace{0.1cm} }\, \hat{x}^{i+1}_{nt} =\, \mathbb{E}\left\{\bsfx_{nt}|\bsfr_{nt}=\hat{r}_{nt}^i,\, \nu^{ri}_{nt}\right\},
        \end{equation}
    \else
        \begin{minipage}{0.45\linewidth}
            \begin{equation}\label{eq:xvar}
                \nu^{x\,i+1}_{nt} =\, \textsc{var}\left\{\bsfx_{nt}|\bsfr_{nt}=\hat{r}_{nt}^i,\, \nu^{ri}_{nt}\right\}
            \end{equation}
        \end{minipage}
        \begin{minipage}{0.45\linewidth}
            \begin{equation}\label{eq:xhat}
            \text{and \hspace{0.1cm} }\, \hat{x}^{i+1}_{nt} =\, \mathbb{E}\left\{\bsfx_{nt}|\bsfr_{nt}=\hat{r}_{nt}^i,\, \nu^{ri}_{nt}\right\}
            \end{equation}
        \end{minipage}
    \fi
    \makeatother

    \vspace{0.25cm}
    \noindent where $\nu^{ri}_{nt}$ and $\hat{r}^i_{nt}$ are\\
    \makeatletter%
    \if@twocolumn
        \begin{equation}\label{eq:rvar}
            \nu^{ri}_{nt} = \left(\sum_{m=1}^M \left(\hat{h}_{mn}^{i}\right)^2 \!\nu^{si}_{mt}\right)^{-1} \text{ and}
        \end{equation}
        \begin{equation}\label{eq:rhat}
            \hat{r}^i_{nt} = \hat{x}^{i}_{nt} \left(\!1 - \nu^{ri}_{nt} \sum_{m=1}^{M} \nu^{hi}_{mn}\, \nu^{si}_{mt}\right)\! + \nu^{ri}_{nt} \sum_{m=1}^{M} \hat{h}^{i\ast}_{mn}\,\hat{s}^i_{mt}.
        \end{equation}
    \else
        \begin{minipage}{0.35\linewidth}
            \begin{equation}\label{eq:rvar}
                \nu^{ri}_{nt} = \left(\sum_{m=1}^M \left(\hat{h}_{mn}^{i}\right)^2 \!\nu^{si}_{mt}\right)^{-1}
            \end{equation}
        \end{minipage}
        \begin{minipage}{0.6\linewidth}
            \begin{equation}\label{eq:rhat}
                \text{and \hspace{0.1cm} }\, \hat{r}^i_{nt} = \hat{x}^{i}_{nt} \left(\!1 - \nu^{ri}_{nt} \sum_{m=1}^{M} \nu^{hi}_{mn}\, \nu^{si}_{mt}\right)\! + \nu^{ri}_{nt} \sum_{m=1}^{M} \hat{h}^{i\ast}_{mn}\,\hat{s}^i_{mt}.
            \end{equation}
        \end{minipage}
    \fi
    \makeatother

    Naturally, these means and variances are approximated values. With the messages based on the GAMP algorithm defined, we describe the messages from the factor node $f_{nt}$ to the decoder and activity estimator.

    \subsection{Decoder and activity estimator}

    According to the general sum-product update rules the corresponding beliefs have to be multiplied point-wise and marginalized. Furthermore, we can directly express the messages from the multiuser detector to the code symbol $\Delta_{f_{nt}\rightarrow c_{nt}}(c_{nt})$ and the message from the code symbol to the decoder function $\Delta_{c_{nt} \rightarrow d_{nt}} (c_{nt})$. However, as the variable code $c_{nt}$ has only two connections, the output message equals the input message. Therefore, we can directly give the message from the multiuser detector to the decoder as
    \makeatletter%
    \if@twocolumn
        \begin{align}
            &\Delta^{i+1}_{f_{nt} \rightarrow d_{nt}} \left(c_{nt}\right) \propto\\ \nonumber &\sum_{x_{nt},s_{nt}} f_{nt} \left(x_{nt}, c_{nt}, s_{nt}\right) \Delta^i_{\xi_{nt} \rightarrow f_{nt}}\left(s_{nt}\right) \prod_{m=1}^{M} \Delta^i_{g_{mt} \rightarrow x_{nt}}\left(x_{nt}\right),
        \end{align}
    \else
        \begin{equation}
            \Delta^{i+1}_{f_{nt} \rightarrow d_{nt}} \left(c_{nt}\right) \propto \sum_{x_{nt},s_{nt}} f_{nt} \left(x_{nt}, c_{nt}, s_{nt}\right) \Delta^i_{\xi_{nt} \rightarrow f_{nt}}\left(s_{nt}\right) \prod_{m=1}^{M} \Delta^i_{g_{mt} \rightarrow x_{nt}}\left(x_{nt}\right)
        \end{equation}
    \fi
    \makeatother

    \noindent where we can see that the message to the decoder already contains information from the activity detector and from the likelihood function. Likewise, the message from the multiuser detector to the activity detector can also directly be formulated as
    \makeatletter%
    \if@twocolumn
        \begin{align}
            &\Delta^{i+1}_{f_{nt} \rightarrow \xi_{nt}} \left(s_{nt}\right) \propto\\ \nonumber &\sum_{x_{nt},c_{nt}}\! f_{nt} \left(x_{nt}, c_{nt}, s_{nt}\right) \Delta^i_{d_{nt} \rightarrow f_{nt}}\!\left(c_{nt}\right) \prod_{m=1}^{M} \Delta^i_{g_{mt} \rightarrow x_{nt}}\left(x_{nt}\right).
        \end{align}
    \else
        \begin{equation}
            \Delta^{i+1}_{f_{nt} \rightarrow \xi_{nt}} \left(s_{nt}\right) \propto \sum_{x_{nt},c_{nt}} f_{nt} \left(x_{nt}, c_{nt}, s_{nt}\right) \Delta^i_{d_{nt} \rightarrow f_{nt}}\left(c_{nt}\right) \prod_{m=1}^{M} \Delta^i_{g_{mt} \rightarrow x_{nt}}\left(x_{nt}\right).
        \end{equation}
    \fi
    \makeatother

    The messages from the function node $f_{nt}$ to the likelihood factor $g_{mt}$ need to be extended to capture the extrinsic information from the channel decoder and the activity detector. Here, we apply the formalism of the sum-product update rules meaning that the messages from the activity detector $\Delta_{\xi_{nt} \rightarrow f_{nt}} \left(x_{nt}\right)$ and the message from the channel decoder $\Delta_{d_{nt} \rightarrow f_{nt}} \left(x_{nt}\right)$ are point-wise multiplied, yielding
    \makeatletter%
    \if@twocolumn
        \begin{align}\nonumber
            &\hspace{-0.25cm}\Delta^{i+1}_{f_{nt} \rightarrow g_{mt}}\left(x_{nt}\right) \propto \sum_{c_{nt},s_{nt}}\! f_{nt}\left(x_{nt},c_{nt},s_{nt}\right) \Delta^i_{d_{nt}\rightarrow f_{nt}}\left(c_{nt}\right)\times\\
            &\hspace{2.5cm}\Delta^i_{\xi_{nt}\rightarrow f_{nt}}\left(s_{nt}\right) \prod_{p\neq m} \Delta^i_{g_{pt}\rightarrow x_{nt}}\left(x_{nt}\right).
        \end{align}
    \else
        \begin{equation}
            \Delta^{i+1}_{f_{nt} \rightarrow g_{mt}}\left(x_{nt}\right) \propto \sum_{c_{nt},s_{nt}} f_{nt}\left(x_{nt},c_{nt},s_{nt}\right) \Delta^i_{d_{nt}\rightarrow f_{nt}}\left(c_{nt}\right)
            \Delta^i_{\xi_{nt}\rightarrow f_{nt}}\left(s_{nt}\right) \prod_{p\neq m} \Delta^i_{g_{pt}\rightarrow x_{nt}}\left(x_{nt}\right).
        \end{equation}
    \fi
    \makeatother

\subsection{LLR conversion}

    In order to detect the activity of devices and decode the transmitted data, the goal is to convert the messages from $f_{nt}$ to $d_{nt}$ and to $\xi_{nt}$ into LLRs. Thus, we study how the beliefs exchanged between multiuser detector, decoder and activity detector influence each other. It is expected that the beliefs from the multiuser detector to the decoder exhibits low magnitude if the activity detector has a high belief toward inactivity. Aditionally, the beliefs from the multiuser detector to the activity detector are also influenced by the beliefs from the decoder about the code symbols.

    Starting with the message from the multiuser detector to the decoder being composed of the message from the activity detector to the multiuser detector and from the message from the likelihood factor. In combination with the definition of the function node $f_{nt}$ we have
    \makeatletter%
    \if@twocolumn
        \begin{flalign}\nonumber
            & \Delta^{i+1}_{f_{nt} \rightarrow d_{nt}} \left(c_{nt}\right)\\ \nonumber
            &\hspace{0.25cm} \propto\! \sum_{x_{nt},s_{nt}}\! f_{nt} \left(x_{nt}, c_{nt}, s_{nt}\right)\, \Delta^i_{\xi_{nt} \rightarrow f_{nt}}\left(s_{nt}\right) \Delta^i_{x_{nt} \rightarrow f_{nt}}\left(x_{nt}\right)\\
            &\hspace{0.25cm} \propto\, \Delta^i_{\xi_{nt} \rightarrow f_{nt}}\left(s_{nt} = 0\right)\, \Delta^i_{x_{nt} \rightarrow f_{nt}}\left(x_{nt} = 0\right) +\\ \nonumber
            &\hspace{0.8cm} (1-c_{nt})\, \Delta^i_{\xi_{nt} \rightarrow f_{nt}}\left(s_{nt} = 1\right)\, \Delta^i_{x_{nt} \rightarrow f_{nt}}\left(x_{nt} = -1\right) +\\ \nonumber &\hspace{0.8cm} c_{nt}\, \Delta^i_{\xi_{nt} \rightarrow f_{nt}}\left(s_{nt} = 1\right)\, \Delta^i_{x_{nt} \rightarrow f_{nt}}\left(x_{nt} = 1\right)
        \end{flalign}
    \else
        \begin{flalign}\nonumber
            & \Delta^{i+1}_{f_{nt} \rightarrow d_{nt}} \left(c_{nt}\right)\\ \nonumber
            &\hspace{0.5cm} \propto\, \sum_{x_{nt},s_{nt}}\, f_{nt} \left(x_{nt}, c_{nt}, s_{nt}\right)\, \Delta^i_{\xi_{nt} \rightarrow f_{nt}}\left(s_{nt}\right) \Delta^i_{x_{nt} \rightarrow f_{nt}}\left(x_{nt}\right)\\
            &\hspace{0.5cm} \propto\, \Delta^i_{\xi_{nt} \rightarrow f_{nt}}\left(s_{nt} = 0\right)\, \Delta^i_{x_{nt} \rightarrow f_{nt}}\left(x_{nt} = 0\right)\\ \nonumber
            &\hspace{1cm} + (1-c_{nt})\, \Delta^i_{\xi_{nt} \rightarrow f_{nt}}\left(s_{nt} = 1\right)\, \Delta^i_{x_{nt} \rightarrow f_{nt}}\left(x_{nt} = -1\right) + c_{nt}\, \Delta^i_{\xi_{nt} \rightarrow f_{nt}}\left(s_{nt} = 1\right)\, \Delta^i_{x_{nt} \rightarrow f_{nt}}\left(x_{nt} = 1\right)
        \end{flalign}
    \fi
    \makeatother

    \noindent where $\Delta^i_{x_{nt} \rightarrow f_{nt}}\left(x_{nt}\right) = \prod_{m=1}^M \Delta^i_{g_{mt}\rightarrow x_{nt}}\left(x_{nt}\right)$. As previously explained, messages are functions reflecting probabilities. In this case, we can summarize the message as code symbol LLR by calculating

    \begin{equation}\label{eq:LLR_ftod}
        L^{i+1}_{f_{nt} \rightarrow d_{nt}}\left(c_{nt}\right) := \log \frac{\Delta^i_{f_{nt}\rightarrow d_{nt}}\left(c_{nt} = 1\right)}{\Delta^i_{f_{nt}\rightarrow d_{nt}}\left(c_{nt} = 0\right)}.
    \end{equation}

    \makeatletter%
    \if@twocolumn
        This expresses the belief of the multiuser detector about the $n,l$-th code symbol as a code symbol LLR that reads as in~(\ref{eq:LLR_ftodbig}). For the sake of completeness, we also look at the activity LLRs from the multiuser to the activity detector and consider how the beliefs from the decoder contribute here. To this end, we consider the message from the multiuser to the activity detector. This message is composed of the beliefs given by the likelihood factors and the beliefs from the decoder. This message reads as

            \setcounter{equation}{39}
            \begin{align}
                &\Delta^{i+1}_{f_{nt} \rightarrow \xi_{nt}} \left(s_{nt}\right) \propto\\ \nonumber &\sum_{x_{nt},c_{nt}} f_{nt} \left(x_{nt}, c_{nt}, s_{nt}\right)  \Delta^i_{d_{nt} \rightarrow f_{nt}}\left(c_{nt}\right) \Delta^i_{x_{nt} \rightarrow f_{nt}}\left(x_{nt}\right).
            \end{align}

            \begin{figure*}
                \setcounter{equation}{38}
                \begin{equation}\label{eq:LLR_ftodbig}
                    L^{i+1}_{f_{nt} \rightarrow d_{nt}}\left(c_{nt}\right) = \log \frac{\Delta^i_{\xi_{nt} \rightarrow f_{nt}}\left(s_{nt} = 0\right) \Delta^i_{x_{nt} \rightarrow f_{nt}}\left(x_{nt} = 0\right) + \Delta^i_{\xi_{nt} \rightarrow f_{nt}}\left(s_{nt} = 1\right) \Delta^i_{x_{nt} \rightarrow f_{nt}}\left(x_{nt} = 1\right)}{\Delta^i_{\xi_{nt} \rightarrow f_{nt}}\left(s_{nt} = 0\right) \Delta_{x_{nt} \rightarrow f_{nt}}\left(x_{nt} = 0\right) + \Delta^i_{\xi_{nt} \rightarrow f_{nt}}\left(s_{nt} = 1\right) \Delta^i_{x_{nt} \rightarrow f_{nt}}\left(x_{nt} = -1\right)}.
                \end{equation}

                \setcounter{equation}{40}
                \begin{flalign} \label{eq:LLR_ftoxi_big}
                    L^{i+1}_{f_{nt} \rightarrow \xi_{nt}}\left(s_{nt}\right) :=& \log \frac{\Delta^i_{f_{nt}\rightarrow \xi_{nt}}\left(s_{nt} = 0\right)}{\Delta^i_{f_{nt}\rightarrow \xi_{nt}}\left(s_{nt} = 1\right)}\\
                    \nonumber
                    :=& \log \frac{\Delta^i_{x_{nt} \rightarrow f_{nt}}\left(x_{nt} = 0\right)}{\Delta^i_{d_{nt} \rightarrow f_{nt}}\left(c_{nt} = 0\right) \Delta^i_{x_{nt} \rightarrow f_{nt}}\left(x_{nt} = -1\right) + \Delta^i_{d_{nt} \rightarrow f_{nt}}\left(c_{nt} = 1\right) \Delta^i_{x_{nt} \rightarrow f_{nt}}\left(x_{nt} = 1\right)},
                \end{flalign}
                \hrulefill
                \vspace*{4pt}
            \end{figure*}

            This message can be compactly summarized as a LLR using the definition of the function node $f_{nt}$ in (\ref{eq:LLR_ftoxi_big}), which is used to the activity detection. One can see in~(\ref{eq:LLR_ftoxi_big}) that the information provided by the decoder does not make difference into the activity detection. Thus, since at this point we already have the means and variances of $\hat{x}$, that is, $\hat{r}$ and $\nu^r$, we can approximate $L^{i+1}_{f_{nt} \rightarrow \xi_{nt}}\left(s_{nt}\right)$ as given by
            \begin{equation}\label{eq:LLR_ftoxi_apprx}
                L^{i+1}_{f_{nt} \rightarrow \xi_{nt}}\left(s_{nt}\right) := \log \frac{\mathcal{N}_c\left(0|\hat{r}_{nt},\nu^r_{nt}\right)}{\mathcal{N}_c\left(0|\hat{r}_{nt},\nu^r_{nt}+\sigma_{X_d}^2\right)}.
            \end{equation}
    \else
       This expresses the belief of the multiuser detector about the $n,l$-th code symbol as a code symbol LLR that reads as

        \begin{equation}\label{eq:LLR_ftodbig}
            L^{i+1}_{f_{nt} \rightarrow d_{nt}}\left(c_{nt}\right) = \log \frac{\Delta^i_{\xi_{nt} \rightarrow f_{nt}}\left(s_{nt} = 0\right) \Delta^i_{x_{nt} \rightarrow f_{nt}}\left(x_{nt} = 0\right) + \Delta^i_{\xi_{nt} \rightarrow f_{nt}}\left(s_{nt} = 1\right) \Delta^i_{x_{nt} \rightarrow f_{nt}}\left(x_{nt} = 1\right)}{\Delta^i_{\xi_{nt} \rightarrow f_{nt}}\left(s_{nt} = 0\right) \Delta_{x_{nt} \rightarrow f_{nt}}\left(x_{nt} = 0\right) + \Delta^i_{\xi_{nt} \rightarrow f_{nt}}\left(s_{nt} = 1\right) \Delta^i_{x_{nt} \rightarrow f_{nt}}\left(x_{nt} = -1\right)}.
        \end{equation}

        For the sake of completeness, we also look at the activity LLRs from the multiuser to the activity detector and consider how the beliefs from the decoder contribute here. To this end, we consider the message from the multiuser to the activity detector. This message is composed of the beliefs given by the likelihood factors and the beliefs from the decoder. This message reads as

        \begin{equation}
            \Delta^{i+1}_{f_{nt} \rightarrow \xi_{nt}} \left(s_{nt}\right) \propto \sum_{x_{nt},c_{nt}} f_{nt} \left(x_{nt}, c_{nt}, s_{nt}\right)  \Delta^i_{d_{nt} \rightarrow f_{nt}}\left(c_{nt}\right) \Delta^i_{x_{nt} \rightarrow f_{nt}}\left(x_{nt}\right)
        \end{equation}

        This message can be compactly summarized as a LLR using the definition of the function node $f_{nt}$,

        \begin{flalign} \label{eq:LLR_ftoxi_big}
            L^{i+1}_{f_{nt} \rightarrow \xi_{nt}}\left(s_{nt}\right) :=& \log \frac{\Delta^i_{f_{nt}\rightarrow \xi_{nt}}\left(s_{nt} = 0\right)}{\Delta^i_{f_{nt}\rightarrow \xi_{nt}}\left(s_{nt} = 1\right)}\\
            \nonumber
            :=& \log \frac{\Delta^i_{x_{nt} \rightarrow f_{nt}}\left(x_{nt} = 0\right)}{\Delta^i_{d_{nt} \rightarrow f_{nt}}\left(c_{nt} = 0\right) \Delta^i_{x_{nt} \rightarrow f_{nt}}\left(x_{nt} = -1\right) + \Delta^i_{d_{nt} \rightarrow f_{nt}}\left(c_{nt} = 1\right) \Delta^i_{x_{nt} \rightarrow f_{nt}}\left(x_{nt} = 1\right)},
        \end{flalign}

        \vspace{0.25cm}
        \noindent which is used to the activity detection. One can see in~(\ref{eq:LLR_ftoxi_big}) that the information provided by the decoder does not make difference into the activity detection. Thus, since at this point we already have the means and variances of $\hat{x}$, that is, $\hat{r}$ and $\nu^r$, we can approximate $L^{i+1}_{f_{nt} \rightarrow \xi_{nt}}\left(s_{nt}\right)$ as given by
        \begin{equation}\label{eq:LLR_ftoxi_apprx}
            L^{i+1}_{f_{nt} \rightarrow \xi_{nt}}\left(s_{nt}\right) := \log \frac{\mathcal{N}_c\left(0|\hat{r}_{nt},\nu^r_{nt}\right)}{\mathcal{N}_c\left(0|\hat{r}_{nt},\nu^r_{nt}+\sigma_{X_D}^2\right)}.
        \end{equation}
    \fi
    \makeatother

\subsection{Sum-Product Algorithm LDPC decoder}

 With the LLRs computed, we use (\ref{eq:LLR_ftod}) for decoding and (\ref{eq:LLR_ftoxi_apprx}) as \textit{a priori} activity probability LLRs into a logarithmic LDPC decoder, as described in~\cite{JMoreiraBook}. Regarding the activity detection, when the evaluated symbol is a pilot, that is, $t \in \mathcal{L}_p$, BiMSGAMP uses the activity prior described in section~\ref{subsection:actprior}, with~(\ref{eq:LLRnleftnm2}) and~(\ref{eq:est_rho}). For data, BiMSGAMP uses the extrinsic LLRs provided by the LDPC decoder to refine the probability of being active of each device, as given by
    \begin{equation}
        \hat{\rho}^i_{nt} =1 \big/\left(1+ \exp{\{L^{i+1}_{f_{nt} \rightarrow \xi_{nt}}\left(s_{nt}\right)\}}\right).
    \end{equation}

    Using the previous LLR values, we have,
    \makeatletter%
    \if@twocolumn
        \begin{align}
            &L^{\text{dec}}_{nt} =\, \text{ Decode }\left[L^{i}_{f_{nt} \rightarrow d_{nt}}\left(c_{nt}\right) - L^{i-1}_{f_{nt} \rightarrow d_{nt}}\left(c_{nt}\right)\right],\\
            &L^{i+1}_{f_{nt} \rightarrow d_{nt}}\left(c_{nt}\right) =\, L^\text{dec}_{{nt}} - L^{i}_{f_{nt} \rightarrow d_{nt}}\left(c_{nt}\right), \text{ and}\\
            \label{eq:rho_end}
            &\hat{\rho}_{nt}^{i+1} =\, \mathbb{E}\left[L^\text{dec}_{nt}, \hat{\rho}_{nt}^{i}\right],
        \end{align}
    \else
        \begin{align}
            L^{\text{dec}}_{nt} &=\, \text{ Decode }\left[L^{i}_{f_{nt} \rightarrow d_{nt}}\left(c_{nt}\right) - L^{i-1}_{f_{nt} \rightarrow d_{nt}}\left(c_{nt}\right)\right],\\
            L^{i+1}_{f_{nt} \rightarrow d_{nt}}\left(c_{nt}\right) &=\, L^\text{dec}_{{nt}} - L^{i}_{f_{nt} \rightarrow d_{nt}}\left(c_{nt}\right),&\\
            \label{eq:rho_end}
            \hat{\rho}_{nt}^{i+1} &=\, \mathbb{E}\left[L^\text{dec}_{nt}, \hat{\rho}_{nt}^{i}\right],&
        \end{align}
    \fi
    \makeatother

    \noindent where $L^{\text{dec}}_{nt}$ is the LLR output of the LDPC decoder and $\hat{\rho}_{nt}^{i+1} = \Delta^i_{\xi_{nt} \rightarrow f_{nt}}\left(s_{nt} = 1\right)$ which closes the loop.

    Since the LDPC decoder decides for bit zero or one, an all-zero frame corresponds to an inactive device. Thus, for a bit matrix $\mathbf{B}$, the final activity detection after the hard decision procedure in $\mathbf{L}^{\text{dec}}$, for the $n$-th device, $\hat{\gamma}_{n} = 0$ for each $L$ bit sequence of zeros i.e., $\mathbf{b}_n = \mathbf{0}$ and $\hat{\gamma}_{n} = 1$, otherwise.
    %
    The procedure, summarized in Algorithm~1, iterates until a predefined threshold condition is satisfied or the iteration $i$ reaches the maximum number of iterations $I$. We consider the threshold given by
    \begin{equation}\label{eq:StopCrit}
        \text{tol} = \left(\frac{\|\hat{\bx}_t^{(i)} -\hat{\bx}_t^{(i-1)}\|}{\|\hat{\bx}_t^{(i)}\|}\right) < 10^{-4},
    \end{equation}
    \noindent that is, if tol reaches a value equal or larger than $10^{-4}$ and/or $i \geq I$, BiMSGAMP stops.

\section{Dynamic Scheduling Strategies} \label{sec:DynSchStr}

In the predicted massive access mMTC scenario in 5G and beyond
mobile communication systems~\cite{Cisco2020}, low complexity
techniques are essential. Unlike previous works, where
message-passing approaches
\cite{bfpeg,dopeg,memd,CasadoTCom2010,vfap,aaidd,kaids,dynmtc,jeadd}
update all messages in parallel, we develop and apply two
message-scheduling strategies that dramatically reduce the
computational cost of the proposed scheme.

    Firstly described in our previous work~\cite{DiRennaWCL2021}, we propose two different criteria to determine a group of nodes $\mathcal{S}^{(i)}$ to be updated. The aim is to renew, at every iteration $i$, only the messages that belong to a group of nodes (that represent the MTCDs) and not all of them, as in the literature. Thus, the stop criterion in~(\ref{eq:StopCrit}) consider not all devices, but only the ones that belongs to the group. As an example, considering a message-scheduling technique that is based on the estimated channels, (\ref{eq:StopCrit}) is given by
    \begin{equation}\label{eq:StopCrit2}
        \text{tol} = \left(\frac{\|\underline{\hat{\bh}}_{m'}^{(i)} -\underline{\hat{\bh}}_{m'}^{(i-1)}\|}{\|\underline{\hat{\bh}}_{m'}^{(i)}\|}\right) < 10^{-4},
    \end{equation}

    \noindent where $\underline{\hat{\bh}}^{(i)}_{m'}$ is a $|\mathcal{S}^{(i)}|\times 1$ vector that corresponds to the estimated channels between the $|\mathcal{S}^{(i)}|$ devices and the $m'$-th BS antenna. With the new stop criterion defined, we explain the two message-scheduling techniques applied in this work.

\subsection{Message-Scheduling based on Activity User Detection}

    The BiMSGAMP-AUD is a BiMSGAMP-type algorithm that has a message-scheduling based on the instantaneous activity user detection. The key idea is to form the group of nodes that the messages are going to be updated only with the nodes that the activity detection $\hat{\rho}_{nt}$, given by~(\ref{eq:rho_end}) surpasses a threshold value, close to 1. That is, if the device is considered as active, it is included in the set $\mathcal{S}^{(i)}$.

    Since in the fist iteration the algorithm only has knowledge of the probability of being active of each device (which is typically much lower than 1), in the first iteration, every node has it messages updated. In the second iteration, the algorithm proceeds judging the $\hat{\rho}_{nt}^{(i)}$ values, thus forming the group $\mathcal{S}^{(i)}$. After that, all messages that belong to $\mathcal{S}^{(i)}$, except for $s^{(i)}_1$ will be updated. Accordingly, the set removes a group of messages that are associated to a specific device one by one, that is

    \begin{equation}
        {\mathcal{S}}^{(i)} = \left[s^{(i-1)}_2, \dots, s^{(i-1)}_{|\mathcal{S}^{(i-1)}|}\right].
    \end{equation}

    In summary, the messages that belong to a group of nodes $\mathcal{S}^{(i)}$ are updated in parallel until the group is empty. The messages that are not associated to a node inside the set are neglected, until the end of update-removal procedure. When the set is finally empty, BiMSGAMP-AUD updates all the messages in parallel, as its happens in the literature. That is, the new set is $\mathcal{S}^{(i)} = \left[1, \dots, N\right]$. After computing new $\hat{\rho}_{nt}$ values, a new set is performed and the procedure continues until the stop criterion is satisfied.

\subsection{Message-Scheduling based on Residual Belief Propagation}
    In this technique, we consider an ordering metric called residual belief propagation (RBP). The residual is the norm (defined over the message space) of the difference between the values of a message before and after an update. A residual is the norm (defined over the message space) of the difference between the values of a message before and after an update. Considering the channel estimation part of BiMSGAMP-RBP, the residual for the belief distribution at $h_{mn}$, is given by

    \begin{equation}\label{eq:Res}
        \text{Res}\left(\Delta_{h_{mn}}\left(h_{mn}\right)\right) = \big|\big|\Delta^{(i+1)}_{h_{mn}}\left(h_{mn}\right) - \Delta^{(i)}_{h_{mn}}\left(h_{mn}\right)\big|\big|.
    \end{equation}

    The idea behind this method is to use the fact that the differences between the messages before and after an update reduces when the factor graph approach converges. Therefore, if a message has a large residual, it can indicates that it is located in a part of the graph that has not converged yet. So, if the messages that have larger residuals are propagated first, the convergence should be accelerated. Based on this idea, the residual values computed in in~(\ref{eq:Res}) are used to form the set $\mathcal{S}^{(i)}$ of messages to be updated in the next iteration. In order to determine the maximum size of the set, we use the fact that the
    activity probability of MTCDs is typically around $5\%$~\cite{DiRennaAccess2020}. Therefore, the set $\mathcal{S}^{(i)}$ is composed by the $0.05\,N$ nodes with highest residual. The update sequence of BiMSGAMP-RBP is the same of BiMSGAMP-AUD, the difference is how both groups are formed. Algorithm 1 summarizes the BiMSGAMP procedure, described in the previous sections. With the main ideas explained, the next section discusses the computational cost and the convergence of the proposed scheme.

\section{Analysis} \label{sec:analysis}

    This section analyses the BiMSGAMP-type schemes in terms of the computational complexity and the convergence in terms of NMSE regarding the activity and data detection, and the channel estimation. All results are discussed and compared with state-of-the-art solutions.

    \makeatletter%
    \if@twocolumn
        \begin{center}
\newcounter{CounterDef}
\newcounter{Counter}
    \begin{table}[t]
    \vspace{0.3cm}
        \scriptsize
            \begin{tabular}{ll}
                \hline
                \multicolumn{2}{l}{\footnotesize\textbf{Algorithm 1} {Bilinear Message-Scheduling GAMP - BiMSGAMP}}           \\ \hline
                \multicolumn{2}{l}{\textbf{definition} }            \\
                \multicolumn{2}{l}{\stepcounter{CounterDef}{{\scriptsize[D\theCounterDef]}} \hspace{0.1cm} $\mathcal{P}\left(z_{mt}|\hat{p}_{mt}, \nu^p_{mt}\right) \triangleq \frac{\mathcal{P}\left(y_{mt}|z_{mt}\right) \mathcal{N}_c\left(z_{mt}|\hat{p}_{mt},\nu^{p}_{mt}\right)}{\int_{z} \mathcal{P}\left(y_{mt}|z\right)\, \mathcal{N}_c\left(z;\hat{p}_{mt},\nu^{p}_{mt}\right)}$}\\[8pt]
                \multicolumn{2}{l}{{\stepcounter{CounterDef}{\scriptsize[D\theCounterDef]}} \hspace{0.1cm} $\mathcal{P}\left(x_{nt}|\hat{r}_{nt}, \nu^r_{nt}\right) \triangleq \frac{\sum_{s}\! \sum_{c}  \mathcal{P}_{\bsfx_{d}}\!\left(x_{nt},c_{nt},s_{nt}\right)\, \mathcal{N}_c\left(x_{nt};\hat{r}_{nt},\nu^{r}_{nt}\right)}{\int_{x} \sum_{s}\! \sum_{c} \mathcal{P}_{\bsfx_{d}}\!\left(x,c_{nt},s_{nt}\right)\, \mathcal{N}_c\left(x;\hat{r}_{nt},\nu^{r}_{nt}\right)}$}\\[8pt]
                \multicolumn{2}{l}{\stepcounter{CounterDef}{{\scriptsize[D\theCounterDef]}} \hspace{0.1cm} $\mathcal{P}\left(h_{mn}|\hat{q}_{mn}, \nu^q_{mn}; \hat{\rho}_{nt}\right) \triangleq \frac{\mathcal{P}\left(h_{mn};\hat{\rho}_{nt}\right)\, \mathcal{N}_c\left(h_{mn};\hat{q}_{mn},\nu^{q}_{mn}\right)}{\int_{h} \mathcal{P}\left(h;\hat{\rho}_{nt}\right)\, \mathcal{N}_c\left(h;\hat{q}_{mn},\nu^{q}_{mn}\right)}$}\\
                \multicolumn{2}{l}{\textbf{initialize}}            \\
                \stepcounter{Counter} 
                {[{\scriptsize A\theCounter}]}  & $i=1$, $\mathcal{S}^{(1)} = \left[1, \dots,N\right]$ and\\
                & $\forall\, m, n, t: \hat{\rho}^{(0)}_{nmt} = \rho_{nt}, \hat{{r}}_{nt}^{(0)} = 0$, $\nu^{r(0)}_{nm} = 1$\\
                \multicolumn{2}{l}{\textbf{repeat}}                \\
                \multicolumn{2}{l}{\hspace{15pt}\fbox{Adapted BiG-AMP approximation}} \\[4pt]
                \multicolumn{2}{l}{\hspace{0.1cm}$\forall\, m$ and $t\in \mathcal{L}_p$,}
                \\
                \stepcounter{Counter} 
                {{[\scriptsize A\theCounter]}}  & $\nu^{pi}_{mt} = \,\sum_{n}^{|\mathcal{S}^{(i)}|}\, |\hat{h}^i_{mn}|^2\, \nu^{x}_{nt} + \nu^{hi}_{nt}\, |{x}_{nt}|^2 + \nu^{hi}_{nt}\,\nu^{x}_{nt}$
                \\[4pt] 
                \stepcounter{Counter}                 
                {{[\scriptsize A\theCounter]}}  & $\hat{p}^i_{mt} = \, \sum_{n}^{|\mathcal{S}^{(i)}|}\, \hat{h}^i_{mn}\, {x}_{nt} - \hat{s}_{mt}^{i-1}\, \nu^{pi}_{mt}$
                \\[4pt]           
                \multicolumn{2}{l}{\hspace{0.1cm}$\forall\, m$ and $t\in \mathcal{L}_d$,}
                \\                
                \stepcounter{Counter} 
                {{[\scriptsize A\theCounter]}}  & $\nu^{pi}_{mt} = \,\sum_{n}^{|\mathcal{S}^{(i)}|}\, |\hat{h}^i_{mn}|^2\, \nu^{xi}_{nt} + \nu^{hi}_{nt}\, |\hat{x}^i_{nt}|^2 + \nu^{hi}_{nt}\,\nu^{xi}_{nt}$
                \\[4pt] 
                \stepcounter{Counter}                 
                {{[\scriptsize A\theCounter]}}  & $\hat{p}^i_{mt} = \, \sum_{n}^{|\mathcal{S}^{(i)}|}\, \hat{h}^i_{mn}\, \hat{x}^i_{nt} - \hat{s}_{mt}^{i-1} \nu^{pi}_{mt}$
                \\[4pt]            
                \multicolumn{2}{l}{\hspace{0.1cm}$\forall\, m$ and $t$,}
                \\                     
                \stepcounter{Counter}                 
                {{[\scriptsize A\theCounter]}}  &$\nu^{zi}_{mt} = \textsc{var}\left\{\bsfz_{mt}| \bsfp_{mt} = \hat{p}^i_{mt}; \nu^{pi}_{mt}\right\}$
                \\[4pt]  
                \stepcounter{Counter}                 
                {{[\scriptsize A\theCounter]}}  & $\hat{z}^i_{mt} = \mathbb{E}\left\{\bsfz_{mt}| \bsfp_{mt} = \hat{p}^i_{mt}; \nu^{pi}_{mt}\right\}$
                \\[4pt]  
                \stepcounter{Counter}                 
                {{[\scriptsize A\theCounter]}}  & $\nu^{si}_{mt} =\, \left(\left(1-\nu^{zi}_{mt}\right)/\nu^{pi}_{mt}\right)/\nu^{pi}_{mt}$
                \\[4pt]           
                \stepcounter{Counter}                 
                {{[\scriptsize A\theCounter]}}  & $\hat{s}^{i}_{mt} =\, \left(\hat{z}^i_{mt}-\hat{p}^i_{mt}\right)/\nu^{pi}_{mt}$
                \\[4pt]  
                \multicolumn{2}{l}{\hspace{0.1cm}$\forall\, m$ and $n \in |\mathcal{S}^{(i)}|$,}
                \\                   
                \stepcounter{Counter}                 
                {{[\scriptsize A\theCounter]}} & $\nu^{qi}_{mn} =\left[\nu^{si}_{mt}\left(\sum_{t \in \mathcal{L}_p} |x_{nt}|^2 + \sum_{t \in \mathcal{L}_d} |\hat{x}^{i}_{nt}|^2 \right)\right]^{-1}$
                \\[4pt]       
                \stepcounter{Counter}                 
                {{[\scriptsize A\theCounter]}}  & $\hat{q}^i_{mn} = \hat{h}^{i}_{mn} \left(\!1 - \nu^{qi}_{mn} \sum_{t \in \mathcal{L}_d} \nu^{xi}_{nt}\, \nu^{si}_{mt}\right) +$\\[4pt]     
                &\hspace{1cm} $\nu^{qi}_{mn} \left(\sum_{t \in \mathcal{L}_p} x^{\ast}_{nt}\,\hat{s}^i_{mt} + \sum_{t \in \mathcal{L}_d} \hat{x}^{i\ast}_{nt}\,\hat{s}^i_{mt}\right)$\\[4pt]     
                \stepcounter{Counter}                 
                {{[\scriptsize A\theCounter]}}  & $\nu^{h\,i+1}_{mn} =\, \textsc{var}\left\{\bsfh_{mn}|\bsfq_{mn}=\hat{q}_{mn}^i, \nu^{qi}_{mn}\right\}$\\[4pt]     
                \stepcounter{Counter}                 
                {{[\scriptsize A\theCounter]}}  & $\hat{h}^{i+1}_{mn} =\, \mathbb{E}\left\{\bsfh_{mn}|\bsfq_{mn}=\hat{q}_{mn}^i, \nu^{qi}_{mn}; \hat{\rho}_{mn}^{i}\right\}$\\[4pt]  
                \stepcounter{Counter}                 
                {{[\scriptsize A\theCounter]}}  &$\nu^{ri}_{nt} = \left(\sum_{m=1}^M |\hat{h}^{i}_{mn}|^2 \nu^{si}_{mt}\right)^{-1}$
                \\[4pt]       
                \stepcounter{Counter}                 
                {{[\scriptsize A\theCounter]}}  &$\hat{r}^i_{nt} = \hat{x}^{i}_{nt} \left(\!1 - \nu^{ri}_{nt} \sum_{m=1}^{M} \nu^{hi}_{mn}\, \nu^{si}_{mt}\right) + \nu^{ri}_{nt} \sum_{m=1}^{M} \hat{h}^{i\ast}_{mn}\,\hat{s}^i_{mt}$\\[4pt]      
                \stepcounter{Counter}                 
                {{[\scriptsize A\theCounter]}}  &$\nu^{x\,i+1}_{mn} =\, \textsc{var}\left\{\bsfx_{nt}|\bsfr_{nt}=\hat{r}_{nt}^i,\, \nu^{ri}_{mn}\right\}$\\[4pt]     
                \stepcounter{Counter}                 
                {{[\scriptsize A\theCounter]}}  &$\hat{x}^{i+1}_{mn} =\, \mathbb{E}\left\{\bsfx_{nt}|\bsfr_{nt}=\hat{r}_{nt}^i,\, \nu^{ri}_{mn}\right\}$\\[4pt]   
                \multicolumn{2}{l}{\hspace{0.1cm}$\forall\, m$ and $t\in \mathcal{L}_p$,}
                \\               
                \stepcounter{Counter}                 
                {{[\scriptsize A\theCounter]}}  & Compute $L_{mn}^{i}$ with~(\ref{eq:LLRnleftnm2}) and $\hat{\rho}^{i+1}_{mnt}$ with~(\ref{eq:est_rho})  \\[4pt]                  
                \multicolumn{2}{l}{\hspace{15pt}\fbox{Joint Activity detection and LDPC decoding}} \\[4pt]
                \multicolumn{2}{l}{\hspace{0.1cm}$\forall\, t\in \mathcal{L}_d$ and $n \in |\mathcal{S}^{(i)}|$,}
                \\                
                \stepcounter{Counter}                 
                {{[\scriptsize A\theCounter]}}  & Compute $L^{i}_{f_{nt} \rightarrow d_{nt}}\left(c_{nt}\right)$ with (\ref{eq:xvar}), (\ref{eq:xhat}), (\ref{eq:LLR_ftod}) and (\ref{eq:LLR_ftoxi_apprx}),  \\[4pt]   
                \stepcounter{Counter}                 
                {{[\scriptsize A\theCounter]}}  &\(\hat{\rho}^i_{nt} =1 \big/\left(1+ \exp{\{L^{i}_{f_{nt} \rightarrow \xi_{nt}}\left(s_{nt}\right)\}}\right)\)  \\[4pt]
                \stepcounter{Counter}                 
                {{[\scriptsize A\theCounter]}}  &$L^{\text{dec}}_{nt} = \text{Decode}\left[L^{i}_{f_{nt} \rightarrow d_{nt}}\left(c_{nt}\right) - L^{i-1}_{f_{nt} \rightarrow d_{nt}}\left(c_{nt}\right)\right]$\\[4pt]  
                \stepcounter{Counter}                 
                {{[\scriptsize A\theCounter]}}  &$L^{i+1}_{f_{nt} \rightarrow d_{nt}}\left(c_{nt}\right) = L^\text{dec}_{{nt}} - L^{i}_{f_{nt} \rightarrow d_{nt}}\left(c_{nt}\right)$\\[4pt]    
                \stepcounter{Counter}                 
                {{[\scriptsize A\theCounter]}} &$\hat{\rho}_{nt}^{i+1} =\mathbb{E}\left[L^\text{dec}_{nt}, \hat{\rho}_{nt}^{i}\right]$\\[4pt]            
                \multicolumn{2}{l}{\hspace{15pt}\fbox{Message-Scheduling update}} \\[4pt]
                \stepcounter{Counter}                 
                {{[\scriptsize A\theCounter]}}  & $\mathcal{S}^{(i)} =$ Update$\left[\mathcal{S}^{(i-1)}\right]$ with chosen message-scheduling technique\\[4pt]   
                \stepcounter{Counter}                 
                {{[\scriptsize A\theCounter]}}  & Update $\text{tol}$ with (\ref{eq:StopCrit2})  and $i = i +1$\\[4pt]                  
                \multicolumn{2}{l}{\textbf{until} $\left(i > I \text{ or tol} < 10^{-4} \right)$} \\ \hline
            \end{tabular}
    \end{table}

        \end{center}
    \else
        \begin{center}
\newcounter{CounterDef}
\newcounter{Counter}
    \begin{table}[h!]
        \scriptsize
            \begin{tabular}{ll}
                \hline
                \multicolumn{2}{l}{\footnotesize\textbf{Algorithm 1} {Bilinear Message-Scheduling GAMP - BiMSGAMP}}           \\ \hline
                \multicolumn{2}{l}{\textbf{definition} }            \\
                \multicolumn{2}{l}{\stepcounter{CounterDef}{{\scriptsize[D\theCounterDef]}} \hspace{0.1cm} $\mathcal{P}\left(z_{mt}|\hat{p}_{mt}, \nu^p_{mt}\right) \triangleq \frac{\mathcal{P}\left(y_{mt}|z_{mt}\right) \mathcal{N}_c\left(z_{mt}|\hat{p}_{mt},\nu^{p}_{mt}\right)}{\int_{z} \mathcal{P}\left(y_{mt}|z\right)\, \mathcal{N}_c\left(z;\hat{p}_{mt},\nu^{p}_{mt}\right)}$}\\[8pt]
                \multicolumn{2}{l}{{\stepcounter{CounterDef}{\scriptsize[D\theCounterDef]}} \hspace{0.1cm} $\mathcal{P}\left(x_{nt}|\hat{r}_{nt}, \nu^r_{nt}\right) \triangleq \frac{\sum_{s}\! \sum_{c}  \mathcal{P}_{\bsfx_{d}}\!\left(x_{nt},c_{nt},s_{nt}\right)\, \mathcal{N}_c\left(x_{nt};\hat{r}_{nt},\nu^{r}_{nt}\right)}{\int_{x} \sum_{s}\! \sum_{c} \mathcal{P}_{\bsfx_{d}}\!\left(x,c_{nt},s_{nt}\right)\, \mathcal{N}_c\left(x;\hat{r}_{nt},\nu^{r}_{nt}\right)}$}\\[8pt]
                \multicolumn{2}{l}{\stepcounter{CounterDef}{{\scriptsize[D\theCounterDef]}} \hspace{0.1cm} $\mathcal{P}\left(h_{mn}|\hat{q}_{mn}, \nu^q_{mn}; \hat{\rho}_{nt}\right) \triangleq \frac{\mathcal{P}\left(h_{mn};\hat{\rho}_{nt}\right)\, \mathcal{N}_c\left(h_{mn};\hat{q}_{mn},\nu^{q}_{mn}\right)}{\int_{h} \mathcal{P}\left(h;\hat{\rho}_{nt}\right)\, \mathcal{N}_c\left(h;\hat{q}_{mn},\nu^{q}_{mn}\right)}$}\\
                \multicolumn{2}{l}{\textbf{initialize}}            \\
                \stepcounter{Counter} 
                {[{\scriptsize A\theCounter}]}  & $i=1$, $\mathcal{S}^{(1)} = \left[1, \dots,N\right]$ and $\forall\, m, n, t: \hat{\rho}^{(0)}_{nmt} = \rho_{nt}, \hat{{r}}_{nt}^{(0)} = 0$, $\nu^{r(0)}_{nm} = 1$\\
                \multicolumn{2}{l}{\textbf{repeat}}                \\
                \multicolumn{2}{l}{\hspace{15pt}\fbox{Adapted BiG-AMP approximation}} \\[2pt]
                \multicolumn{2}{l}{\hspace{0.1cm}$\forall\, m$ and $t\in \mathcal{L}_p$,}
                \\
                \stepcounter{Counter} 
                {{[\scriptsize A\theCounter]}}  & $\nu^{pi}_{mt} = \,\sum_{n}^{|\mathcal{S}^{(i)}|}\, |\hat{h}^i_{mn}|^2\, \nu^{x}_{nt} + \nu^{hi}_{nt}\, |{x}_{nt}|^2 + \nu^{hi}_{nt}\,\nu^{x}_{nt}$
                \\[1pt] 
                \stepcounter{Counter}                 
                {{[\scriptsize A\theCounter]}}  & $\hat{p}^i_{mt} = \, \sum_{n}^{|\mathcal{S}^{(i)}|}\, \hat{h}^i_{mn}\, {x}_{nt} - \hat{s}_{mt}^{i-1}\, \nu^{pi}_{mt}$
                \\[1pt]           
                \multicolumn{2}{l}{\hspace{0.1cm}$\forall\, m$ and $t\in \mathcal{L}_d$,}
                \\                
                \stepcounter{Counter} 
                {{[\scriptsize A\theCounter]}}  & $\nu^{pi}_{mt} = \,\sum_{n}^{|\mathcal{S}^{(i)}|}\, |\hat{h}^i_{mn}|^2\, \nu^{xi}_{nt} + \nu^{hi}_{nt}\, |\hat{x}^i_{nt}|^2 + \nu^{hi}_{nt}\,\nu^{xi}_{nt}$
                \\[1pt] 
                \stepcounter{Counter}                 
                {{[\scriptsize A\theCounter]}}  & $\hat{p}^i_{mt} = \, \sum_{n}^{|\mathcal{S}^{(i)}|}\, \hat{h}^i_{mn}\, \hat{x}^i_{nt} - \hat{s}_{mt}^{i-1} \nu^{pi}_{mt}$
                \\[1pt]            
                \multicolumn{2}{l}{\hspace{0.1cm}$\forall\, m$ and $t$,}
                \\                     
                \stepcounter{Counter}                 
                {{[\scriptsize A\theCounter]}}  &$\nu^{zi}_{mt} = \textsc{var}\left\{\bsfz_{mt}| \bsfp_{mt} = \hat{p}^i_{mt}; \nu^{pi}_{mt}\right\}$
                \\[1pt]  
                \stepcounter{Counter}                 
                {{[\scriptsize A\theCounter]}}  & $\hat{z}^i_{mt} = \mathbb{E}\left\{\bsfz_{mt}| \bsfp_{mt} = \hat{p}^i_{mt}; \nu^{pi}_{mt}\right\}$
                \\[1pt]  
                \stepcounter{Counter}                 
                {{[\scriptsize A\theCounter]}}  & $\nu^{si}_{mt} =\, \left(\left(1-\nu^{zi}_{mt}\right)/\nu^{pi}_{mt}\right)/\nu^{pi}_{mt}$
                \\[1pt]           
                \stepcounter{Counter}                 
                {{[\scriptsize A\theCounter]}}  & $\hat{s}^{i}_{mt} =\, \left(\hat{z}^i_{mt}-\hat{p}^i_{mt}\right)/\nu^{pi}_{mt}$
                \\[1pt]  
                \multicolumn{2}{l}{\hspace{0.1cm}$\forall\, m$ and $n \in |\mathcal{S}^{(i)}|$,}
                \\                   
                \stepcounter{Counter}                 
                {{[\scriptsize A\theCounter]}} & $\nu^{qi}_{mn} =\left[\nu^{si}_{mt}\left(\sum_{t \in L_P} |x_{nt}|^2 + \sum_{t \in L_D} |\hat{x}^{i}_{nt}|^2 \right)\right]^{-1}$
                \\[1pt]       
                \stepcounter{Counter}                 
                {{[\scriptsize A\theCounter]}}  & $\hat{q}^i_{mn} = \hat{h}^{i}_{mn} \left(\!1 - \nu^{qi}_{mn} \sum_{t \in L_D} \nu^{xi}_{nt}\, \nu^{si}_{mt}\right) + \nu^{qi}_{mn} \left(\sum_{t \in L_P} x^{\ast}_{nt}\,\hat{s}^i_{mt} + \sum_{t \in L_D} \hat{x}^{i\ast}_{nt}\,\hat{s}^i_{mt}\right)$\\[1pt]     
                \stepcounter{Counter}                 
                {{[\scriptsize A\theCounter]}}  & $\nu^{h\,i+1}_{mn} =\, \textsc{var}\left\{\bsfh_{mn}|\bsfq_{mn}=\hat{q}_{mn}^i, \nu^{qi}_{mn}\right\}$\\[1pt]     
                \stepcounter{Counter}                 
                {{[\scriptsize A\theCounter]}}  & $\hat{h}^{i+1}_{mn} =\, \mathbb{E}\left\{\bsfh_{mn}|\bsfq_{mn}=\hat{q}_{mn}^i, \nu^{qi}_{mn}; \hat{\rho}_{mn}^{i}\right\}$\\[1pt]  
                \stepcounter{Counter}                 
                {{[\scriptsize A\theCounter]}}  &$\nu^{ri}_{nt} = \left(\sum_{m=1}^M |\hat{h}^{i}_{mn}|^2 \nu^{si}_{mt}\right)^{-1}$
                \\[1pt]       
                \stepcounter{Counter}                 
                {{[\scriptsize A\theCounter]}}  &$\hat{r}^i_{nt} = \hat{x}^{i}_{nt} \left(\!1 - \nu^{ri}_{nt} \sum_{m=1}^{M} \nu^{hi}_{mn}\, \nu^{si}_{mt}\right) + \nu^{ri}_{nt} \sum_{m=1}^{M} \hat{h}^{i\ast}_{mn}\,\hat{s}^i_{mt}$\\[1pt]      
                \stepcounter{Counter}                 
                {{[\scriptsize A\theCounter]}}  &$\nu^{x\,i+1}_{mn} =\, \textsc{var}\left\{\bsfx_{nt}|\bsfr_{nt}=\hat{r}_{nt}^i,\, \nu^{ri}_{mn}\right\}$\\[1pt]     
                \stepcounter{Counter}                 
                {{[\scriptsize A\theCounter]}}  &$\hat{x}^{i+1}_{mn} =\, \mathbb{E}\left\{\bsfx_{nt}|\bsfr_{nt}=\hat{r}_{nt}^i,\, \nu^{ri}_{mn}\right\}$\\[1pt]   
                \multicolumn{2}{l}{\hspace{0.1cm}$\forall\, m$ and $t\in \mathcal{L}_p$,}
                \\               
                \stepcounter{Counter}                 
                {{[\scriptsize A\theCounter]}}  & Compute $L_{mn}^{i}$ with~(\ref{eq:LLRnleftnm2}) and $\hat{\rho}^{i+1}_{mnt}$ with~(\ref{eq:est_rho})  \\[1pt]                  
                \multicolumn{2}{l}{\hspace{15pt}\fbox{Joint Activity detection and LDPC decoding}} \\[1pt]
                \multicolumn{2}{l}{\hspace{0.1cm}$\forall\, t\in \mathcal{L}_d$ and $n \in |\mathcal{S}^{(i)}|$,}
                \\                
                \stepcounter{Counter}                 
                {{[\scriptsize A\theCounter]}}  & Compute $L^{i}_{f_{nt} \rightarrow d_{nt}}\left(c_{nt}\right)$ with (\ref{eq:xvar}), (\ref{eq:xhat}), (\ref{eq:LLR_ftod}) and (\ref{eq:LLR_ftoxi_apprx}),  \\[1pt]   
                \stepcounter{Counter}                 
                {{[\scriptsize A\theCounter]}}  &\(\hat{\rho}^i_{nt} =1 \big/\left(1+ \exp{\{L^{i}_{f_{nt} \rightarrow \xi_{nt}}\left(s_{nt}\right)\}}\right)\)  \\[1pt]
                \stepcounter{Counter}                 
                {{[\scriptsize A\theCounter]}}  &$L^{\text{dec}}_{nt} = \text{Decode}\left[L^{i}_{f_{nt} \rightarrow d_{nt}}\left(c_{nt}\right) - L^{i-1}_{f_{nt} \rightarrow d_{nt}}\left(c_{nt}\right)\right]$\\[1pt]  
                \stepcounter{Counter}                 
                {{[\scriptsize A\theCounter]}}  &$L^{i+1}_{f_{nt} \rightarrow d_{nt}}\left(c_{nt}\right) = L^\text{dec}_{{nt}} - L^{i}_{f_{nt} \rightarrow d_{nt}}\left(c_{nt}\right)$\\[1pt]    
                \stepcounter{Counter}                 
                {{[\scriptsize A\theCounter]}} &$\hat{\rho}_{nt}^{i+1} =\mathbb{E}\left[L^\text{dec}_{nt}, \hat{\rho}_{nt}^{i}\right]$\\[1pt]            
                \multicolumn{2}{l}{\hspace{15pt}\fbox{Message-Scheduling update}} \\[1pt]
                \stepcounter{Counter}                 
                {{[\scriptsize A\theCounter]}}  & $\mathcal{S}^{(i)} =$ Update$\left[\mathcal{S}^{(i-1)}\right]$ with chosen message-scheduling technique\\[1pt]   
                \stepcounter{Counter}                 
                {{[\scriptsize A\theCounter]}}  & Update $\text{tol}$ with (\ref{eq:StopCrit2})  and $i = i +1$\\[1pt]                  
                \multicolumn{2}{l}{\textbf{until} $\left(i > I \text{ or tol} < 10^{-4} \right)$} \\ \hline
            \end{tabular}
    \end{table}

        \end{center}
    \fi
    \makeatother

\subsection{Computational Cost}

    \begin{figure}[t]
        \centering
\begin{tikzpicture}
\begin{axis}[%
width=5.4cm,
height=4.2cm,
scale only axis,
xmin=0,
xmax=251,
mark repeat = 5,
xlabel style={font=\color{white!15!black}},
xlabel={$N$ devices},
ymode=log,
ymin=1,
ymax=5000000,
yminorticks=true,
ylabel style={font=\color{white!15!black}},
ylabel={FLOP},
axis background/.style={fill=white},
title style={font=\bfseries, align=center},
xmajorgrids,
ymajorgrids,
yminorgrids,
legend style={at={(0.45,0.015)}, font=\tiny, anchor=south west, legend cell align=left, align=left, draw=white!15!black}
]
 
\addplot [color=\AMPcolor, line width=1.0pt, mark=x,mark size = 3pt, mark options={solid, \AMPcolor}]
  table[row sep=crcr]{%
10	174\\
20	697\\
30	1389\\
40	2572\\
50	3804\\
60	5647\\
70	7419\\
80	9922\\
90	12234\\
100	15397\\
110	18249\\
120	22072\\
130	25464\\
140	29947\\
150	33879\\
160	39022\\
170	43494\\
180	49297\\
190	54309\\
200	60772\\
210	66324\\
220	73447\\
230	79539\\
240	87322\\
250	93954\\
};
\addlegendentry{AMP~\cite{Donoho2009}}

\addplot [color=\BIGAMPcolor, line width=1.0pt, mark=+, mark size = 3pt, mark options={solid, \BIGAMPcolor}]
  table[row sep=crcr]{%
10	7265\\
20	16200\\
30	25570\\
40	37475\\
50	49275\\
60	64150\\
70	78380\\
80	96225\\
90	112885\\
100	133700\\
110	152790\\
120	176575\\
130	198095\\
140	224850\\
150	248800\\
160	278525\\
170	304905\\
180	337600\\
190	366410\\
200	402075\\
210	433315\\
220	471950\\
230	505620\\
240	547225\\
250	583325\\
};
\addlegendentry{BiG-AMP~\cite{JTParkerTSP2014}}

\addplot [color=\TBiGAMPcolor, line width=1.0pt, mark=diamond, mark size = 2pt,mark options={solid, \TBiGAMPcolor}]
  table[row sep=crcr]{%
10	8133\\
20	18168\\
30	29038\\
40	42843\\
50	56943\\
60	74518\\
70	91848\\
80	113193\\
90	133753\\
100	158868\\
110	182658\\
120	211543\\
130	238563\\
140	271218\\
150	301468\\
160	337893\\
170	371373\\
180	411568\\
190	448278\\
200	492243\\
210	532183\\
220	579918\\
230	623088\\
240	674593\\
250	720993\\
};
\addlegendentry{Turbo-BiG-AMP~\cite{TDingTWC2019}}

\addplot [color=\JEMAMPcolor, line width=1.0pt, mark=star, mark size = 3pt, mark options={solid, \JEMAMPcolor}]
  table[row sep=crcr]{%
10	7865\\
20	18843\\
30	31525\\
40	48883\\
50	66785\\
60	90523\\
70	113645\\
80	143763\\
90	172105\\
100	208603\\
110	242165\\
120	285043\\
130	323825\\
140	373083\\
150	417085\\
160	472723\\
170	521945\\
180	583963\\
190	638405\\
200	706803\\
210	766465\\
220	841243\\
230	906125\\
240	987283\\
250	1057385\\
};
\addlegendentry{Joint-EM-AMP~\cite{CWeiCommLet2017}}

\addplot [color=\HyGAMPcolor, line width=1.0pt, mark=diamond,mark size = 3pt, mark options={solid, \HyGAMPcolor}]
  table[row sep=crcr]{%
10	3998\\
20	14585\\
30	28393\\
40	50970\\
50	74588\\
60	109155\\
70	142583\\
80	189140\\
90	232378\\
100	290925\\
110	343973\\
120	414510\\
130	477368\\
140	559895\\
150	632563\\
160	727080\\
170	809558\\
180	916065\\
190	1008353\\
200	1126850\\
210	1228948\\
220	1359435\\
230	1471343\\
240	1613820\\
250	1735538\\
};
\addlegendentry{HyGAMP~\cite{SRanganTSP2017}}

\addplot [color=\RBPcolor, line width=1.0pt, mark=triangle,mark size = 3pt, mark options={solid, \RBPcolor}]
  table[row sep=crcr]{%
10	1894\\
20	4552\\
30	7192\\
40	11159\\
50	14870\\
60	20146\\
70	24928\\
80	31513\\
90	37366\\
100	45260\\
110	52184\\
120	61387\\
130	69382\\
140	79894\\
150	88960\\
160	100781\\
170	110918\\
180	124048\\
190	135256\\
200	149695\\
210	161974\\
220	177722\\
230	191072\\
240	208129\\
250	222550\\
};
\addlegendentry{BiMSGAMP-RBP}

\addplot [color=\AUDcolor, line width=1.0pt, mark=o,mark size = 2.3pt, mark options={solid, \AUDcolor}]
  table[row sep=crcr]{%
10	1894\\
20	3376\\
30	5778\\
40	7617\\
50	8843\\
60	13048\\
70	14512\\
80	19669\\
90	21371\\
100	23924\\
110	29420\\
120	32330\\
130	38659\\
140	41926\\
150	44104\\
160	52712\\
170	55128\\
180	64688\\
190	67342\\
200	71323\\
210	80746\\
220	85084\\
230	95340\\
240	100035\\
250	103165\\
};
\addlegendentry{BiMSGAMP-AUD}

\end{axis}
\end{tikzpicture}%
         \vspace{-1em}
        \caption{Floating-point operation (FLOP) counting per iteration. Each operation has a weight as defined in the Lightspeed toolbox~\cite{TMinkaLightSpeed}.}
        \label{fig:ch6_flop}
    \end{figure}
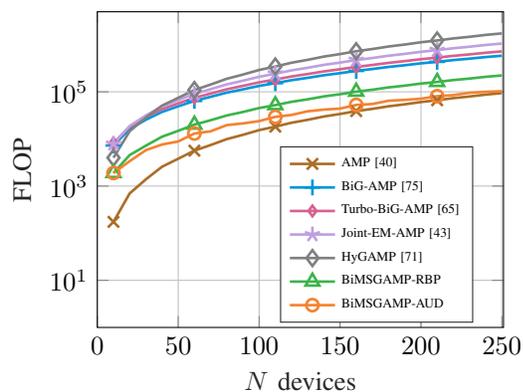

    %
    \makeatletter%
    \if@twocolumn
        %
\begin{table*}[t]
    \caption{Operations counting of considering techniques per iteration. Approaches that originally considered just the joint activity and signal detection have a separate channel estimation part, adapted using the same solution.}
    \label{tab:ch6_complexity}
    \centering
    \scriptsize
   %
    \begin{tabular}{p{1.4cm}llp{2.0cm}l} \hline
        Algorithm & \begin{tabular}[c]{@{}l@{}}Additions, Subtractions\\ and Multiplications\end{tabular} & Divisions and Square roots & Modulus & Sine, exp and log \\ \hline 
        AMP~\cite{Donoho2009}   & $L\left[N\left(6M+1\right)+3M-2\right]$ & $L \left(M+3\right)$ & N/A & N/A\\
         \arrayrulecolor{llgray}\hline \\[-0.2cm]
       \begin{tabular}[c]{@{}l@{}} Joint-EM-\\ AMP~\cite{CWeiCommLet2017}\end{tabular}   & \begin{tabular}[c]{@{}l@{}}$ L_p\left[N(6M\!+\!1)\!+\!3M\!-\!2\right] \!+ L_d\left[(N\!+\!L)(10|\mathcal{A}|^2\!+\!16|\mathcal{A}|\!+\!\right.$\\ $7M\!+\!27)\!+\!(M\!+\!L)(9N\!+-\!2) \!+\!\left.N L(6|\mathcal{A}|^2\!+\!17|\mathcal{A}|\!+\!11)\!+-\!N)\right]$\end{tabular} & \begin{tabular}[c]{@{}l@{}} $L_p\left[M\!+\!3\right]+$\\$L_d\left[N(M\!+\!L\!+\!1)\!+\!\right.$\\$\left.2(N\!+\!L)(M\!+\!|\mathcal{A}|\!+\!1)\right]$\end{tabular}  & \begin{tabular}[c]{@{}l@{}}$(N\!+\!L)\times$\\$(3|\mathcal{A}|\!+\!M\!+\!1)\!+\!$\\$2N(|\mathcal{A}|\!+\!M)\!+$\\$N(\!L\!+\!1)$\end{tabular} & N/A\\
         \arrayrulecolor{llgray}\hline \\[-0.2cm]       
        \begin{tabular}[c]{@{}l@{}} BiG-\\AMP~\cite{JTParkerTSP2014}\end{tabular} & \begin{tabular}[c]{@{}l@{}}$ (M+L)(12N+10)+ (N+L)(7M+55)+(M+N)(7L+47)$\end{tabular} & $12M+11(N+L)$
        & \begin{tabular}[c]{@{}l@{}}$L(3N\!+\!M\!+\!8)+$\\$2M(N\!+\!8)\!+\!16N$\end{tabular} & $7\left(M\!+\!L \!+\!2N\right)$\\ 
         \arrayrulecolor{llgray}\hline \\[-0.2cm]      
        \begin{tabular}[c]{@{}l@{}} Turbo-BiG-\\AMP~\cite{TDingTWC2019}\end{tabular} 
        & \begin{tabular}[c]{@{}l@{}}$(M\!+\!L)(12N\!+\!10)+ (N\!+\!L)(7M\!+\!55)\!+\!$\\$(M\!+\!N)(7L\!+\!47) + (3(L\!+\!T')\!+\!2N\!-\!1)(N\!+\!T')$\end{tabular} & \begin{tabular}[c]{@{}l@{}}$12M\!+\!11(N\!+\!L)\!+\!$\\$3(N\!+\!T')$\end{tabular} & \begin{tabular}[c]{@{}l@{}}$L(3N\!+\!M\!+\!8)\!+$\\$\!2M(N\!+\!8)\!+$\\$\!16N\!+\!(N\!+\!T')L$\end{tabular} & $7(M\!+\!L\!+\!2N)$\\
         \arrayrulecolor{llgray}\hline \\[-0.2cm]           
        HyGAMP~\cite{SRanganTSP2017}  & $L\left(N\left(17M\!+\!72\right)\!+\!9M\right)$ & $L\left(3N\left(M\!+\!2\right)\!+\!5M\right)$ & $L\left(2N\left(M\!+\!4\right)\right)$ & $L\left(N\left(3M\!+\!1\right)\right)$\\
         \arrayrulecolor{llgray}\hline \\[-0.2cm]           
        \begin{tabular}[c]{@{}l@{}}BiMSGAMP-\\ type\end{tabular} & \begin{tabular}[c]{@{}l@{}}$(M+L)(12|\mathcal{S}|+10)+(|\mathcal{S}|+L)(7M+55)+$\\$(M+|\mathcal{S}|)(7L+47)+8|\mathcal{S}|(M+2)$\end{tabular} & \begin{tabular}[c]{@{}l@{}}$12M\!+\!3|\mathcal{S}|M+\!$\\ $11(|\mathcal{S}|\!+\!L)$\end{tabular} & \begin{tabular}[c]{@{}l@{}}$L(3|\mathcal{S}|\!+\!M\!+\!8)+$\\$2M(|\mathcal{S}|\!+\!8)\!+\!16|\mathcal{S}|$\end{tabular}& \begin{tabular}[c]{@{}l@{}}$7(M\!+\!L\!+\!2|\mathcal{S}|)$\\$\!+\!3|\mathcal{S}|M$\end{tabular}\\ \arrayrulecolor{black}\hline
    \end{tabular}
\end{table*}

    \else
        %
\begin{table}[t]
    \caption{Operations counting of considering techniques per iteration. Approaches that originally considered just the joint activity and signal detection have a separate channel estimation part, adapted using the same solution.}
    \label{tab:ch6_complexity}
    \centering
    \scriptsize
    \begin{tabular}{p{2cm}p{5cm}lp{2.0cm}l} \hline
        Algorithm & \begin{tabular}[c]{@{}l@{}}Additions, Subtractions\\[-0.25cm] and Multiplications\end{tabular} & Divisions and Square roots & Modulus & Sine, exp and log \\ \hline
        AMP~\cite{Donoho2009}   & $L\left[N\left(6M+1\right)+3M-2\right]$ & $L \left(M+3\right)$ & N/A & N/A\\
       \begin{tabular}[c]{@{}l@{}} Joint-EM-\\[-0.25cm] AMP~\cite{CWeiCommLet2017}\end{tabular}   & \begin{tabular}[c]{@{}l@{}}$ L_p\left[N(6M\!+\!1)\!+\!3M\!-\!2\right] \!+\!$\\ $L_d\left[(N\!+\!L)(10|\mathcal{A}|^2\!+\!16|\mathcal{A}|\!+\!\right.$\\ $7M\!+\!27)\!+\!(M\!+\!L)(9N\!+-\!2) \!+\!$\\
       $\left.N L(6|\mathcal{A}|^2\!+\!17|\mathcal{A}|\!+\!11)\!+-\!N)\right]$\end{tabular} & \begin{tabular}[c]{@{}l@{}} $L_p\left[M\!+\!3\right]+$\\$L_d\left[N(M\!+\!L\!+\!1)\!+\!\right.$\\$\left.2(N\!+\!L)(M\!+\!|\mathcal{A}|\!+\!1)\right]$\end{tabular}  & \begin{tabular}[c]{@{}l@{}}$(N\!+\!L)\times$\\$(3|\mathcal{A}|\!+\!M\!+\!1)\!+\!$\\$2N(|\mathcal{A}|\!+\!M)\!+$\\$N(\!L\!+\!1)$\end{tabular} & N/A\\
        BiG-AMP~\cite{JTParkerTSP2014} & \begin{tabular}[c]{@{}l@{}}$ (M+L)(12N+10)+$\\ $(N+L)(7M+55)+$\\$(M+N)(7L+47)$\end{tabular} & $12M+11(N+L)$
        & \begin{tabular}[c]{@{}l@{}}$L(3N\!+\!M\!+\!8)+$\\$2M(N\!+\!8)\!+\!16N$\end{tabular} & $7\left(M\!+\!L \!+\!2N\right)$\\        
        Turbo-BiG-AMP~\cite{TDingTWC2019}  & \begin{tabular}[c]{@{}l@{}}$(M\!+\!L)(12N\!+\!10)+$\\$(N\!+\!L)(7M\!+\!55)\!+\!$\\$(M\!+\!N)(7L\!+\!47)\!+\!$\\$(3(L\!+\!T')\!+\!2N\!-\!1)(N\!+\!T')$\end{tabular} & \begin{tabular}[c]{@{}l@{}}$12M\!+\!11(N\!+\!L)\!+\!$\\$3(N\!+\!T')$\end{tabular} & \begin{tabular}[c]{@{}l@{}}$L(3N\!+\!M\!+\!8)\!+$\\$\!2M(N\!+\!8)\!+$\\$\!16N\!+\!(N\!+\!T')L$\end{tabular} & $7(M\!+\!L\!+\!2N)$\\
        HyGAMP~\cite{SRanganTSP2017}  & $L\left(N\left(17M\!+\!72\right)\!+\!9M\right)$ & $L\left(3N\left(M\!+\!2\right)\!+\!5M\right)$ & $L\left(2N\left(M\!+\!4\right)\right)$ & $L\left(N\left(3M\!+\!1\right)\right)$\\
        \begin{tabular}[c]{@{}l@{}}BiMSGAMP-\\[-0.25cm] type\end{tabular} & \begin{tabular}[c]{@{}l@{}}$(M+L)(12|\mathcal{S}|+10)+$\\$(|\mathcal{S}|+L)(7M+55)+$\\$(M+|\mathcal{S}|)(7L+47)$\\$+8|\mathcal{S}|(M+2)$\end{tabular} & \begin{tabular}[c]{@{}l@{}}$12M\!+\!3|\mathcal{S}|M+\!$\\ $11(|\mathcal{S}|\!+\!L)$\end{tabular} & \begin{tabular}[c]{@{}l@{}}$L(3|\mathcal{S}|\!+\!M\!+\!8)+$\\$2M(|\mathcal{S}|\!+\!8)\!+\!16|\mathcal{S}|$\end{tabular}& \begin{tabular}[c]{@{}l@{}}$7(M\!+\!L\!+\!2|\mathcal{S}|)$\\$\!+\!3|\mathcal{S}|M$\end{tabular}\\
        \hline
    \end{tabular}
\end{table}

    \fi
    \makeatother

    The computational cost of BiMSGAMP-type schemes is analyzed below by counting each required numerical operation in terms of complex FLOPs. In particular, to provide a more precise comparison, Table~\ref{tab:ch6_complexity} separates the number of operations in four groups, since the number of required FLOPs is different, depending of the operation type. Thus, for a different number of devices $N$, proposed and state-of-the-art algorithms are compared. In the case of joint activity and data detection algorithms, in order to try to provide a fair comparison, a separate channel estimation part has been considered, where an adapted version of the same solution is considered. This approach has also been used in order to verify the performance of each solution.

    As it is clearly shown in Fig.~\ref{fig:ch6_flop}, a key benefit of using message-scheduling approaches is the computational cost saving. As explained before, while the state-of-the-art algorithms as BiG-AMP~\cite{JTParkerTSP2014} and HyGAMP~\cite{SRanganTSP2017} have $O(MN)$ messages to be computed, BiMSGAMP-type schemes demands $O(M|\mathcal{S}^{(i)}|)$. With the prediction that the mMTC scenario need to handle up to $300,000$ devices per cell~\cite{3GPPTR36888}, the gain of BiMSGAMP is evident since $|\mathcal{S}^{(i)}| << N$. In order to highlight this benefit, Table~\ref{tab:ch6_complexity} provides the number of operations needed for each state-of-the-art algorithm in terms of $N$ devices, $M$ BS antennas and $L = L_p + L_d$ frame size. For $N = [10,\, 250]$, $M = N/4$, $L_p = 64$ and $L_d = 128$, Fig.~\ref{fig:ch6_flop} shows that message-scheduling techniques dramatically reduce the computational cost, where BiMSGAMP-type schemes are less costly than most approaches. Note that since $|\mathcal{S}^{(i)}|$ of BiMSGAMP-AUD varies with each iteration, in order to compare the computational cost of every BiMSGAMP-type scheme we considered the mean values of each set size acquired in our simulations. As at each new iteration BiMSGAMP-RBP updates $0.1 N$ nodes, it requires a computational cost slightly higher than BiMSGAMP-AUD.

\subsection{Convergence}%

    In order to analyze the convergence of BiMSGAMP-type schemes, we devise an SE analysis under the large system limit. The MSE of BiMSGAMP is characterized via a set of simple one-dimensional equations that allow us to validate and compare the numerical results and the theoretical analysis.

    \subsubsection{Adaptive damping}
    We remark that the approximations made in the BiMSGAMP derivation presented in Section~\ref{sec:prob_form} are justified in the large system limit, i.e., the case where $M$, $N$, and $T \rightarrow \infty$ with fixed $M/N$ and $T/N$. However, the algorithm may diverge in practical applications. as these dimensions are finite. Thus, in order to avoid this issue, we use in our simulations the adaptive damping strategy, similar to the one described in~\cite{PSchniterTSP2015}. As seen in the literature~\cite{Rangan2012,JTParkerTSP2014}, the use of ``damping'' with GAMP/BiGAMP yields provable convergence guarantees with arbitrary matrices. As an example, let $\mu^{i} \in (0,1]$ be the damping factor applying to the parameters $\hat{p}_{mt}^{i}$ and $\nu^{pi}_{mt}$. With $\vartheta^{i}$ as the parameter to be updated, the damping factor is used as $\vartheta^{i} = \mu^{i}\vartheta^{i} + [1-\mu^{i}]\vartheta^{i-1}$, where we use $\mu^{i} = 0.95$ as in~\cite{Rangan2012,JTParkerTSP2014}.

    \subsubsection{State evolution}
    We characterize the SE of the BiMSGAMP algorithm. The main idea is to study its behaviour by evaluating its asymptotic MSE performance. Specifically, under the large system limit BiMSGAMP-type schemes efficiency can be fully described via a set of simple one-dimensional SE equations with the main derivation steps described below. Under the bilinear generalized model, we give a detailed SE derivation 
    that highlights the gains obtained by the message-scheduling techniques and the activity detection procedures.

    Following the assumptions of the SE analysis for AMP-like algorithms as in~\cite{Rangan2012} and~\cite{MBayatiTIT2011}, we consider the BiMSGAMP-scheme with scalar variances as $\nu_{nt}^{xi} \approx \frac{1}{|\mathcal{S}^{i}|T} \sum_{n=1}^{|\mathcal{S}^{i}|} \sum_{t=1}^T$ $\nu_{nt}^{xi} = \onu^{xi}$ and, similarly, $\nu_{nt}^{hi} \approx \onu^{hi}$ and $\nu_{mt}^{\hat{z}i} \approx \onu^{\hat{z}i}$. Thus, we can include these new values in order to rewrite the variance parameters in Algorithm 1 as given by
    \makeatletter%
    \if@twocolumn
       \begin{align}
            \nu_{mt}^{si} \approx& \, \left(\frac{1 - \nu^{\hat{z}i}}{\onu^{pi}}\right) \left(\frac{1}{\onu^{pi}}\right) = \onu^{si}
        \end{align}
        \begin{align}
            \nu^{ri}_{nt} \approx& \, \left(\frac{\onu^{si}}{|\mathcal{S}^{i}|}\, \sum_{m=1}^M \sum_{n=1}^{|\mathcal{S}^{i}|} |\hat{h}^{i}_{mn}|^2\right)^{-1} = \onu^{ri}, \text{ and }
        \end{align}
        \begin{align}
            \nu_{mt}^{pi} \approx& \, \frac{\onu^{hi}}{T} \sum_{n=1}^{|\mathcal{S}^{i}|} \sum_{t=1}^T\,  |\hat{x}^{i}_{nt}|^2\, +\\ \nonumber
            & \frac{\onu^{xi}}{M} \sum_{m=1}^{M} \sum_{t=1}^T\, |\hat{h}^i_{mn}|^2\, + |\mathcal{S}^{i}| \onu^{hi}\,\onu^{xi} = \onu^{pi},
        \end{align}

        \noindent where analogously, $\nu^{qi}_{mn} \approx \onu^{qi}$. The means are given by
        \begin{align}
            \hat{r}_{nt}^{i} = &\, \hat{x}_{nt}^i \left(\!1 - M\, \onu^{ri}\,\nu^{si}\, \onu^{hi}\,\right) + \onu^{ri} \sum_{m=1}^{M} \hat{h}^{i\ast}_{mn}\,\hat{s}^i_{mt},\\
            \hat{q}_{mn}^{i} = &\, \hat{h}^{i}_{mn} \left(\!1 - L_d\, \onu^{qi}\, \nu^{si}\,\onu^{xi}\right) + \\ \nonumber
            & \onu^{qi} \left(\sum_{t \in L_P} x^{\ast}_{nt}\,\hat{s}^i_{mt} + \sum_{t \in L_D} \hat{x}^{i\ast}_{nt}\,\hat{s}^i_{mt}\right),
        \end{align}
    \else
        \begin{align}
            \nu_{mt}^{pi} \approx &\, \frac{\onu^{hi}}{T} \sum_{n=1}^{|\mathcal{S}^{i}|} \sum_{t=1}^T\,  |\hat{x}^{i}_{nt}|^2\, + \frac{\onu^{xi}}{M} \sum_{m=1}^{M} \sum_{t=1}^T\, |\hat{h}^i_{mn}|^2\, + |\mathcal{S}^{i}| \onu^{hi}\,\onu^{xi} = \onu^{pi},
        \end{align}
        \begin{minipage}{0.45\linewidth}
            \begin{equation}
               \nu_{mt}^{si} \approx \, \left(\frac{1 - \nu^{\hat{z}i}}{\onu^{pi}}\right) \left(\frac{1}{\onu^{pi}}\right) = \onu^{si},
            \end{equation}
        \end{minipage}
        \begin{minipage}{0.5\linewidth}
            \begin{equation}
                \text{and \hspace{0.1cm}}\,  \nu^{ri}_{nt} \approx \, \left(\frac{\onu^{si}}{|\mathcal{S}^{i}|}\, \sum_{m=1}^M \sum_{n=1}^{|\mathcal{S}^{i}|} |\hat{h}^{i}_{mn}|^2\right)^{-1} = \onu^{ri},
            \end{equation}
        \end{minipage}

        \noindent where analogously, $\nu^{qi}_{mn} \approx \onu^{qi}$. The means are given by
        \begin{align}
            \hat{r}_{nt}^{i} = &\, \hat{x}_{nt}^i \left(\!1 - M\, \onu^{ri}\,\nu^{si}\, \onu^{hi}\,\right) + \onu^{ri} \sum_{m=1}^{M} \hat{h}^{i\ast}_{mn}\,\hat{s}^i_{mt},\\
            \hat{q}_{mn}^{i} = &\, \hat{h}^{i}_{mn} \left(\!1 - L_d\, \onu^{qi}\, \nu^{si}\,\onu^{xi}\right) + \onu^{qi} \left(\sum_{t \in L_P} x^{\ast}_{nt}\,\hat{s}^i_{mt} + \sum_{t \in L_D} \hat{x}^{i\ast}_{nt}\,\hat{s}^i_{mt}\right),
        \end{align}
    \fi
    \makeatother

\noindent which builds the scalar-variance BiMSGAMP algorithm. Considering two \textit{pseudo-Lipschitz} functions, $\varphi\left(\cdot\right)$ and $\psi\left(\cdot\right)$, we state the first main assumption:

\noindent \textbf{Assumption 1} \textit{The mean-related parameters} $y_{mt}, p_{mt}^i, z_{mt}, \hat{r}_{nt}^{i}, x_{nt}^{i}, \hat{q}_{mn}^{i}$ and $h_{mn}^{i}$ \textit{empirically converge to the following random variables with second order moments}
    \makeatletter%
    \if@twocolumn
        \begin{align}\label{eq:AppC_Assump1}
            & \underset{T, |\mathcal{S}^{i}| \rightarrow \infty}{\text{lim}} \left\{y_{mt}, p_{mt}^i, z_{mt}, \hat{r}_{nt}^{i}, x_{nt}^{i}, \hat{q}_{mn}^{i}, h_{mn}^{i}\right\} \overset{\text{PL(2)}}{=}\\ \nonumber
            & \hspace{1.5cm}\left\{\bsfy, \bsfp^{i}, \bsfz^{i}, \bsfr^{i}, \bsfx^{i}, \bsfq^{i}, \bsfh^{i}\right\}
        \end{align}
    \else
        \begin{equation}\label{eq:AppC_Assump1}
            \underset{T, |\mathcal{S}^{i}| \rightarrow \infty}{\text{lim}} \left\{y_{mt}, p_{mt}^i, z_{mt}, \hat{r}_{nt}^{i}, x_{nt}^{i}, \hat{q}_{mn}^{i}, h_{mn}^{i}\right\} \overset{\text{PL(2)}}{=} \left\{\bsfy, \bsfp^{i}, \bsfz^{i}, \bsfr^{i}, \bsfx^{i}, \bsfq^{i}, \bsfh^{i}\right\}
        \end{equation}
    \fi
    \makeatother

Based on this assumption, the goal is to compute the asymptotic MSE of $i-$th iteration of $\hat{\bX}^i$, $\hat{\bH}^i$ and $\hat{\bZ}^i$. Thus, the next steps are the particularization of the \textit{pseudo-Lipchitz} continuous functions to compute the equivalent mean and variances.

\noindent \textbf{Proposition 1} \textit{Equivalently for $\hat{\bH}^{i}$ and $\hat{\bZ}^{i}$, the asymptotic MSE of the $i$-th iteration of $\hat{\bX}^{i}$, is almost sure identical to $\onu^{xi}$ and $\mathbb{E}_{\bsfr^{i-1}}\left\{\psi^{(x)}(\bsfx^{i-1})\right\}$.}

Recalling that the approximate posterior distribution $\mathcal{P}\left(x_{nt}^{i}|y\right)$ has as mean and variance $\hat{x}^{i}_{nt}$ and $\nu^{xi}_{nt}$, in the $\textsf{MSE}\left(\bX^i\right)$ expression they are computed as\\

    \begin{minipage}{0.45\linewidth}
        \begin{equation}
            \varphi^{(x)}\left(r_{nt}^{i-1}\right) = \hat{x}^{i}_{nt},
        \end{equation}
    \end{minipage}
    \begin{minipage}{0.5\linewidth}
        \begin{equation} \label{eq:AppC_pseudoLip}
            \text{and \hspace{0.1cm}}\,  \psi^{(x)}\left(r_{nt}^{i-1}\right) = \nu^{xi}_{nt}.
        \end{equation}
    \end{minipage}

As for the asymptotic MSE of $\hat{\bX}^{i}$, we can write
    \makeatletter%
    \if@twocolumn
        \begin{align}
            &\textsf{mse}\left(\bX^i\right) =\, \underset{T, |\mathcal{S}^{i}| \rightarrow \infty}{\text{lim}} \frac{1}{T |\mathcal{S}^{i}|} \|\hat{\bX}^{i} - \bX\|^2_\text{F}\\ \nonumber &\hspace{0.5cm} =\, \underset{T, |\mathcal{S}^{i}| \rightarrow \infty}{\text{lim}} \sum_{n=1}^{|\mathcal{S}^{i}|} \sum_{t=1}^{T} \left(\hat{x}^{i}_{nt} - x_{nt}^{i}\right)^2\, =\, \mathbb{E}_{\bsfr^{i-1}}\left\{\psi^{(x)}\left(\bsfr^{i-1}\right)\right\}
        \end{align}
    \else
        \begin{align}
            \textsf{MSE}\left(\bX^i\right) =&\, \underset{T, |\mathcal{S}^{i}| \rightarrow \infty}{\text{lim}} \frac{1}{T |\mathcal{S}^{i}|} \|\hat{\bX}^{i} - \bX\|^2_\text{F} \,=\, \underset{T, |\mathcal{S}^{i}| \rightarrow \infty}{\text{lim}} \sum_{n=1}^{|\mathcal{S}^{i}|} \sum_{t=1}^{T} \left(\hat{x}^{i}_{nt} - x_{nt}^{i}\right)^2\, =\, \mathbb{E}_{\bsfr^{i-1}}\left\{\psi^{(x)}\left(\bsfr^{i-1}\right)\right\}
        \end{align}
    \fi
    \makeatother

\noindent where the last equality can be obtained by empirical convergence. Rewriting and using (\ref{eq:AppC_pseudoLip}), we have
    \makeatletter%
    \if@twocolumn
        \begin{align}
            \textsf{mse}\left(\bX^i\right) =&\, \frac{1}{|\mathcal{S}^{i}| T} \sum_{n=1}^{|\mathcal{S}^{i}|} \sum_{t=1}^{T} \varphi^{(x)}\left(r_{nt}^{i-1}\right)\\
            \nonumber
            =&\,\mathbb{E}_{\bsfr^{i-1}}\left\{\psi^{(x)}\left(\bsfr^{i-1}\right)\right\} = \overline{\hat{\nu}}^{xi}
        \end{align}
    \else
        \begin{align}
            \textsf{MSE}\left(\bX^i\right) =&\, \frac{1}{|\mathcal{S}^{i}| T} \sum_{n=1}^{|\mathcal{S}^{i}|} \sum_{t=1}^{T} \varphi^{(x)}\left(r_{nt}^{i-1}\right) =\,\mathbb{E}_{\bsfr^{i-1}}\left\{\psi^{(x)}\left(\bsfr^{i-1}\right)\right\} = \overline{\hat{\nu}}^{xi}
        \end{align}
    \fi
    \makeatother

\noindent and, similarly for the asymptotic MSE of $\hat{\bH}^i$, $\hat{\bZ}^i$ and $\hat{\brho}$, we have:
    \makeatletter%
    \if@twocolumn
        \begin{align}\nonumber
            \textsf{mse}\left(\bH^i\right) =&\, \underset{M, |\mathcal{S}^{i}| \rightarrow \infty}{\text{lim}} \frac{1}{M |\mathcal{S}^{i}|} \|\hat{\bH}^{i} - \bH\|^2_\text{F} = \overline{\hat{\nu}}^{hi}\\
            =&\, \mathbb{E}_{\bsfz^{i},\bsfq^{i}}\left\{\psi^{(h)}\left(\bsfh^{i}, \bsfq^{i}\right)\right\} \\
            \textsf{mse}\left(\bZ^i\right) =&\, \underset{M, T \rightarrow \infty}{\text{lim}} \frac{1}{M T} \|\hat{\bZ}^{i} - \bZ\|^2_\text{F} = \overline{\hat{\nu}}^{zi} = \mathbb{E}_{\bsfz^{i}}\left\{\psi^{(z)}\left(\bsfp^{i}\right)\right\}
        \end{align}

    \noindent where

        \begin{equation*}
            \psi^{h}\left(h_{mn}, q^{i}_{mn}\right) = \nu_{mn}^{hi} \hspace{0.3cm} \textrm{(\the\numexpr\value{equation}+1\relax)} \hspace{0.2cm}\text{and}\hspace{0.2cm}  \psi^{z}\left(\hat{p}_{mt}^{i}\right) = \nu_{mt}^{\hat{z}i}  \hspace{0.3cm} \textrm{(\the\numexpr\value{equation}+2\relax)}.
        \end{equation*}
    \else
        \begin{align}
            \textsf{MSE}\left(\bH^i\right) =&\, \underset{M, |\mathcal{S}^{i}| \rightarrow \infty}{\text{lim}} \frac{1}{M |\mathcal{S}^{i}|} \|\hat{\bH}^{i} - \bH\|^2_\text{F} = \overline{\hat{\nu}}^{hi} = \mathbb{E}_{\bsfz^{i},\bsfq^{i}}\left\{\psi^{(h)}\left(\bsfh^{i}, \bsfq^{i}\right)\right\} \\
            \textsf{MSE}\left(\bZ^i\right) =&\, \underset{M, T \rightarrow \infty}{\text{lim}} \frac{1}{M T} \|\hat{\bZ}^{i} - \bZ\|^2_\text{F} = \overline{\hat{\nu}}^{zi} = \mathbb{E}_{\bsfz^{i}}\left\{\psi^{(z)}\left(\bsfp^{i}\right)\right\}
        \end{align}

    \vspace{0.5cm}
    \noindent where           \hspace{-0.5cm}
        \begin{minipage}{0.45\linewidth}
            \begin{equation}
              \psi^{h}\left(h_{mn}, q^{i}_{mn}\right) = \nu_{mn}^{hi}
            \end{equation}
        \end{minipage}
        \begin{minipage}{0.5\linewidth}
            \begin{equation}
                \text{and \hspace{0.1cm}}\,  \psi^{z}\left(\hat{p}_{mt}^{i}\right) = \nu_{mt}^{\hat{z}i}.
            \end{equation}
        \end{minipage}
        \vspace{0.5cm}
    \fi
    \makeatother

The next step is to derive the asymptotic MSEs of those MMSE estimators. Omitting the iteration index $i$ for simplicity, we start with the variance $\overline{\nu}^{\hat{z}}$, as given by
\setcounter{equation}{64}
    \begin{align} \label{eq:AppC_dif_meanZ}
        \overline{\nu}^{\hat{z}} =& \mathbb{E}_{\bsfz}\left\{\psi^{(z)}\left(\bsfz\right)\right\} = \mathbb{E}_{\bsfz}\left\{\mathbb{E}\left\{|z|^2\right\} - |\mathbb{E}\left\{z\right\}|^2\right\} = \chi_z - \varrho_z
    \end{align}

\noindent where the inner expectation is taken over the approximate posterior distribution $\mathcal{P}\!\left(z|\hat{p}\right)$
    \begin{align}
        \zeta_{mt} = \mathcal{P}\!\left(z|\hat{p}\right) = \frac{\mathcal{P}\!\left(y_{mt}|z_{mt}\right)\mathcal{N}_c\left(z_{mt}|p_{mt},\nu_{mt}^{\hat{p}}\right)}{\int \mathcal{P}\!\left(y_{mt}|z\right)\mathcal{N}_c\left(z|p_{mt},\nu_{mt}^{p}\right)\text{d}z}.
    \end{align}

Moreover, the distribution $\mathcal{P}(\hat{p})$ can be obtained by solving the following equation
    \begin{equation}\label{eq:AppC_pZP}
        \int \mathcal{P}\left(\hat{p}\right)\, \mathcal{P}\left(z|\hat{p}\right)\,\text{d}z = \mathcal{P}\left(z\right).
    \end{equation}
In the large system limit, $z$ can be seen as a Gaussian random variable with zero mean and variance
    \makeatletter%
    \if@twocolumn
        \begin{equation} \label{eq:AppCMeanz}
            \begin{split}
                \chi_z =&\, \mathbb{E}\left\{\left(\sum_{n=1}^{|\mathcal{S}^{i}|} h_{mn}\,x_{nt}\right)\left(\sum_{r=1}^{|\mathcal{S}^{i}|} h_{mr}\,x_{rt}\right)\right\} \\
                =&\, \sum_{n=1}^{|\mathcal{S}^{i}|} \mathbb{E}\left\{\left(h_{mn}\right)^2\left(x_{nt}\right)^2\right\} = |\mathcal{S}^{i}|\,\chi_h\,\chi_x,
            \end{split}
        \end{equation}

        \noindent where
        \vspace{-0.2cm}
            \begin{equation}\label{eq:AppCMeanh}
                   \chi_h =\, \int h^2\, \mathcal{P}\!\left(h;\hat{\rho}\right)\text{d}h
            \end{equation}

        \vspace{-0.2cm}
            and
            \begin{equation} \label{eq:AppCMeanx}
                \chi_x =\, \int x^2\, \sum_c \sum_s \mathcal{P}\!\left(x,c,s\right)\text{d}x.
            \end{equation}
        As a result, solving (\ref{eq:AppC_pZP}) yields $\mathcal{P}\!\left(\hat{p}\right) = \mathcal{N}_c\left(\hat{p} | 0,\chi_z - \nu^p\right)$ and, with $\text{D}{\iota} = \mathcal{N}_c\left(\iota|0,1\right)$, $\varrho_z$ is given by

            \begin{equation}\label{eq:AppCVarz}
            \resizebox{.90\hsize}{!}{$
                \varrho_z =\! \int \frac{\left[\int z\,p_{y|z}\left(y|z\right)\mathcal{N}_c\left(z|\sqrt{|\mathcal{S}^{i}|\varrho_x\varrho_h}\,\iota,\nu^p\right)\text{d}z\right]^2}{\int p_{y|z}\left(y|z\right)\mathcal{N}_c\left(z|\sqrt{|\mathcal{S}^{i}|\varrho_x\varrho_h}\,\iota,\nu^p\right)\text{d}z}\text{D}\iota\,\text{d}y.
                $}
            \end{equation}

        Naturally, the computation of $\onu^x$ and $\onu^h$ follows $\onu^z$. Thus, remembering that $\onu^{x}$ refers to the MSE associated with the approximate posterior $\mathcal{P}(x|y)$ and $\onu^h$ to $\mathcal{P}(h;\hat{\rho}|y)$ and depends on the known prior $\mathcal{P}(h;\hat{\rho}) = \hat{\rho}\mathcal{N}_c\left(h|0,\sigma^2_h\right) + \left(1-\hat{\rho}\right)\delta(h)$, we can get

            \begin{align} \label{eq:AppCVarx}
                \varrho_x = \int \frac{\left[\int x\, \sum_c \sum_s \mathcal{P}\!\left(x,c,s\right)\mathcal{N}_c\left(x|\zeta,\nu^q\right)\text{d}x\right]^2}{\int \sum_c \sum_s \mathcal{P}\!\left(x,c,s\right)\mathcal{N}_c\left(x|\zeta,\nu^q\right)\text{d}x}\, \text{d}\zeta,
            \end{align}

        \noindent and
        \vspace{-0.2cm}
            \begin{align}\label{eq:AppCVarh}
                \varrho_h = \int \frac{\left[\int h\, \mathcal{P}\!\left(h;\hat{\rho}\right)\mathcal{N}_c\left(h|\zeta,\nu^r\right)\text{d}h\right]^2}{\int \mathcal{P}\!\left(h|\hat{\rho}\right)\mathcal{N}_c\left(h|\zeta,\nu^r\right)\text{d}h}\, \text{d}\zeta.
            \end{align}

        One can notice that the variance related parameters $\nu^{p}$, $\nu^r$ and $\nu^q$ and the activity detection $\hat{\rho}$ have impact on $\onu^{\hat{z}}$, $\onu^x$ and $\onu^h$. We thus apply the results above to represent those variance related parameters, which yields

            \begin{align}\label{eq:AppCVarp}
                \nu^{p} =& |\mathcal{S}^{i}|\left(\chi_x \chi_h - \varrho_x \varrho_h\right), \\
                \label{eq:AppCVars}
                \nu^s =& \frac{\varrho_z - |\mathcal{S}^{i}|\varrho_x \varrho_h}{|\mathcal{S}^{i}|^2\left(\chi_x \chi_h - \varrho_x \varrho_h\right)^2},\\
                \label{eq:AppCVarr}
                \nu^r =& \frac{|\mathcal{S}^{i}|^2 \left(\chi_x \chi_h - \varrho_x \varrho_h\right)^2}{M \varrho_h\left(\varrho_z - |\mathcal{S}^{i}| \varrho_x \varrho_h\right)},\\
                \label{eq:AppCVarq}
                \nu^q =& \frac{|\mathcal{S}^{i}|^2 \left(\chi_x \chi_h - \varrho_x \varrho_h\right)^2}{T \varrho_x\left(\varrho_z - |\mathcal{S}^{i}| \varrho_x \varrho_h\right)}
            \end{align}

        \vspace{0.5cm}
         \noindent and for the activity detection part, using~(\ref{eq:AppCVarr}), we have
            \begin{align}\label{eq:AppCrho}
                \hat{\rho} = \frac{\mathcal{N}_c\left(0 | \hat{r}, \nu^r + \chi_x\right)}{\mathcal{N}_c\left(0 | \hat{r}, \nu^r + \chi_x\right) + \mathcal{N}_c\left(0 | \hat{r}, \nu^r\right)}.
            \end{align}

            Therefore, the SE of the proposed BiMSGAMP-schemes is given by (\ref{eq:AppCMeanz})-(\ref{eq:AppCrho}). Since each message-scheduling technique consider just a set $|S^i|$ of the $N$ devices (nodes), instead of compute the mean values with $N$, we considered the size of the set $|S^i|$. As the analysis is based on the large system limit, that is, when $N,\, T \rightarrow \infty$, the assumption still valid. Another important point is the inclusion of the instantaneous activity detection in the procedure. Present in the prior density of the channel, as long as the iteration marker grows, the estimated probability of being active $\hat{\rho}$ is refined and, consequently, a more accurate channel and signal means and variances are obtained, i.e., smaller MSE.

    \else
         \begin{align} \label{eq:AppCMeanz}
            \chi_z =&\, \mathbb{E}\left\{\left(\sum_{n=1}^{|\mathcal{S}^{i}|} h_{mn}\,x_{nt}\right)\left(\sum_{r=1}^{|\mathcal{S}^{i}|} h_{mr}\,x_{rt}\right)\right\} \,=\, \sum_{n=1}^{|\mathcal{S}^{i}|} \mathbb{E}\left\{\left(h_{mn}\right)^2\left(x_{nt}\right)^2\right\} = |\mathcal{S}^{i}|\,\chi_h\,\chi_x,
        \end{align}

        \noindent where

            \begin{minipage}{0.45\linewidth}
                \begin{equation}\label{eq:AppCMeanh}
                    \chi_h =\, \int h^2\, \mathcal{P}\!\left(h;\hat{\rho}\right)\text{d}h
                \end{equation}
            \end{minipage}
            \begin{minipage}{0.5\linewidth}
                \begin{equation} \label{eq:AppCMeanx}
                    \text{and \hspace{0.1cm}}\, \chi_x =\, \int x^2\, \sum_c \sum_s \mathcal{P}\!\left(x,c,s\right)\text{d}x.
                \end{equation}
            \end{minipage}

        As a result, solving (\ref{eq:AppC_pZP}) yields $\mathcal{P}\!\left(\hat{p}\right) = \mathcal{N}_c\left(\hat{p} | 0,\chi_z - \nu^p\right)$ and, with $\text{D}{\iota} = \mathcal{N}_c\left(\iota|0,1\right)$, $\varrho_z$ is given by
            \begin{align}\label{eq:AppCVarz}
                \varrho_z = \int \frac{\left[\int z\,p_{y|z}\left(y|z\right)\mathcal{N}_c\left(z|\sqrt{|\mathcal{S}^{i}|\varrho_x\varrho_h}\,\iota,\nu^p\right)\text{d}z\right]^2}{\int p_{y|z}\left(y|z\right)\mathcal{N}_c\left(z|\sqrt{|\mathcal{S}^{i}|\varrho_x\varrho_h}\,\iota,\nu^p\right)\text{d}z}\text{D}\iota\,\text{d}y.
            \end{align}

        Naturally, the computation of $\onu^x$ and $\onu^h$ follows $\onu^z$. Thus, recalling that $\onu^{x}$ refers to the MSE associated with the approximate posterior $\mathcal{P}(x|y)$ and $\onu^h$ to $\mathcal{P}(h;\hat{\rho}|y)$ and depends on the known prior $\mathcal{P}(h;\hat{\rho}) = \hat{\rho}\mathcal{N}_c\left(h|0,\sigma^2_h\right) + \left(1-\hat{\rho}\right)\delta(h)$, we can get

            \begin{align} \label{eq:AppCVarx}
                \varrho_x = \int \frac{\left[\int x\, \sum_c \sum_s \mathcal{P}\!\left(x,c,s\right)\mathcal{N}_c\left(x|\zeta,\nu^q\right)\text{d}x\right]^2}{\int \sum_c \sum_s \mathcal{P}\!\left(x,c,s\right)\mathcal{N}_c\left(x|\zeta,\nu^q\right)\text{d}x}\, \text{d}\zeta,
            \end{align}

        \noindent and
            \begin{align}\label{eq:AppCVarh}
                \varrho_h = \int \frac{\left[\int h\, \mathcal{P}\!\left(h;\hat{\rho}\right)\mathcal{N}_c\left(h|\zeta,\nu^r\right)\text{d}h\right]^2}{\int \mathcal{P}\!\left(h|\hat{\rho}\right)\mathcal{N}_c\left(h|\zeta,\nu^r\right)\text{d}h}\, \text{d}\zeta.
            \end{align}

        One can notice that the variance related parameters $\nu^{p}$, $\nu^r$ and $\nu^q$ and the activity detection $\hat{\rho}$ have impact on $\onu^{\hat{z}}$, $\onu^x$ and $\onu^h$. We thus apply the results above to represent those variance related parameters, which yields

            \begin{minipage}{0.45\linewidth}
                \begin{equation}\label{eq:AppCVarp}
                    \nu^{p} = |\mathcal{S}^{i}|\left(\chi_x \chi_h - \varrho_x \varrho_h\right),
                \end{equation}
            \end{minipage}
            \begin{minipage}{0.45\linewidth}
                \begin{equation} \label{eq:AppCVars}
                    \nu^s = \frac{\varrho_z - |\mathcal{S}^{i}|\varrho_x \varrho_h}{|\mathcal{S}^{i}|^2\left(\chi_x \chi_h - \varrho_x \varrho_h\right)^2},
                \end{equation}
            \end{minipage}

            \begin{minipage}{0.45\linewidth}
                \begin{equation}\label{eq:AppCVarr}
                    \nu^r = \frac{|\mathcal{S}^{i}|^2 \left(\chi_x \chi_h - \varrho_x \varrho_h\right)^2}{M \varrho_h\left(\varrho_z - |\mathcal{S}^{i}| \varrho_x \varrho_h\right)},
                \end{equation}
            \end{minipage}
            \begin{minipage}{0.45\linewidth}
                \begin{equation} \label{eq:AppCVarq}
                    \nu^q = \frac{|\mathcal{S}^{i}|^2 \left(\chi_x \chi_h - \varrho_x \varrho_h\right)^2}{T \varrho_x\left(\varrho_z - |\mathcal{S}^{i}| \varrho_x \varrho_h\right)}
                \end{equation}
            \end{minipage}

        \vspace{0.5cm}
         \noindent and for the activity detection part, using~(\ref{eq:AppCVarr}), we have
            \begin{align}\label{eq:AppCrho}
                \hat{\rho} = \frac{\mathcal{N}_c\left(0 | \hat{r}, \nu^r + \chi_x\right)}{\mathcal{N}_c\left(0 | \hat{r}, \nu^r + \chi_x\right) + \mathcal{N}_c\left(0 | \hat{r}, \nu^r\right)}.
            \end{align}

            Therefore, the SE of the proposed BiMSGAMP-schemes is given by (\ref{eq:AppCMeanz})-(\ref{eq:AppCrho}). Since each message-scheduling technique consider just a set $|S^i|$ of the $N$ devices (nodes), instead of compute the mean values with $N$, we considered the size of the set $|S^i|$. As the analysis is based on the large system limit, that is, when $N,\, T \rightarrow \infty$, the assumption is still valid. Another important point is the inclusion of the instantaneous activity detection in the procedure. Present in the prior density of the channel, as long as the iteration marker grows, the estimated probability of being active $\hat{\rho}$ is refined and, consequently, a more accurate channel and signal means and variances are obtained, i.e., smaller MSE.
    \fi
    \makeatother

\section{Numerical results} \label{sec:Num_res}

    In this section we provide numerical results in order to evaluate and compare the BiMSGAMP schemes with the literature. We start with the convergence analysis, where we study in Figs.~\ref{fig:conv_figures}(a)-(c) the behaviour of BiMSGAMP in terms of the NMSE of the channel estimation, activity and data detection in the asynchronous scenario. For this study, besides the BiMSGAMP-type schemes, we considered a bilinear version of the HyGAMP algorithm~\cite{SRanganTSP2017}.  For channel estimation, notice in Figs.~\ref{fig:conv_figures}(a) that the residual-based metric displays jumps on the convergence due to the group update. The neglected nodes have a considerable influence in low-SNRs scenarios, as when the set $\mathcal{S}$ is empty all nodes are updated, which fits with the ``jumps'' on iterations. We remark that for the convergence analysis all the channels are considered and that HyGAMP's and BiMSGAMP-AUD's performances are almost the same. Regarding the activity detection, the jumping behaviour of BiMSGAMP disappears since it is computed only after the procedure. Furthermore, although for SNR values less than $5$ dB the convergence of HyGAMP and BiMSGAMP-AUD is quite similar, from SNR $= 10$ dB, BiMSGAMP-AUD outperforms HyGAMP up to SNR $= 20$ dB, where all schemes perform equally.  The convergence performance in channel NMSE is similar but the ``jumps'' of BiMSGAMP-RBP vanishes from SNR $= 10$ dB.
    In most scenarios BiMSGAMP-type solutions converge equally or faster than HyGAMP, but with a considerable computational cost saving. Even with more iterations to reach convergence, using BiMSGAMP-RBP with the expected massive number of devices requiring connection, dynamic scheduling approaches outperform algorithms with message passing in parallel.
    Regarding the data NMSE, we also evaluate the SE of BiMSGAMP-AUD, comparing its simulation results with the theory, under the synchronous and the asynchronous mMTC scenarios, beyond different SNR values. One can notice in Fig.~\ref{fig:NMSE_it} that the asymptotic prediction given by the iterative equations given by the steady-evolution derived in the last section matches the simulation results.
    \makeatletter%
    \if@twocolumn
        \begin{figure*}[t]
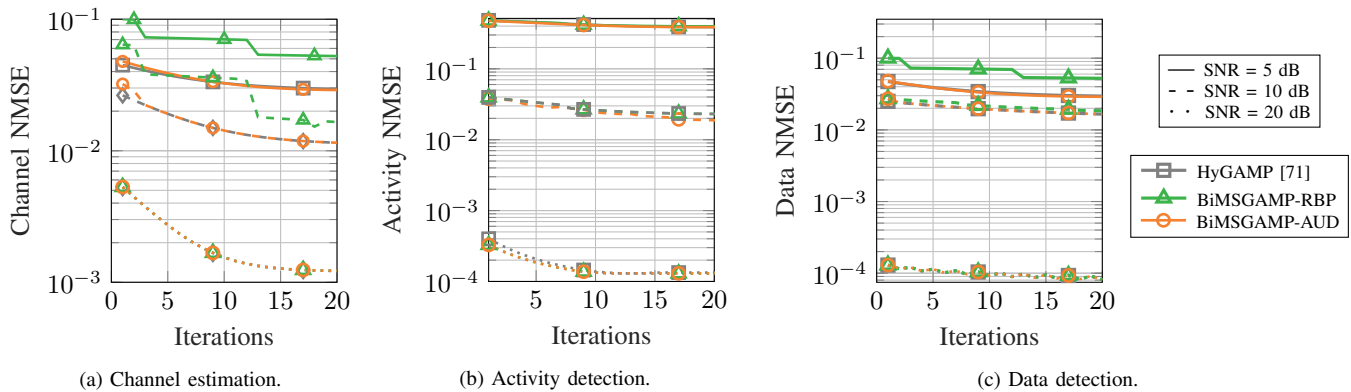

            \begin{minipage}[b]{0.25\linewidth}
                \centerline{\input{conv_channel.tex}}
                \centerline{\footnotesize (a) Channel estimation.}\medskip
            \end{minipage}
            \hspace{0.2cm}
            \begin{minipage}[b]{0.25\linewidth}
                \centerline{\input{conv_aud.tex}}
                \centerline{\footnotesize (b) Activity detection.}\medskip
            \end{minipage}
            \hspace{1cm}
            \begin{minipage}[b]{0.35\linewidth}
                \centerline{\input{conv_data.tex}}
                \centerline{\footnotesize (c) Data detection.}\medskip
            \end{minipage}
             \vspace{-0.5em}
            \caption{Convergence of channel estimation and data detection in terms of NMSE for different SNR values.}
            \label{fig:conv_figures}%
        \end{figure*}
    \else
        \begin{figure}[t]
            \begin{minipage}[b]{0.25\linewidth}
                \centerline{\input{conv_channel.tex}}
                \centerline{\footnotesize (a) Channel estimation.}\medskip
            \end{minipage}
            \hspace{0.2cm}
            \begin{minipage}[b]{0.25\linewidth}
                \centerline{\input{conv_aud.tex}}
                \centerline{\footnotesize (b) Activity detection.}\medskip
            \end{minipage}
            \hspace{1cm}
            \begin{minipage}[b]{0.35\linewidth}
                \centerline{\input{conv_data.tex}}
                \centerline{\footnotesize (c) Data detection.}\medskip
            \end{minipage}
             \vspace{-2em}
            \caption{Convergence of channel estimation and data detection in terms of NMSE for different SNR values.}
            \label{fig:conv_figures}%
        \end{figure}
    \fi
    \makeatother

    In order to assess the proposed schemes, the well-known approximate message passing (AMP)~\cite{Donoho2009}, and state-of-the-art solutions as the joint expectation-maximization AMP (Joint-EM-AMP)~\cite{CWeiCommLet2017}, a bilinear version of AMP (BiG-AMP)~\cite{JTParkerTSP2014}, HyGAMP~\cite{SRanganTSP2017} and the Turbo-BiG-AMP~\cite{TDingTWC2019} are used for comparison. HyGAMP includes a loopy belief propagation (LBP) part for user activity detection before the GAMPs factor graph, refining the AUD. The Joint-EM-AMP uses expectation maximization (EM) algorithm to perform the activity detection, while the means and variances for signal detection are provided by AMP. Turbo-BiG-AMP is a modification of BiG-AMP that is designed for an asynchronous scenario and aims to locate the beginning of the each received frame. As a lower bound, we consider the oracle HyGAMP (OHyGAMP), a version of HyGAMP with perfect activity detection.

    Averaging the results over $10^4$ runs, we consider an uplink under-determined mMTC system with $N=100$ devices with a single antenna each transmitting to a BS equipped with $M=32$ antennas. All the simulated schemes experience a block-fading channel model as described in (\ref{eq:Hasy}). In each observation window, the number of active devices vary, but this number is limited by 10\% of $N$. The channel coding considered is LDPC with rate $1/2$ and the modulation scheme is QPSK. The balance between pilots and data is $L_p = 64$, $L_d = 128$, where the pilots are given in Section~\ref{sec:SysMod} and the symbols are modulated after channel coding with block length of 256 bits. The average SNR is given by $10\log\left(NR \sigma^2_x/\sigma^2_w\right)$.

    \makeatletter%
    \if@twocolumn
            \begin{figure}[t]
                \centering
\begin{tikzpicture}
\begin{axis}[%
width=5.4cm,
height=4.2cm,
at={(0.758in,0.481in)},
scale only axis,
xmin=0.5,
xmax=20,
xlabel style={font=\color{white!15!black}},
xlabel={Iterations},
ymin=-40,
ymax=15,
yminorticks=true,
ylabel style={font=\color{white!15!black}},
ylabel={NMSE (dB)},
axis background/.style={fill=white},
xmajorgrids,
ymajorgrids,
yminorgrids,
legend style={at={(1.2,0.1)}, anchor=south west,  font=\scriptsize, legend cell align=left, align=left, draw=white!15!black}
]

\draw (2.0,-3.6) -- (2.0,1);   
\node[coordinate] (A) at (axis cs:2.0,-7) {};  
\draw[black] (A) ellipse (0.4 and 3.5);           
\node[coordinate,pin={[pin distance=0cm]above:{\tiny Synchronous}}] at (axis cs:2.9,0.05){};  
\draw (3.0,-28.5) -- (3.0,-35);   
\node[coordinate] (B) at (axis cs:3,-21.5) {};        
\draw[black] (B) ellipse (0.4 and 7);           
\node[coordinate,pin={[pin distance=0cm]below:{\tiny Asynchronous}}] at (axis cs:3.0,-33){};  

\addplot [color=\AUDcolor, line width=1.0pt]
  table[row sep=crcr]{%
1	-7.05653615676392\\
2	-12.4976510583977\\
3	-15.4196376618839\\
4	-16.1268361937189\\
5	-16.2468220651174\\
6	-16.2735231475593\\
7	-16.2733430388164\\
8	-16.2794538139838\\
9	-16.2934705164771\\
10	-16.2991764293592\\
11	-16.2981573201258\\
12	-16.2969280332425\\
13	-16.2969009965628\\
14	-16.296534942621\\
15	-16.2958589017621\\
16	-16.2954375903616\\
17	-16.2953159629034\\
18	-16.2954509949638\\
19	-16.2956827533673\\
20	-16.2958050685182\\
};
\label{0dBSimA}

\addplot [color=\AUDcolor, only marks, line width=1.0pt, mark=o,mark size = 2.0pt, mark options={solid, \AUDcolor}]
  table[row sep=crcr]{%
1	-7.41015645005781\\
2	-13.3131230490127\\
3	-15.8637958632638\\
4	-16.4036459770331\\
5	-16.4876980659147\\
6	-16.4999869187872\\
7	-16.5017662489079\\
8	-16.5020235169744\\
9	-16.5020607069663\\
10	-16.5020660828936\\
11	-16.5020668599971\\
12	-16.5020669723293\\
13	-16.5020669885671\\
14	-16.5020669909144\\
15	-16.5020669912536\\
16	-16.5020669913027\\
17	-16.5020669913098\\
18	-16.5020669913108\\
19	-16.5020669913109\\
20	-16.502066991311\\
};
\label{0dBTheA}

\addplot [color=\AUDcolor, dashed, line width=1.0pt]
  table[row sep=crcr]{%
1	-7.14926358299793\\
2	-14.3575458200581\\
3	-19.5134474355358\\
4	-21.3111609057399\\
5	-21.7130577767893\\
6	-21.7887001115347\\
7	-21.7870892029482\\
8	-21.7876154247081\\
9	-21.7940534139163\\
10	-21.797986010932\\
11	-21.7980394758936\\
12	-21.7973379131285\\
13	-21.7971062928673\\
14	-21.7971095051645\\
15	-21.7971329854122\\
16	-21.79714162293\\
17	-21.7971436898855\\
18	-21.7971551370563\\
19	-21.7971686833329\\
20	-21.7971679270932\\
};
\label{5dBSimA}

\addplot [color=\AUDcolor, only marks, line width=1.0pt, mark=diamond,mark size = 2.0pt, mark options={solid, \AUDcolor}]
  table[row sep=crcr]{%
1	-7.85438766906154\\
2	-15.6427091966653\\
3	-20.7913548201673\\
4	-22.2035477040039\\
5	-22.4062250495602\\
6	-22.4308170702269\\
7	-22.4337314845836\\
8	-22.4340758908977\\
9	-22.4341165768366\\
10	-22.4341213830192\\
11	-22.4341219507653\\
12	-22.434122017832\\
13	-22.4341220257546\\
14	-22.4341220266904\\
15	-22.4341220268009\\
16	-22.4341220268141\\
17	-22.4341220268156\\
18	-22.4341220268157\\
19	-22.4341220268157\\
20	-22.4341220268157\\
};
\label{5dBTheA}

\addplot [color=\AUDcolor, dotted, line width=1.0pt]
  table[row sep=crcr]{%
1	-7.51498571271837\\
2	-15.8633695658748\\
3	-24.1092836194969\\
4	-30.5258582801552\\
5	-32.8237435088396\\
6	-33.2143093090731\\
7	-33.2598813119969\\
8	-33.2544392749027\\
9	-33.2563335429961\\
10	-33.2606101707047\\
11	-33.2621922312473\\
12	-33.261981670787\\
13	-33.2614318633408\\
14	-33.2612084811642\\
15	-33.2612946103137\\
16	-33.2614565360346\\
17	-33.261526260381\\
18	-33.2614934440594\\
19	-33.2614432315364\\
20	-33.2614322362372\\
};
\label{10dBSimA}

\addplot [color=\AUDcolor, only marks, line width=1.0pt, mark=square, mark size = 1.5pt, mark options={solid, \AUDcolor}]
  table[row sep=crcr]{%
1	-8.05510610608479\\
2	-17.0653950311554\\
3	-26.4789505219372\\
4	-32.402263051897\\
5	-33.6496429523025\\
6	-33.773085242536\\
7	-33.7836422040837\\
8	-33.7845322678463\\
9	-33.784607218392\\
10	-33.7846135291865\\
11	-33.7846140605477\\
12	-33.784614105287\\
13	-33.7846141090546\\
14	-33.7846141093704\\
15	-33.7846141093981\\
16	-33.7846141094004\\
17	-33.7846141094004\\
18	-33.7846141094004\\
19	-33.7846141094004\\
20	-33.7846141094004\\
};
\label{10dBTheA}


\addplot [color=dark_green, line width=1.0pt]
  table[row sep=crcr]{%
1	-3.62264637412592\\
2	-5.99162387624991\\
3	-7.61965439603659\\
4	-8.57752789476958\\
5	-9.07151418881023\\
6	-9.30517763639439\\
7	-9.42343418644522\\
8	-9.47975583302774\\
9	-9.51303050453288\\
10	-9.52829720104249\\
11	-9.53511207897576\\
12	-9.54010474198674\\
13	-9.54373939044988\\
14	-9.54509587397027\\
15	-9.54461538589734\\
16	-9.54474467584602\\
17	-9.54543560208558\\
18	-9.54554408567703\\
19	-9.54553262369369\\
20	-9.54558075741257\\
};
\label{0dBSimS}

\addplot [color=dark_green, only marks, line width=1.0pt, mark=o, mark size = 1.5pt, mark options={solid, dark_green}]
  table[row sep=crcr]{%
1	-3.54381613044377\\
2	-5.93335241831239\\
3	-7.49563045425854\\
4	-8.40483709060501\\
5	-8.87930402136629\\
6	-9.10954707819702\\
7	-9.21688628911487\\
8	-9.26594113771461\\
9	-9.28815030949481\\
10	-9.29816212757338\\
11	-9.30266661402187\\
12	-9.30469147349423\\
13	-9.30560132857508\\
14	-9.3060100920757\\
15	-9.30619371935091\\
16	-9.30627620656438\\
17	-9.30631326004853\\
18	-9.30632990445503\\
19	-9.30633738108967\\
20	-9.30634073957432\\
};
\label{0dBTheS}

\addplot [color=dark_green, dashed, line width=1.0pt]
  table[row sep=crcr]{%
1	-3.91827823250228\\
2	-7.04569884497134\\
3	-9.81252297891391\\
4	-12.0792550570735\\
5	-13.5658156455491\\
6	-14.3797542185458\\
7	-14.8367579887435\\
8	-15.0638063577109\\
9	-15.1915755375862\\
10	-15.2684107712295\\
11	-15.2973539348736\\
12	-15.3028606812638\\
13	-15.3051796465958\\
14	-15.3077097340296\\
15	-15.3105381474648\\
16	-15.3118284102275\\
17	-15.3111962842512\\
18	-15.3107962233595\\
19	-15.3112373320628\\
20	-15.3119219955273\\
};
\label{5dBSimS}

\addplot [color=dark_green, only marks, line width=1.0pt, mark=diamond, mark size = 2pt, mark options={solid, dark_green}]
  table[row sep=crcr]{%
1	-3.86482892381151\\
2	-7.02843904210955\\
3	-9.78530198993636\\
4	-12.0046921264819\\
5	-13.530856698434\\
6	-14.4102582173713\\
7	-14.8505360154668\\
8	-15.0528268136902\\
9	-15.1417751128833\\
10	-15.1800976702334\\
11	-15.1964609058195\\
12	-15.2034207636873\\
13	-15.2063761367287\\
14	-15.2076301965321\\
15	-15.2081621752259\\
16	-15.2083878147037\\
17	-15.2084835148366\\
18	-15.2085241030629\\
19	-15.2085413171262\\
20	-15.2085486178335\\
};
\label{5dBTheS}

\addplot [color=dark_green, dotted, line width=1.0pt]
  table[row sep=crcr]{%
1	-4.05161316492962\\
2	-7.61917330447923\\
3	-11.3208271564135\\
4	-15.3275816725211\\
5	-19.2029030766076\\
6	-22.5528831127286\\
7	-25.0016952912472\\
8	-26.2266671040756\\
9	-26.7023157739672\\
10	-26.9078229625032\\
11	-26.9752501407731\\
12	-27.0033894056553\\
13	-27.0260915006603\\
14	-27.0335697938513\\
15	-27.0329699965755\\
16	-27.0334413838694\\
17	-27.034840461588\\
18	-27.0359217995947\\
19	-27.0365498739971\\
20	-27.0366065663325\\
};
\label{10dBSimS}

\addplot [color=dark_green, only marks, line width=1.0pt, mark=square, mark size = 1.5pt, mark options={solid, dark_green}]
  table[row sep=crcr]{%
1	-4.0127118724824\\
2	-7.6154661175609\\
3	-11.342154249057\\
4	-15.315149192294\\
5	-19.3621050414254\\
6	-22.9341998036681\\
7	-25.3434740911503\\
8	-26.5075326736725\\
9	-26.9428591314113\\
10	-27.0861984360416\\
11	-27.1311966446626\\
12	-27.145102816066\\
13	-27.1493792381218\\
14	-27.1506923220619\\
15	-27.1510953184376\\
16	-27.1512189836003\\
17	-27.151256930339\\
18	-27.1512685741634\\
19	-27.1512721470156\\
20	-27.1512732433277\\
};
\label{10dBTheS}
\end{axis}

\node [draw,fill=white] at (rel axis cs: 1.1,1.105) {\shortstack[l]{ 
    {\scriptsize SE}            {\scriptsize Sim.}   \hspace{0.2cm}{\scriptsize SNR}\\
    \hspace{0.1cm}\ref{0dBTheS} \hspace{0.05cm}\ref{0dBSimS} \hfill{\scriptsize 0 dB} \\    
    \hspace{0.1cm}\ref{5dBTheS} \hspace{0.09cm}\ref{5dBSimS} \hfill {\scriptsize 5 dB} \\
    \hspace{0.1cm}\ref{10dBTheS} \hspace{0.07cm}\ref{10dBSimS} {\scriptsize 10 dB}
}};

\end{tikzpicture}%
                \vspace{-0.5em}
                \caption{\footnotesize Normalized mean squared error vs. Iterations of BiMSGAMP-AUD. Markers indicate the state evolution results and lines the simulated ones, for the same SNR value. For example, for SNR $= 5$ dB, the diamond marker depicts the SE and the dashed line for the simulated results.}
                \label{fig:NMSE_it}
            \end{figure}
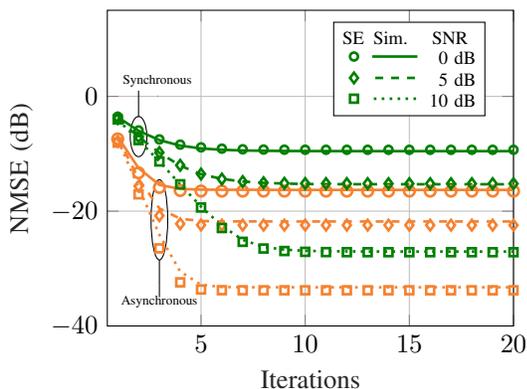

            \begin{figure}[t]
                \centering
\begin{tikzpicture}
\begin{axis}[%
width=5.4cm,
height=4.2cm,
at={(0.758in,0.481in)},
scale only axis,
xmin=0,
xmax=20,
xlabel style={font=\color{white!15!black}},
xlabel={Average SNR (dB)},
ymode=log,
ymin=0.005,
ymax=0.3,
yminorticks=true,
ylabel style={font=\color{white!15!black}},
ylabel={NMSE},
axis background/.style={fill=white},
xmajorgrids,
ymajorgrids,
yminorgrids,
legend style={at={(0.01,0.02)}, font=\tiny , anchor=south west, legend cell align=left, align=left, draw=white!15!black}
]

\addplot [color=\AMPcolor, line width=1.0pt, mark=x,mark size = 2.0pt, mark options={solid, \AMPcolor}]
  table[row sep=crcr]{%
    0	    0.339368611432585\\
    2.5	    0.247785054766128\\
    5	    0.187276169971292\\
    7.5	    0.144658336704716\\
    10	    0.0968100987642\\
    12.5	0.08479287320259\\
    15	    0.0800922102392\\
    17.5	0.077298756333049\\
    20	    0.073422514728563\\ 
};
\addlegendentry{AMP~\cite{Donoho2009}}

\addplot [color=\BIGAMPcolor, line width=1.0pt, mark=+, mark size = 2.0pt, mark options={solid, \BIGAMPcolor}]
 table[row sep=crcr]{%
    0	    0.339368611432585\\
    2.5	    0.247785054766128\\
    5	    0.187276169971292\\
    7.5	    0.144658336704716\\
    10	    0.10968100987642\\
    12.5	0.07779287320259\\
    15	    0.06068922102392\\
    17.5	0.044298756333049\\
    20	    0.035422514728563\\
};
\addlegendentry{BiG-AMP~\cite{JTParkerTSP2014}}

\addplot [color=\TBiGAMPcolor, line width=1.0pt, mark=diamond, mark size = 2pt,mark options={solid, \TBiGAMPcolor}]
  table[row sep=crcr]{%
    0	    0.339368611432585\\
    2.5	    0.247785054766128\\
    5	    0.187276169971292\\
    7.5	    0.144658336704716\\
    10	    0.10968100987642\\
    12.5	0.07779287320259\\
    15	    0.06068922102392\\
    17.5	0.044298756333049\\
    20	    0.035422514728563\\
};
\addlegendentry{Turbo-BiG-AMP~\cite{TDingTWC2019}}

\addplot [color=\HyGAMPcolor, line width=1.0pt, mark=square, mark size = 2.0pt, mark options={solid, \HyGAMPcolor}]
  table[row sep=crcr]{%
	0	    0.233589723082056\\
    2.5	    0.168156207445745\\
    5	    0.127188772523005\\
    7.5	    0.095848517877148\\
    10	    0.066490485783424\\
    12.5	0.041934189009815\\
    15	    0.02700595272274\\
    17.5	0.013503184424646\\
    20	    0.006829070154634\\
};
\addlegendentry{HyGAMP~\cite{SRanganTSP2017}}

\addplot [color=\RBPcolor, line width=1.0pt, mark=triangle,mark size = 2.0pt, mark options={solid, \RBPcolor}]
  table[row sep=crcr]{%
	0	    0.233589723082056\\
    2.5	    0.168156207445745\\
    5	    0.137188772523005\\
    7.5	    0.106848517877148\\
    10	    0.078490485783424\\
    12.5	0.056934189009815\\
    15	    0.0328884420214621\\
    17.5	0.013503184424646\\
    20	    0.006829070154634\\
};
\addlegendentry{BiMSGAMP-RBP}

\addplot [color=\AUDcolor, line width=1.0pt, mark=o,mark size = 2.0pt, mark options={solid, \AUDcolor}]
  table[row sep=crcr]{%
    0	    0.229988010351578\\
    2.5	    0.161730326763752\\
    5	    0.11877704776262\\
    7.5	    0.0863963907911693\\
    10	    0.0598986465205961\\
    12.5	0.0375901283749599\\
    15	    0.0234321954055599\\
    17.5	0.0135031844246462\\
    20	    0.00682907015463444\\  
};
\addlegendentry{BiMSGAMP-AUD}

\end{axis}
\end{tikzpicture}%
                \vspace{-0.5em}
                \caption{\footnotesize Normalized mean squared error vs. Average SNR (dB).}
                \label{fig:NMSE_channel}
            \end{figure}
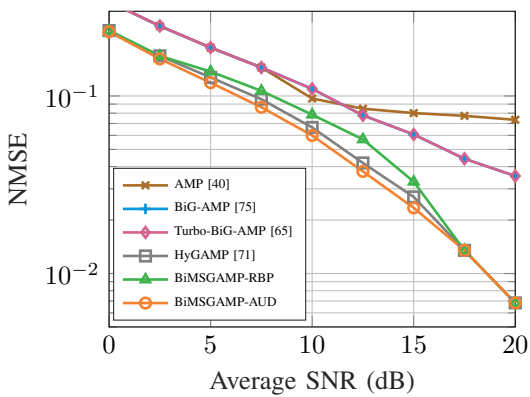

    \else
        \begin{figure}[!tbp]
            \begin{minipage}[b]{0.45\textwidth}
                \centering
\begin{tikzpicture}
\begin{axis}[%
width=5.4cm,
height=4.2cm,
at={(0.758in,0.481in)},
scale only axis,
xmin=0.5,
xmax=20,
xlabel style={font=\color{white!15!black}},
xlabel={Iterations},
ymin=-40,
ymax=15,
yminorticks=true,
ylabel style={font=\color{white!15!black}},
ylabel={NMSE (dB)},
axis background/.style={fill=white},
xmajorgrids,
ymajorgrids,
yminorgrids,
legend style={at={(1.2,0.1)}, anchor=south west,  font=\scriptsize, legend cell align=left, align=left, draw=white!15!black}
]

\draw (2.0,-3.6) -- (2.0,1);   
\node[coordinate] (A) at (axis cs:2.0,-7) {};  
\draw[black] (A) ellipse (0.4 and 3.5);           
\node[coordinate,pin={[pin distance=0cm]above:{\tiny Synchronous}}] at (axis cs:2.9,0.05){};  
\draw (3.0,-28.5) -- (3.0,-35);   
\node[coordinate] (B) at (axis cs:3,-21.5) {};        
\draw[black] (B) ellipse (0.4 and 7);           
\node[coordinate,pin={[pin distance=0cm]below:{\tiny Asynchronous}}] at (axis cs:3.0,-33){};  

\addplot [color=\AUDcolor, line width=1.0pt]
  table[row sep=crcr]{%
1	-7.05653615676392\\
2	-12.4976510583977\\
3	-15.4196376618839\\
4	-16.1268361937189\\
5	-16.2468220651174\\
6	-16.2735231475593\\
7	-16.2733430388164\\
8	-16.2794538139838\\
9	-16.2934705164771\\
10	-16.2991764293592\\
11	-16.2981573201258\\
12	-16.2969280332425\\
13	-16.2969009965628\\
14	-16.296534942621\\
15	-16.2958589017621\\
16	-16.2954375903616\\
17	-16.2953159629034\\
18	-16.2954509949638\\
19	-16.2956827533673\\
20	-16.2958050685182\\
};
\label{0dBSimA}

\addplot [color=\AUDcolor, only marks, line width=1.0pt, mark=o,mark size = 2.0pt, mark options={solid, \AUDcolor}]
  table[row sep=crcr]{%
1	-7.41015645005781\\
2	-13.3131230490127\\
3	-15.8637958632638\\
4	-16.4036459770331\\
5	-16.4876980659147\\
6	-16.4999869187872\\
7	-16.5017662489079\\
8	-16.5020235169744\\
9	-16.5020607069663\\
10	-16.5020660828936\\
11	-16.5020668599971\\
12	-16.5020669723293\\
13	-16.5020669885671\\
14	-16.5020669909144\\
15	-16.5020669912536\\
16	-16.5020669913027\\
17	-16.5020669913098\\
18	-16.5020669913108\\
19	-16.5020669913109\\
20	-16.502066991311\\
};
\label{0dBTheA}

\addplot [color=\AUDcolor, dashed, line width=1.0pt]
  table[row sep=crcr]{%
1	-7.14926358299793\\
2	-14.3575458200581\\
3	-19.5134474355358\\
4	-21.3111609057399\\
5	-21.7130577767893\\
6	-21.7887001115347\\
7	-21.7870892029482\\
8	-21.7876154247081\\
9	-21.7940534139163\\
10	-21.797986010932\\
11	-21.7980394758936\\
12	-21.7973379131285\\
13	-21.7971062928673\\
14	-21.7971095051645\\
15	-21.7971329854122\\
16	-21.79714162293\\
17	-21.7971436898855\\
18	-21.7971551370563\\
19	-21.7971686833329\\
20	-21.7971679270932\\
};
\label{5dBSimA}

\addplot [color=\AUDcolor, only marks, line width=1.0pt, mark=diamond,mark size = 2.0pt, mark options={solid, \AUDcolor}]
  table[row sep=crcr]{%
1	-7.85438766906154\\
2	-15.6427091966653\\
3	-20.7913548201673\\
4	-22.2035477040039\\
5	-22.4062250495602\\
6	-22.4308170702269\\
7	-22.4337314845836\\
8	-22.4340758908977\\
9	-22.4341165768366\\
10	-22.4341213830192\\
11	-22.4341219507653\\
12	-22.434122017832\\
13	-22.4341220257546\\
14	-22.4341220266904\\
15	-22.4341220268009\\
16	-22.4341220268141\\
17	-22.4341220268156\\
18	-22.4341220268157\\
19	-22.4341220268157\\
20	-22.4341220268157\\
};
\label{5dBTheA}

\addplot [color=\AUDcolor, dotted, line width=1.0pt]
  table[row sep=crcr]{%
1	-7.51498571271837\\
2	-15.8633695658748\\
3	-24.1092836194969\\
4	-30.5258582801552\\
5	-32.8237435088396\\
6	-33.2143093090731\\
7	-33.2598813119969\\
8	-33.2544392749027\\
9	-33.2563335429961\\
10	-33.2606101707047\\
11	-33.2621922312473\\
12	-33.261981670787\\
13	-33.2614318633408\\
14	-33.2612084811642\\
15	-33.2612946103137\\
16	-33.2614565360346\\
17	-33.261526260381\\
18	-33.2614934440594\\
19	-33.2614432315364\\
20	-33.2614322362372\\
};
\label{10dBSimA}

\addplot [color=\AUDcolor, only marks, line width=1.0pt, mark=square, mark size = 1.5pt, mark options={solid, \AUDcolor}]
  table[row sep=crcr]{%
1	-8.05510610608479\\
2	-17.0653950311554\\
3	-26.4789505219372\\
4	-32.402263051897\\
5	-33.6496429523025\\
6	-33.773085242536\\
7	-33.7836422040837\\
8	-33.7845322678463\\
9	-33.784607218392\\
10	-33.7846135291865\\
11	-33.7846140605477\\
12	-33.784614105287\\
13	-33.7846141090546\\
14	-33.7846141093704\\
15	-33.7846141093981\\
16	-33.7846141094004\\
17	-33.7846141094004\\
18	-33.7846141094004\\
19	-33.7846141094004\\
20	-33.7846141094004\\
};
\label{10dBTheA}


\addplot [color=dark_green, line width=1.0pt]
  table[row sep=crcr]{%
1	-3.62264637412592\\
2	-5.99162387624991\\
3	-7.61965439603659\\
4	-8.57752789476958\\
5	-9.07151418881023\\
6	-9.30517763639439\\
7	-9.42343418644522\\
8	-9.47975583302774\\
9	-9.51303050453288\\
10	-9.52829720104249\\
11	-9.53511207897576\\
12	-9.54010474198674\\
13	-9.54373939044988\\
14	-9.54509587397027\\
15	-9.54461538589734\\
16	-9.54474467584602\\
17	-9.54543560208558\\
18	-9.54554408567703\\
19	-9.54553262369369\\
20	-9.54558075741257\\
};
\label{0dBSimS}

\addplot [color=dark_green, only marks, line width=1.0pt, mark=o, mark size = 1.5pt, mark options={solid, dark_green}]
  table[row sep=crcr]{%
1	-3.54381613044377\\
2	-5.93335241831239\\
3	-7.49563045425854\\
4	-8.40483709060501\\
5	-8.87930402136629\\
6	-9.10954707819702\\
7	-9.21688628911487\\
8	-9.26594113771461\\
9	-9.28815030949481\\
10	-9.29816212757338\\
11	-9.30266661402187\\
12	-9.30469147349423\\
13	-9.30560132857508\\
14	-9.3060100920757\\
15	-9.30619371935091\\
16	-9.30627620656438\\
17	-9.30631326004853\\
18	-9.30632990445503\\
19	-9.30633738108967\\
20	-9.30634073957432\\
};
\label{0dBTheS}

\addplot [color=dark_green, dashed, line width=1.0pt]
  table[row sep=crcr]{%
1	-3.91827823250228\\
2	-7.04569884497134\\
3	-9.81252297891391\\
4	-12.0792550570735\\
5	-13.5658156455491\\
6	-14.3797542185458\\
7	-14.8367579887435\\
8	-15.0638063577109\\
9	-15.1915755375862\\
10	-15.2684107712295\\
11	-15.2973539348736\\
12	-15.3028606812638\\
13	-15.3051796465958\\
14	-15.3077097340296\\
15	-15.3105381474648\\
16	-15.3118284102275\\
17	-15.3111962842512\\
18	-15.3107962233595\\
19	-15.3112373320628\\
20	-15.3119219955273\\
};
\label{5dBSimS}

\addplot [color=dark_green, only marks, line width=1.0pt, mark=diamond, mark size = 2pt, mark options={solid, dark_green}]
  table[row sep=crcr]{%
1	-3.86482892381151\\
2	-7.02843904210955\\
3	-9.78530198993636\\
4	-12.0046921264819\\
5	-13.530856698434\\
6	-14.4102582173713\\
7	-14.8505360154668\\
8	-15.0528268136902\\
9	-15.1417751128833\\
10	-15.1800976702334\\
11	-15.1964609058195\\
12	-15.2034207636873\\
13	-15.2063761367287\\
14	-15.2076301965321\\
15	-15.2081621752259\\
16	-15.2083878147037\\
17	-15.2084835148366\\
18	-15.2085241030629\\
19	-15.2085413171262\\
20	-15.2085486178335\\
};
\label{5dBTheS}

\addplot [color=dark_green, dotted, line width=1.0pt]
  table[row sep=crcr]{%
1	-4.05161316492962\\
2	-7.61917330447923\\
3	-11.3208271564135\\
4	-15.3275816725211\\
5	-19.2029030766076\\
6	-22.5528831127286\\
7	-25.0016952912472\\
8	-26.2266671040756\\
9	-26.7023157739672\\
10	-26.9078229625032\\
11	-26.9752501407731\\
12	-27.0033894056553\\
13	-27.0260915006603\\
14	-27.0335697938513\\
15	-27.0329699965755\\
16	-27.0334413838694\\
17	-27.034840461588\\
18	-27.0359217995947\\
19	-27.0365498739971\\
20	-27.0366065663325\\
};
\label{10dBSimS}

\addplot [color=dark_green, only marks, line width=1.0pt, mark=square, mark size = 1.5pt, mark options={solid, dark_green}]
  table[row sep=crcr]{%
1	-4.0127118724824\\
2	-7.6154661175609\\
3	-11.342154249057\\
4	-15.315149192294\\
5	-19.3621050414254\\
6	-22.9341998036681\\
7	-25.3434740911503\\
8	-26.5075326736725\\
9	-26.9428591314113\\
10	-27.0861984360416\\
11	-27.1311966446626\\
12	-27.145102816066\\
13	-27.1493792381218\\
14	-27.1506923220619\\
15	-27.1510953184376\\
16	-27.1512189836003\\
17	-27.151256930339\\
18	-27.1512685741634\\
19	-27.1512721470156\\
20	-27.1512732433277\\
};
\label{10dBTheS}
\end{axis}

\node [draw,fill=white] at (rel axis cs: 1.1,1.105) {\shortstack[l]{ 
    {\scriptsize SE}            {\scriptsize Sim.}   \hspace{0.2cm}{\scriptsize SNR}\\
    \hspace{0.1cm}\ref{0dBTheS} \hspace{0.05cm}\ref{0dBSimS} \hfill{\scriptsize 0 dB} \\    
    \hspace{0.1cm}\ref{5dBTheS} \hspace{0.09cm}\ref{5dBSimS} \hfill {\scriptsize 5 dB} \\
    \hspace{0.1cm}\ref{10dBTheS} \hspace{0.07cm}\ref{10dBSimS} {\scriptsize 10 dB}
}};

\end{tikzpicture}%
                \vspace{-1.5em}
                \caption{\footnotesize Normalized mean squared error vs. Iterations of BiMSGAMP-AUD. Markers indicate the state evolution results and lines the simulated ones, for the same SNR value. For example, for SNR $= 5$ dB, the diamond marker depicts the SE and the dashed line for the simulated results.}
                \label{fig:NMSE_it}
            \end{minipage}\hfill
            \begin{minipage}[b]{0.45\linewidth}
                \centering
\begin{tikzpicture}
\begin{axis}[%
width=5.4cm,
height=4.2cm,
at={(0.758in,0.481in)},
scale only axis,
xmin=0,
xmax=20,
xlabel style={font=\color{white!15!black}},
xlabel={Average SNR (dB)},
ymode=log,
ymin=0.005,
ymax=0.3,
yminorticks=true,
ylabel style={font=\color{white!15!black}},
ylabel={NMSE},
axis background/.style={fill=white},
xmajorgrids,
ymajorgrids,
yminorgrids,
legend style={at={(0.01,0.02)}, font=\tiny , anchor=south west, legend cell align=left, align=left, draw=white!15!black}
]

\addplot [color=\AMPcolor, line width=1.0pt, mark=x,mark size = 2.0pt, mark options={solid, \AMPcolor}]
  table[row sep=crcr]{%
    0	    0.339368611432585\\
    2.5	    0.247785054766128\\
    5	    0.187276169971292\\
    7.5	    0.144658336704716\\
    10	    0.0968100987642\\
    12.5	0.08479287320259\\
    15	    0.0800922102392\\
    17.5	0.077298756333049\\
    20	    0.073422514728563\\ 
};
\addlegendentry{AMP~\cite{Donoho2009}}

\addplot [color=\BIGAMPcolor, line width=1.0pt, mark=+, mark size = 2.0pt, mark options={solid, \BIGAMPcolor}]
 table[row sep=crcr]{%
    0	    0.339368611432585\\
    2.5	    0.247785054766128\\
    5	    0.187276169971292\\
    7.5	    0.144658336704716\\
    10	    0.10968100987642\\
    12.5	0.07779287320259\\
    15	    0.06068922102392\\
    17.5	0.044298756333049\\
    20	    0.035422514728563\\
};
\addlegendentry{BiG-AMP~\cite{JTParkerTSP2014}}

\addplot [color=\TBiGAMPcolor, line width=1.0pt, mark=diamond, mark size = 2pt,mark options={solid, \TBiGAMPcolor}]
  table[row sep=crcr]{%
    0	    0.339368611432585\\
    2.5	    0.247785054766128\\
    5	    0.187276169971292\\
    7.5	    0.144658336704716\\
    10	    0.10968100987642\\
    12.5	0.07779287320259\\
    15	    0.06068922102392\\
    17.5	0.044298756333049\\
    20	    0.035422514728563\\
};
\addlegendentry{Turbo-BiG-AMP~\cite{TDingTWC2019}}

\addplot [color=\HyGAMPcolor, line width=1.0pt, mark=square, mark size = 2.0pt, mark options={solid, \HyGAMPcolor}]
  table[row sep=crcr]{%
	0	    0.233589723082056\\
    2.5	    0.168156207445745\\
    5	    0.127188772523005\\
    7.5	    0.095848517877148\\
    10	    0.066490485783424\\
    12.5	0.041934189009815\\
    15	    0.02700595272274\\
    17.5	0.013503184424646\\
    20	    0.006829070154634\\
};
\addlegendentry{HyGAMP~\cite{SRanganTSP2017}}

\addplot [color=\RBPcolor, line width=1.0pt, mark=triangle,mark size = 2.0pt, mark options={solid, \RBPcolor}]
  table[row sep=crcr]{%
	0	    0.233589723082056\\
    2.5	    0.168156207445745\\
    5	    0.137188772523005\\
    7.5	    0.106848517877148\\
    10	    0.078490485783424\\
    12.5	0.056934189009815\\
    15	    0.0328884420214621\\
    17.5	0.013503184424646\\
    20	    0.006829070154634\\
};
\addlegendentry{BiMSGAMP-RBP}

\addplot [color=\AUDcolor, line width=1.0pt, mark=o,mark size = 2.0pt, mark options={solid, \AUDcolor}]
  table[row sep=crcr]{%
    0	    0.229988010351578\\
    2.5	    0.161730326763752\\
    5	    0.11877704776262\\
    7.5	    0.0863963907911693\\
    10	    0.0598986465205961\\
    12.5	0.0375901283749599\\
    15	    0.0234321954055599\\
    17.5	0.0135031844246462\\
    20	    0.00682907015463444\\  
};
\addlegendentry{BiMSGAMP-AUD}

\end{axis}
\end{tikzpicture}%
                \vspace{-1.5em}
                \caption{\footnotesize Normalized mean squared error vs. Average SNR (dB).}
                \label{fig:NMSE_channel}
            \end{minipage}
        \end{figure}
    \fi
    \makeatother

    \makeatletter%
    \if@twocolumn
        \begin{figure*}[t]
            \begin{minipage}[b]{0.6\linewidth}
                \centerline{
%
\begin{tikzpicture}

\begin{axis}[%
width=5.4cm,
height=4.2cm,
at={(0.758in,0.481in)},
scale only axis,
xmin=0,
xmax=20,
xlabel style={font=\color{white!15!black}},
xlabel={Average SNR (dB)},
ymode=log,
ymin=0.01,
ymax=1,
yminorticks=true,
ylabel style={font=\color{white!15!black}},
ylabel={FER},
axis background/.style={fill=white},
xmajorgrids,
ymajorgrids,
yminorgrids,
legend style={at={(1.2,0.14)}, anchor=south west,  font=\scriptsize, legend cell align=left, align=left, draw=white!15!black}
]

\addplot [color=\AMPcolor, dashed,line width=1.0pt, mark=x,mark size = 2pt, mark options={solid, \AMPcolor}]
  table[row sep=crcr]{%
    0	    0.99\\
    2.5	    0.99\\
    5	    0.87\\
    7.5	    0.85\\
    10	    0.84\\
    12.5	0.825\\
    15	    0.822\\
    17.5	0.81\\
    20	    0.8\\  
};
\label{AMP_s}

\addplot [color=\AMPcolor, line width=1.0pt, mark=x,mark size = 2pt, mark options={solid, \AMPcolor}]
  table[row sep=crcr]{%
    0	    0.998571428571429\\
    2.5	    0.99\\
    5	    0.95\\
    7.5	    0.90\\
    10	    0.89\\
    12.5	0.88\\
    15	    0.88\\
    17.5	0.87\\
    20	    0.85\\ 
};
\label{AMP_a}

\addplot [color=\BIGAMPcolor, dashed,line width=1.0pt, mark=triangle, mark size = 2pt, mark options={solid, rotate=90, \BIGAMPcolor}]
  table[row sep=crcr]{%
    0	    0.972920634920635\\
    2.5	    0.959428571428571\\
    5	    0.842873015873016\\
    7.5	    0.712626984126984\\
    10	    0.629825396825397\\
    12.5	0.52285714285714\\
    15	    0.424801587301587\\
    17.5	0.371949603174603\\
    20	    0.303037301587302\\
};
\label{BiGAMP_s}

\addplot [color=\BIGAMPcolor, line width=1.0pt, mark=triangle, mark size = 2pt, mark options={solid, rotate=90, \BIGAMPcolor}]
  table[row sep=crcr]{%
    0	    0.986666666666667\\
    2.5	    0.93\\
    5	    0.743333333333333\\
    7.5	    0.47\\
    10	    0.333333333333333\\
    12.5	0.216666666666667\\
    15	    0.107\\
    17.5	0.026666666666667\\
    20	    0.008\\
};
\label{BiGAMP_a}

\addplot [color=\TBiGAMPcolor, line width=1.0pt, mark=diamond, mark size = 2pt,mark options={solid, \TBiGAMPcolor}]
  table[row sep=crcr]{%
    0	    1\\
    2.5	    1\\
    5	    0.876666666666667\\
    7.5	    0.446666666666667\\
    10	    0.176666666666667\\
    12.5	0.052666666666667\\
    15	    0.01533333333333\\
    17.5	0.0052\\
    20	    0\\
};
\label{Turbo_BiG_AMP_a}

\addplot [color=\JEMAMPcolor, dashed,line width=1.0pt, mark=asterisk, mark size = 2pt, mark options={solid, \JEMAMPcolor}]
  table[row sep=crcr]{%
    0	    0.977777777777778\\
    2.5	    0.939142857142857\\
    5	    0.839246031746032\\
    7.5	    0.747753968253968\\
    10	    0.625885714285714\\
    12.5	0.553079365079365\\
    15	    0.463039682539683\\
    17.5	0.421738095238095\\
    20	    0.361793650793651\\
};
\label{JEMAMP_s}

\addplot [color=\JEMAMPcolor, line width=1.0pt,  mark=asterisk, mark size = 2pt, mark options={solid, \JEMAMPcolor}]
  table[row sep=crcr]{%
    0	    0.993333333333333\\
    2.5	    0.91\\
    5	    0.76\\
    7.5	    0.503333333333333\\
    10	    0.395\\
    12.5	0.308333333333333\\
    15	    0.21333333333333\\
    17.5	0.124666666666667\\
    20	    0.046666666666667\\
};
\label{JEMAMP_a}

\addplot [color=\HyGAMPcolor, dashed,line width=1.0pt, mark=square, mark size = 2.0pt, mark options={solid, \HyGAMPcolor}]
  table[row sep=crcr]{%
    0	1\\
    2.5	1\\
    5	0.94\\
    7.5	0.7985\\
    10	0.667404761904762\\
    12.5	0.482452380952381\\
    15	0.284873015873016\\
    17.5	0.181738095238095\\
    20	0.0617936507936508\\
};
\label{HYGAMP_s}

\addplot [color=\HyGAMPcolor, line width=1.0pt, mark=square, mark size = 2.0pt, mark options={solid, \HyGAMPcolor}]
  table[row sep=crcr]{%
    0	    1\\
    2.5	    1\\
    5	    0.753333333333333\\
    7.5	    0.286666666666667\\
    10	    0.092\\
    12.5	0.035\\
    15	    0.01083333333333\\
    17.5	0.005\\
    20	    0\\
};
\label{HYGAMP_a}

\addplot [color=\RBPcolor, dashed,line width=1.0pt, mark=triangle,mark size = 2pt, mark options={solid, \RBPcolor}]
  table[row sep=crcr]{%
    0	    0.9875\\
    2.5	    0.883111111111111\\
    5	    0.843460317460317\\
    7.5	    0.777984126984127\\
    10	    0.702134920634921\\
    12.5	0.427412698412699\\
    15	    0.359253968253968\\
    17.5	0.191380952380952\\
    20	    0.061793650793651\\
};
\label{RBP_s}

\addplot [color=\RBPcolor, line width=1.0pt, mark=triangle,mark size = 2pt, mark options={solid, \RBPcolor}]
  table[row sep=crcr]{%
    0	    0.966666666666667\\
    2.5	    0.756666666666667\\
    5	    0.553333333333333\\
    7.5	    0.32666666666667\\
    10	    0.124333333333333\\
    12.5	0.0311084594\\
    15	    0.009243333333333\\
    17.5	0.005\\
    20	    0\\
};
\label{RBP_a}

\addplot [color=\AUDcolor, dashed,line width=1.0pt, mark=o,mark size = 2.0pt, mark options={solid, \AUDcolor}]
  table[row sep=crcr]{%
    0	    0.940833333333333\\
    2.5	    0.804626984126984\\
    5	    0.699\\
    7.5	    0.61018253968254\\
    10	    0.537888888888889\\
    12.5	0.442301587301587\\
    15	    0.308801587301587\\
    17.5	0.156380952380952\\
    20	    0.0617936507936508\\
};
\label{AUD_s}

\addplot [color=\AUDcolor, line width=1.0pt, mark=o,mark size = 2.0pt, mark options={solid, \AUDcolor}]
  table[row sep=crcr]{%
    0	    0.983333333333333\\
    2.5	    0.693333333333333\\
    5	    0.353333333333333\\
    7.5	    0.136666666666667\\
    10	    0.06333333333333\\
    12.5	0.025\\
    15	    0.01\\
    17.5	0\\
    20	    0\\
};
\label{AUD_a}

\addplot [color=black, dashed,line width=1.0pt, mark=square*, mark size = 1.1pt, mark options={solid, black}]
  table[row sep=crcr]{%
    0	    0.849285714285714\\
    2.5	    0.772325396825397\\
    5	    0.698309523809524\\
    7.5	    0.588746031746032\\
    10	    0.506785714285714\\
    12.5	0.386873015873016\\
    15	    0.230222222222222\\
    17.5	0.102714285714286\\
    20	    0.058579365079365\\
};
\label{OHYGAMP_s}

\addplot [color=black, line width=1.0pt, mark=square*, mark size = 1.1pt, mark options={solid, black}]
  table[row sep=crcr]{%
    0	    0.553333333333333\\
    2.5	    0.443333333333333\\
    5	    0.22\\
    7.5	    0.11\\
    10	    0.033333333333333\\
    12.5	0.01\\
    15	    0.006333333333333\\
    17.5	0\\
    20	    0\\
};
\label{OHYGAMP_a}
\end{axis}

\node [draw,fill=white] at (rel axis cs: 1.75,1.6) {\shortstack[l]{ 
    \ref{AMP_a} {\scriptsize AMP~\cite{Donoho2009}} \\
    \ref{BiGAMP_a} {\scriptsize BiG-AMP~\cite{JTParkerTSP2014}}\\     
    \ref{JEMAMP_a} {\scriptsize Joint-EM-AMP~\cite{CWeiCommLet2017}}\\    
    \ref{HYGAMP_a} {\scriptsize HyGAMP~\cite{SRanganTSP2017}}\\   
    \ref{Turbo_BiG_AMP_a} {\scriptsize Turbo-BiG-AMP~\cite{TDingTWC2019}}\\ 
    \ref{RBP_a} {\scriptsize BiMSGAMP-RBP}\\    
    \ref{AUD_a} {\scriptsize BiMSGAMP-AUD}\\
    \ref{OHYGAMP_a}  {\scriptsize OHyGAMP}
}};

\node [draw,fill=white] at (rel axis cs: 1.75,2.1) {\shortstack[l]{ 
    - -  {\scriptsize Synchronous GFRA}\\    
    ---  {\scriptsize Asynchronous GFRA}
}};
\end{tikzpicture}%
}
                \centerline{\footnotesize (a) Frame error rate vs. Average SNR (dB).}\medskip
            \end{minipage}
            \begin{minipage}[b]{0.39\linewidth}
                \centerline{
\begin{tikzpicture}
\begin{axis}[%
width=5.4cm,
height=4.2cm,
at={(0.758in,0.481in)},
scale only axis,
mark repeat = 2,
xmin=0.0005,
xmax=0.025,
xlabel style={font=\color{white!15!black}},
xlabel={False Alarm Rate},
xmode=log,
ymode=log,
ymin=0.0001,
ymax=0.1,
yminorticks=true,
ylabel style={font=\color{white!15!black}},
ylabel={Missed Detection Rate},
axis background/.style={fill=white},
xmajorgrids,
ymajorgrids,
yminorgrids,
legend style={at={(0.05,0.01)}, font=\tiny, anchor=south west, legend cell align=left, align=left, draw=white!15!black}
]

\addplot [color=\HyGAMPcolor, line width=1.0pt, mark=square, mark size = 2.0pt, mark options={solid, \HyGAMPcolor}]
  table[row sep=crcr]{%
0.000489314199255675 0.0002370986542942092 \\
0.0005344254854083944 0.00048545816215143635 \\
0.0005597880067102694 0.0006936075852880272\\
0.0005151584608058517 0.0003469434211324178\\
0.0006058556272344318 0.0010808821231000584\\
0.0006465477351102133 0.0015086667471699033\\
0.0006867851158972012 0.001954978177911692\\
0.0007458210307331696 0.002650545909727248\\
0.0008193228206173793 0.0035888148925309147\\
0.001055417222582518 0.005820152209473489\\
0.000930014427319673 0.00460637150083731\\
0.0011504266429771268 0.007042372703048723\\
0.0012704688972187727 0.008089612334074979\\
0.0014575410189970643 0.009932869196928073\\
0.0017090940860712528 0.011876720749748543\\
0.0020703020576246815 0.015428936208461919\\
0.0026373836724160713 0.017939057776147602\\
%
0.003924860524003047 0.02688917420523749\\
0.00539003330787633 0.029636777926206077\\
0.00681913386623994 0.031367425359552156\\
0.0087220300215399 0.03293105420660645\\
0.011718812629856622 0.03429345704895215\\
0.016539688115358848 0.03513780658027035\\
0.02308981363216598 0.036002945095621866\\
0.03498992349252213 0.036002945095621866\\
};

\addplot [color=\HyGAMPcolor, dashed,line width=1.0pt, mark=square, mark size = 2.0pt, mark options={solid, \HyGAMPcolor}]
  table[row sep=crcr]{%
0.000153871	0.00010197	\\
0.000171736	0.000106012	\\
0.000203356	0.000120994	\\
0.000218190 0.0013692576\\
0.000238771	0.000377415	\\
0.000310272	0.000668033	\\
0.000337629	0.000911106	\\
0.000386506	0.001305879	\\
0.000461552	0.001895082	\\
0.000537372	0.007584633	\\
0.000630957	0.009750801	\\
0.000792652	0.0105579955	\\
0.000915099	0.0127423692	\\
0.00105646	0.017251364	\\
0.001163242	0.020565049	\\
0.001416447	0.0220536746	\\
0.001658051	0.0302005246	\\
0.00180972	0.032362239	\\
0.001855896	0.038162361	\\
0.002257262	0.045928651	\\
0.002849526	0.055394287	\\
0.004570512	0.061258487	\\
0.005036486	0.056563687	\\
0.005244396	0.065389385	\\
0.005882855	0.065389385	\\
0.007283603	0.067281351	\\
0.009266378	0.070022424	\\
0.010739831	0.067494921	\\
0.012508599	0.070519798	\\
0.01427825	0.075645857	\\
0.016463181	0.073641332	\\
0.01861744	0.074876865	\\
0.023350696	0.077512755	\\
0.029477492	0.079084299	\\
};

\addplot [color=\TBiGAMPcolor, line width=1.0pt, mark=diamond, mark size = 2pt,mark options={solid, \TBiGAMPcolor}]
  table[row sep=crcr]{%
0.010562165521799834 0.01928462808226607\\
0.014113791761820533 0.019599884906767766\\
0.02465607772879353 0.020082459574328908\\
0.0367553586699007 0.020576915858077048\\
0.01885968530647384 0.019759440716631396\\
0.008123416436245057 0.018821225033879722\\
0.0079152910982998 0.018583620893795233\\
0.004079152910982998 0.016583620893795233\\
0.0028310808264524114 0.014872495533018885\\
0.0018707776368885789 0.012223967592066643\\
0.0012960928096145206 0.009669965014361391\\
0.0009558829084235898 0.007334447826352913\\
0.0007145007503796613 0.0056875054639171675\\
0.0005842222848016939 0.004411721297710888\\
0.0004886504958817522 0.0035203714769220075\\
0.0003770199159517425 0.002534167527656916\\
0.0002930502854577394 0.001912502460246219\\
0.00021819408602348275 0.001369053833492576\\
0.00016401215428119738 0.0009687970127115359\\
0.00013052763804649034 0.000685559346811387\\
0.00011209513822936698 0.0005635580702568055\\
0.00010000000000000021 0.0004579584033357946\\
};

\addplot [color=\TBiGAMPcolor, line width=1.0pt, dashed,mark=diamond, mark size = 2pt,mark options={solid, \TBiGAMPcolor}]
  table[row sep=crcr]{%
0.00010433979537295824 0.0036910256943108047\\
0.00012972388347720604 0.004438002730903954\\
0.00015516619321348065 0.005169059693251489\\
0.00018459391711095983 0.005731385694168232\\
0.00022199921228929656 0.006486835335728294\\
0.0002673466226953171 0.007222815187696189\\
0.00031453093843953706 0.007922461503647258\\
0.0003818701947801542 0.008978384874588824\\
0.00045360147861706323 0.010045191105396735\\
0.0005379867492383172 0.011025950778809869\\
0.0006501512733949092 0.01218581134137075\\
0.0007885915873624638 0.01355933850794409\\
0.0009233937962242302 0.014695245966210087\\
0.0011186720651933408 0.01618535296920849\\
0.001333723572433873 0.017578500633231894\\
0.0015918430202787597 0.019148589781701024\\
0.0019040458474518715 0.0207213254907294\\
0.0022774799669403618 0.02271492090432738\\
0.002724154466530398 0.02537383712242499\\
0.003258433736076785 0.02698012643762072\\
0.003897499404990083 0.02885825569606045\\
0.004661902877972199 0.030437346138390067\\
0.0055762262377306955 0.03151725406864262\\
0.00940055773273313 0.03491195135769875\\
0.007293750314525081 0.03395246218269532\\
0.011402684968987213 0.036296032483100565\\
0.013435877428667295 0.03749243627386817\\
0.01583160481668601 0.038415548835854275\\
0.01885968530647384 0.03810534653338214\\
0.02246694090572183 0.03904354974340707\\
0.028893995509091164 0.039361389630166284\\
0.035959982973931506 0.04098982426133143\\
};

\addplot [color=\RBPcolor, line width=1.0pt, mark=triangle,mark size = 2pt, mark options={solid, \RBPcolor}]
  table[row sep=crcr]{%
0.010239755435542348 0.0038094400097139105\\
0.01166294027534609 0.004842582467003001\\
0.012871420677812773 0.0060271612251486945\\
0.014246102062944197 0.0073143691735023005\\
0.015689781051531085 0.008598363230979296\\
0.018110753015188403 0.010830201817741742\\
0.02293000693937404 0.01564079444653699\\
0.00926706371008957 0.003103593481454677\\
0.008346177242389764 0.0024503975280369397\\
0.0074811255312683955 0.0018689147295323722\\
0.006705733366210035 0.001369053833492576\\
0.006155398839877056 0.0009913917639001112\\
0.005491227306649114 0.0006935077427080176\\
0.005064603543891595 0.0004907536964516966\\
0.004783569257132789 0.0003413237104286842\\
0.0045613238628900935 0.0002471674283834213\\ 
0.0044119247773244 0.00019514748279407556\\
0.004328761281082938 0.0001446106407142123\\
0.004247165392508449 0.00010233224823075555\\
};

\addplot [color=\RBPcolor, line width=1.0pt, dashed,mark=triangle,mark size = 2pt, mark options={solid, \RBPcolor}]
  table[row sep=crcr]{%
0.002533429	0.000105446\\
0.002776701	0.00013798\\
0.002806059	0.000187603\\
0.003752424	0.000612438\\
0.004708125	0.001429951\\
0.006567085	0.003045947\\
0.009551042	0.00686908\\
0.013652008	0.011424254\\
0.020178992	0.020046377\\
0.037097337	0.041099959\\
};

\addplot [color=\AUDcolor, line width=1.0pt, mark=o,mark size = 2.0pt, mark options={solid, \AUDcolor}]
  table[row sep=crcr]{%
0.00712	0.0001\\
0.00725	0.00025\\
0.0075	0.0005\\
0.00775	0.00075\\
0.008	0.001\\
0.0085	0.0015\\
0.009	0.002\\
0.011	0.004\\
0.012	0.005\\
0.013	0.006\\
0.014	0.007\\
0.015	0.008\\
%
0.021	0.014\\
0.022	0.015\\
0.023	0.016\\
0.02375	0.01675\\
};

\addplot [color=\AUDcolor, line width=1.0pt, dashed,mark=o,mark size = 2.0pt, mark options={solid, \AUDcolor}]
  table[row sep=crcr]{%
0.00306902	0.000114812	\\
0.003282698	0.000141524	\\
0.003406059	0.000187603	\\
0.003504842	0.00021709	\\
0.003644056	0.000338451	\\
0.003862626	0.000508086	\\
0.004259492	0.000737776	\\
0.005129697	0.001284028	\\
0.006609726	0.00225082	\\
0.010084857	0.005899552	\\
0.014021323	0.009652328	\\
0.01965975	0.01977993	\\
0.025982378	0.025792129	\\
0.035219971	0.034660974	\\
};
\end{axis}
\end{tikzpicture}%
}
                \centerline{\footnotesize (b) Missed detection rate vs. False alarm rate.}\medskip
            \end{minipage} \hfill
             \vspace{-0.5em}
            \caption{Key performance indicators for synchronous and asynchronous mMTC scenarios with $N=100, M=32$ and $L=192$, by $10^4$ Monte Carlo trials.}
            \label{fig:AER_FER}%
        \end{figure*}

    \else
        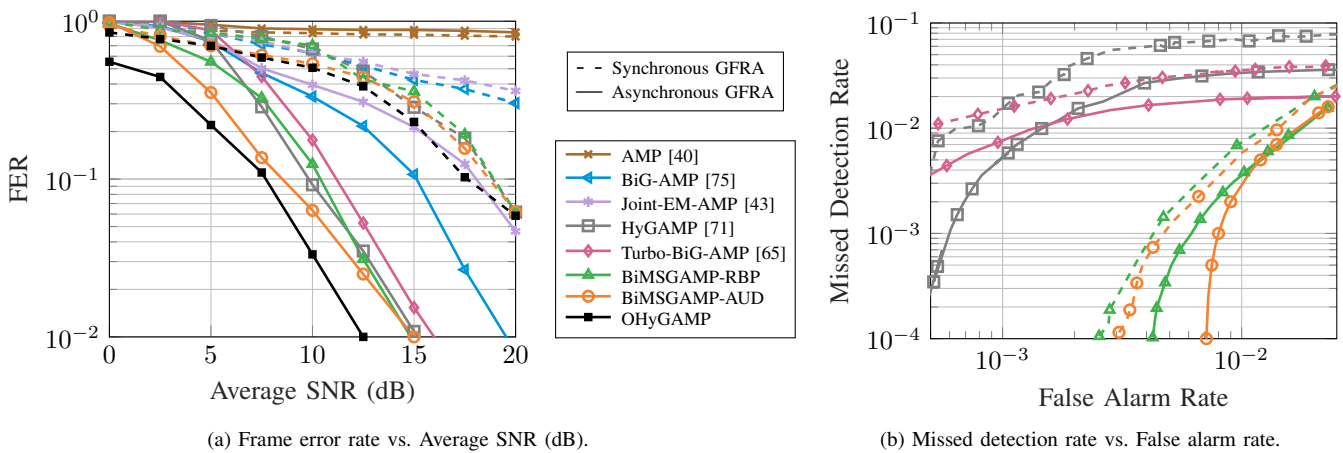
\begin{figure}[t]
            \begin{minipage}[b]{0.6\linewidth}
                \centerline{
%
\begin{tikzpicture}

\begin{axis}[%
width=5.4cm,
height=4.2cm,
at={(0.758in,0.481in)},
scale only axis,
xmin=0,
xmax=20,
xlabel style={font=\color{white!15!black}},
xlabel={Average SNR (dB)},
ymode=log,
ymin=0.01,
ymax=1,
yminorticks=true,
ylabel style={font=\color{white!15!black}},
ylabel={FER},
axis background/.style={fill=white},
xmajorgrids,
ymajorgrids,
yminorgrids,
legend style={at={(1.2,0.14)}, anchor=south west,  font=\scriptsize, legend cell align=left, align=left, draw=white!15!black}
]

\addplot [color=\AMPcolor, dashed,line width=1.0pt, mark=x,mark size = 2pt, mark options={solid, \AMPcolor}]
  table[row sep=crcr]{%
    0	    0.99\\
    2.5	    0.99\\
    5	    0.87\\
    7.5	    0.85\\
    10	    0.84\\
    12.5	0.825\\
    15	    0.822\\
    17.5	0.81\\
    20	    0.8\\  
};
\label{AMP_s}

\addplot [color=\AMPcolor, line width=1.0pt, mark=x,mark size = 2pt, mark options={solid, \AMPcolor}]
  table[row sep=crcr]{%
    0	    0.998571428571429\\
    2.5	    0.99\\
    5	    0.95\\
    7.5	    0.90\\
    10	    0.89\\
    12.5	0.88\\
    15	    0.88\\
    17.5	0.87\\
    20	    0.85\\ 
};
\label{AMP_a}

\addplot [color=\BIGAMPcolor, dashed,line width=1.0pt, mark=triangle, mark size = 2pt, mark options={solid, rotate=90, \BIGAMPcolor}]
  table[row sep=crcr]{%
    0	    0.972920634920635\\
    2.5	    0.959428571428571\\
    5	    0.842873015873016\\
    7.5	    0.712626984126984\\
    10	    0.629825396825397\\
    12.5	0.52285714285714\\
    15	    0.424801587301587\\
    17.5	0.371949603174603\\
    20	    0.303037301587302\\
};
\label{BiGAMP_s}

\addplot [color=\BIGAMPcolor, line width=1.0pt, mark=triangle, mark size = 2pt, mark options={solid, rotate=90, \BIGAMPcolor}]
  table[row sep=crcr]{%
    0	    0.986666666666667\\
    2.5	    0.93\\
    5	    0.743333333333333\\
    7.5	    0.47\\
    10	    0.333333333333333\\
    12.5	0.216666666666667\\
    15	    0.107\\
    17.5	0.026666666666667\\
    20	    0.008\\
};
\label{BiGAMP_a}

\addplot [color=\TBiGAMPcolor, line width=1.0pt, mark=diamond, mark size = 2pt,mark options={solid, \TBiGAMPcolor}]
  table[row sep=crcr]{%
    0	    1\\
    2.5	    1\\
    5	    0.876666666666667\\
    7.5	    0.446666666666667\\
    10	    0.176666666666667\\
    12.5	0.052666666666667\\
    15	    0.01533333333333\\
    17.5	0.0052\\
    20	    0\\
};
\label{Turbo_BiG_AMP_a}

\addplot [color=\JEMAMPcolor, dashed,line width=1.0pt, mark=asterisk, mark size = 2pt, mark options={solid, \JEMAMPcolor}]
  table[row sep=crcr]{%
    0	    0.977777777777778\\
    2.5	    0.939142857142857\\
    5	    0.839246031746032\\
    7.5	    0.747753968253968\\
    10	    0.625885714285714\\
    12.5	0.553079365079365\\
    15	    0.463039682539683\\
    17.5	0.421738095238095\\
    20	    0.361793650793651\\
};
\label{JEMAMP_s}

\addplot [color=\JEMAMPcolor, line width=1.0pt,  mark=asterisk, mark size = 2pt, mark options={solid, \JEMAMPcolor}]
  table[row sep=crcr]{%
    0	    0.993333333333333\\
    2.5	    0.91\\
    5	    0.76\\
    7.5	    0.503333333333333\\
    10	    0.395\\
    12.5	0.308333333333333\\
    15	    0.21333333333333\\
    17.5	0.124666666666667\\
    20	    0.046666666666667\\
};
\label{JEMAMP_a}

\addplot [color=\HyGAMPcolor, dashed,line width=1.0pt, mark=square, mark size = 2.0pt, mark options={solid, \HyGAMPcolor}]
  table[row sep=crcr]{%
    0	1\\
    2.5	1\\
    5	0.94\\
    7.5	0.7985\\
    10	0.667404761904762\\
    12.5	0.482452380952381\\
    15	0.284873015873016\\
    17.5	0.181738095238095\\
    20	0.0617936507936508\\
};
\label{HYGAMP_s}

\addplot [color=\HyGAMPcolor, line width=1.0pt, mark=square, mark size = 2.0pt, mark options={solid, \HyGAMPcolor}]
  table[row sep=crcr]{%
    0	    1\\
    2.5	    1\\
    5	    0.753333333333333\\
    7.5	    0.286666666666667\\
    10	    0.092\\
    12.5	0.035\\
    15	    0.01083333333333\\
    17.5	0.005\\
    20	    0\\
};
\label{HYGAMP_a}

\addplot [color=\RBPcolor, dashed,line width=1.0pt, mark=triangle,mark size = 2pt, mark options={solid, \RBPcolor}]
  table[row sep=crcr]{%
    0	    0.9875\\
    2.5	    0.883111111111111\\
    5	    0.843460317460317\\
    7.5	    0.777984126984127\\
    10	    0.702134920634921\\
    12.5	0.427412698412699\\
    15	    0.359253968253968\\
    17.5	0.191380952380952\\
    20	    0.061793650793651\\
};
\label{RBP_s}

\addplot [color=\RBPcolor, line width=1.0pt, mark=triangle,mark size = 2pt, mark options={solid, \RBPcolor}]
  table[row sep=crcr]{%
    0	    0.966666666666667\\
    2.5	    0.756666666666667\\
    5	    0.553333333333333\\
    7.5	    0.32666666666667\\
    10	    0.124333333333333\\
    12.5	0.0311084594\\
    15	    0.009243333333333\\
    17.5	0.005\\
    20	    0\\
};
\label{RBP_a}

\addplot [color=\AUDcolor, dashed,line width=1.0pt, mark=o,mark size = 2.0pt, mark options={solid, \AUDcolor}]
  table[row sep=crcr]{%
    0	    0.940833333333333\\
    2.5	    0.804626984126984\\
    5	    0.699\\
    7.5	    0.61018253968254\\
    10	    0.537888888888889\\
    12.5	0.442301587301587\\
    15	    0.308801587301587\\
    17.5	0.156380952380952\\
    20	    0.0617936507936508\\
};
\label{AUD_s}

\addplot [color=\AUDcolor, line width=1.0pt, mark=o,mark size = 2.0pt, mark options={solid, \AUDcolor}]
  table[row sep=crcr]{%
    0	    0.983333333333333\\
    2.5	    0.693333333333333\\
    5	    0.353333333333333\\
    7.5	    0.136666666666667\\
    10	    0.06333333333333\\
    12.5	0.025\\
    15	    0.01\\
    17.5	0\\
    20	    0\\
};
\label{AUD_a}

\addplot [color=black, dashed,line width=1.0pt, mark=square*, mark size = 1.1pt, mark options={solid, black}]
  table[row sep=crcr]{%
    0	    0.849285714285714\\
    2.5	    0.772325396825397\\
    5	    0.698309523809524\\
    7.5	    0.588746031746032\\
    10	    0.506785714285714\\
    12.5	0.386873015873016\\
    15	    0.230222222222222\\
    17.5	0.102714285714286\\
    20	    0.058579365079365\\
};
\label{OHYGAMP_s}

\addplot [color=black, line width=1.0pt, mark=square*, mark size = 1.1pt, mark options={solid, black}]
  table[row sep=crcr]{%
    0	    0.553333333333333\\
    2.5	    0.443333333333333\\
    5	    0.22\\
    7.5	    0.11\\
    10	    0.033333333333333\\
    12.5	0.01\\
    15	    0.006333333333333\\
    17.5	0\\
    20	    0\\
};
\label{OHYGAMP_a}
\end{axis}

\node [draw,fill=white] at (rel axis cs: 1.75,1.6) {\shortstack[l]{ 
    \ref{AMP_a} {\scriptsize AMP~\cite{Donoho2009}} \\
    \ref{BiGAMP_a} {\scriptsize BiG-AMP~\cite{JTParkerTSP2014}}\\     
    \ref{JEMAMP_a} {\scriptsize Joint-EM-AMP~\cite{CWeiCommLet2017}}\\    
    \ref{HYGAMP_a} {\scriptsize HyGAMP~\cite{SRanganTSP2017}}\\   
    \ref{Turbo_BiG_AMP_a} {\scriptsize Turbo-BiG-AMP~\cite{TDingTWC2019}}\\ 
    \ref{RBP_a} {\scriptsize BiMSGAMP-RBP}\\    
    \ref{AUD_a} {\scriptsize BiMSGAMP-AUD}\\
    \ref{OHYGAMP_a}  {\scriptsize OHyGAMP}
}};

\node [draw,fill=white] at (rel axis cs: 1.75,2.1) {\shortstack[l]{ 
    - -  {\scriptsize Synchronous GFRA}\\    
    ---  {\scriptsize Asynchronous GFRA}
}};
\end{tikzpicture}%
}
                \centerline{\footnotesize (a) Frame error rate vs. Average SNR (dB).}\medskip
            \end{minipage}
            \begin{minipage}[b]{0.39\linewidth}
                \centerline{
\begin{tikzpicture}
\begin{axis}[%
width=5.4cm,
height=4.2cm,
at={(0.758in,0.481in)},
scale only axis,
mark repeat = 2,
xmin=0.0005,
xmax=0.025,
xlabel style={font=\color{white!15!black}},
xlabel={False Alarm Rate},
xmode=log,
ymode=log,
ymin=0.0001,
ymax=0.1,
yminorticks=true,
ylabel style={font=\color{white!15!black}},
ylabel={Missed Detection Rate},
axis background/.style={fill=white},
xmajorgrids,
ymajorgrids,
yminorgrids,
legend style={at={(0.05,0.01)}, font=\tiny, anchor=south west, legend cell align=left, align=left, draw=white!15!black}
]

\addplot [color=\HyGAMPcolor, line width=1.0pt, mark=square, mark size = 2.0pt, mark options={solid, \HyGAMPcolor}]
  table[row sep=crcr]{%
0.000489314199255675 0.0002370986542942092 \\
0.0005344254854083944 0.00048545816215143635 \\
0.0005597880067102694 0.0006936075852880272\\
0.0005151584608058517 0.0003469434211324178\\
0.0006058556272344318 0.0010808821231000584\\
0.0006465477351102133 0.0015086667471699033\\
0.0006867851158972012 0.001954978177911692\\
0.0007458210307331696 0.002650545909727248\\
0.0008193228206173793 0.0035888148925309147\\
0.001055417222582518 0.005820152209473489\\
0.000930014427319673 0.00460637150083731\\
0.0011504266429771268 0.007042372703048723\\
0.0012704688972187727 0.008089612334074979\\
0.0014575410189970643 0.009932869196928073\\
0.0017090940860712528 0.011876720749748543\\
0.0020703020576246815 0.015428936208461919\\
0.0026373836724160713 0.017939057776147602\\
%
0.003924860524003047 0.02688917420523749\\
0.00539003330787633 0.029636777926206077\\
0.00681913386623994 0.031367425359552156\\
0.0087220300215399 0.03293105420660645\\
0.011718812629856622 0.03429345704895215\\
0.016539688115358848 0.03513780658027035\\
0.02308981363216598 0.036002945095621866\\
0.03498992349252213 0.036002945095621866\\
};

\addplot [color=\HyGAMPcolor, dashed,line width=1.0pt, mark=square, mark size = 2.0pt, mark options={solid, \HyGAMPcolor}]
  table[row sep=crcr]{%
0.000153871	0.00010197	\\
0.000171736	0.000106012	\\
0.000203356	0.000120994	\\
0.000218190 0.0013692576\\
0.000238771	0.000377415	\\
0.000310272	0.000668033	\\
0.000337629	0.000911106	\\
0.000386506	0.001305879	\\
0.000461552	0.001895082	\\
0.000537372	0.007584633	\\
0.000630957	0.009750801	\\
0.000792652	0.0105579955	\\
0.000915099	0.0127423692	\\
0.00105646	0.017251364	\\
0.001163242	0.020565049	\\
0.001416447	0.0220536746	\\
0.001658051	0.0302005246	\\
0.00180972	0.032362239	\\
0.001855896	0.038162361	\\
0.002257262	0.045928651	\\
0.002849526	0.055394287	\\
0.004570512	0.061258487	\\
0.005036486	0.056563687	\\
0.005244396	0.065389385	\\
0.005882855	0.065389385	\\
0.007283603	0.067281351	\\
0.009266378	0.070022424	\\
0.010739831	0.067494921	\\
0.012508599	0.070519798	\\
0.01427825	0.075645857	\\
0.016463181	0.073641332	\\
0.01861744	0.074876865	\\
0.023350696	0.077512755	\\
0.029477492	0.079084299	\\
};

\addplot [color=\TBiGAMPcolor, line width=1.0pt, mark=diamond, mark size = 2pt,mark options={solid, \TBiGAMPcolor}]
  table[row sep=crcr]{%
0.010562165521799834 0.01928462808226607\\
0.014113791761820533 0.019599884906767766\\
0.02465607772879353 0.020082459574328908\\
0.0367553586699007 0.020576915858077048\\
0.01885968530647384 0.019759440716631396\\
0.008123416436245057 0.018821225033879722\\
0.0079152910982998 0.018583620893795233\\
0.004079152910982998 0.016583620893795233\\
0.0028310808264524114 0.014872495533018885\\
0.0018707776368885789 0.012223967592066643\\
0.0012960928096145206 0.009669965014361391\\
0.0009558829084235898 0.007334447826352913\\
0.0007145007503796613 0.0056875054639171675\\
0.0005842222848016939 0.004411721297710888\\
0.0004886504958817522 0.0035203714769220075\\
0.0003770199159517425 0.002534167527656916\\
0.0002930502854577394 0.001912502460246219\\
0.00021819408602348275 0.001369053833492576\\
0.00016401215428119738 0.0009687970127115359\\
0.00013052763804649034 0.000685559346811387\\
0.00011209513822936698 0.0005635580702568055\\
0.00010000000000000021 0.0004579584033357946\\
};

\addplot [color=\TBiGAMPcolor, line width=1.0pt, dashed,mark=diamond, mark size = 2pt,mark options={solid, \TBiGAMPcolor}]
  table[row sep=crcr]{%
0.00010433979537295824 0.0036910256943108047\\
0.00012972388347720604 0.004438002730903954\\
0.00015516619321348065 0.005169059693251489\\
0.00018459391711095983 0.005731385694168232\\
0.00022199921228929656 0.006486835335728294\\
0.0002673466226953171 0.007222815187696189\\
0.00031453093843953706 0.007922461503647258\\
0.0003818701947801542 0.008978384874588824\\
0.00045360147861706323 0.010045191105396735\\
0.0005379867492383172 0.011025950778809869\\
0.0006501512733949092 0.01218581134137075\\
0.0007885915873624638 0.01355933850794409\\
0.0009233937962242302 0.014695245966210087\\
0.0011186720651933408 0.01618535296920849\\
0.001333723572433873 0.017578500633231894\\
0.0015918430202787597 0.019148589781701024\\
0.0019040458474518715 0.0207213254907294\\
0.0022774799669403618 0.02271492090432738\\
0.002724154466530398 0.02537383712242499\\
0.003258433736076785 0.02698012643762072\\
0.003897499404990083 0.02885825569606045\\
0.004661902877972199 0.030437346138390067\\
0.0055762262377306955 0.03151725406864262\\
0.00940055773273313 0.03491195135769875\\
0.007293750314525081 0.03395246218269532\\
0.011402684968987213 0.036296032483100565\\
0.013435877428667295 0.03749243627386817\\
0.01583160481668601 0.038415548835854275\\
0.01885968530647384 0.03810534653338214\\
0.02246694090572183 0.03904354974340707\\
0.028893995509091164 0.039361389630166284\\
0.035959982973931506 0.04098982426133143\\
};

\addplot [color=\RBPcolor, line width=1.0pt, mark=triangle,mark size = 2pt, mark options={solid, \RBPcolor}]
  table[row sep=crcr]{%
0.010239755435542348 0.0038094400097139105\\
0.01166294027534609 0.004842582467003001\\
0.012871420677812773 0.0060271612251486945\\
0.014246102062944197 0.0073143691735023005\\
0.015689781051531085 0.008598363230979296\\
0.018110753015188403 0.010830201817741742\\
0.02293000693937404 0.01564079444653699\\
0.00926706371008957 0.003103593481454677\\
0.008346177242389764 0.0024503975280369397\\
0.0074811255312683955 0.0018689147295323722\\
0.006705733366210035 0.001369053833492576\\
0.006155398839877056 0.0009913917639001112\\
0.005491227306649114 0.0006935077427080176\\
0.005064603543891595 0.0004907536964516966\\
0.004783569257132789 0.0003413237104286842\\
0.0045613238628900935 0.0002471674283834213\\ 
0.0044119247773244 0.00019514748279407556\\
0.004328761281082938 0.0001446106407142123\\
0.004247165392508449 0.00010233224823075555\\
};

\addplot [color=\RBPcolor, line width=1.0pt, dashed,mark=triangle,mark size = 2pt, mark options={solid, \RBPcolor}]
  table[row sep=crcr]{%
0.002533429	0.000105446\\
0.002776701	0.00013798\\
0.002806059	0.000187603\\
0.003752424	0.000612438\\
0.004708125	0.001429951\\
0.006567085	0.003045947\\
0.009551042	0.00686908\\
0.013652008	0.011424254\\
0.020178992	0.020046377\\
0.037097337	0.041099959\\
};

\addplot [color=\AUDcolor, line width=1.0pt, mark=o,mark size = 2.0pt, mark options={solid, \AUDcolor}]
  table[row sep=crcr]{%
0.00712	0.0001\\
0.00725	0.00025\\
0.0075	0.0005\\
0.00775	0.00075\\
0.008	0.001\\
0.0085	0.0015\\
0.009	0.002\\
0.011	0.004\\
0.012	0.005\\
0.013	0.006\\
0.014	0.007\\
0.015	0.008\\
%
0.021	0.014\\
0.022	0.015\\
0.023	0.016\\
0.02375	0.01675\\
};

\addplot [color=\AUDcolor, line width=1.0pt, dashed,mark=o,mark size = 2.0pt, mark options={solid, \AUDcolor}]
  table[row sep=crcr]{%
0.00306902	0.000114812	\\
0.003282698	0.000141524	\\
0.003406059	0.000187603	\\
0.003504842	0.00021709	\\
0.003644056	0.000338451	\\
0.003862626	0.000508086	\\
0.004259492	0.000737776	\\
0.005129697	0.001284028	\\
0.006609726	0.00225082	\\
0.010084857	0.005899552	\\
0.014021323	0.009652328	\\
0.01965975	0.01977993	\\
0.025982378	0.025792129	\\
0.035219971	0.034660974	\\
};
\end{axis}
\end{tikzpicture}%
}
                \centerline{\footnotesize (b) Missed detection rate vs. False alarm rate.}\medskip
            \end{minipage} \hfill
             \vspace{-2.5em}
            \caption{Key performance indicators for synchronous and asynchronous mMTC scenarios with $N=100, M=32$ and $L=192$, by $10^4$ Monte Carlo trials.}
            \label{fig:AER_FER}%
        \end{figure}
    \fi
    \makeatother

    Since there are not many works in the literature that perform joint channel estimation, activity detection and signal decoding in the mMTC scenario, the algorithms that do not consider the channel estimation part (AMP, Joint-EM-AMP and HyGAMP), have a separate channel estimation stage, which is adapted using the same solution considered. Additionally, in order to give a fair comparison, all evaluated algorithms used the LLR conversion presented in Section~\ref{sec:jointAUDCEDATA}. Thus, despite the fact that it estimates the channels in a separate stage, we can consider that HyGAMP is an equivalent parallel version of the BiMSGAMP. The maximum number of iterations employed for AMP and Joint-EM-AMP is $N/2$ while $20$ for BiG-AMP, Turbo-BiG-AMP, HyGAMP and BiMSGAMP-type schemes.

    Considering the asynchronous GFRA scenario, Fig.~\ref{fig:NMSE_channel} depicts the NMSE versus different signal-to-noise ratio values. We notice that in this new scenario the metric that considers the activity detection as message-scheduling, BiMSGAMP-AUD, reached the oracle HyGAMP performance, outperforming all other approaches. It is a good indication that the use of the channel decoding LLRs to refine the AUD improves not only the data detection, but also the channel estimation. Nevertheless, HyGAMP outperforms the residual-based solution, for SNR values $ > 7.5$ dB. Since BiMSGAMP-RBP updates all nodes in the end of the set procedure, it is possible that the neglected nodes had a considerable influence in this scenario. Besides that, the well-known AMP exhibits a poor performance even though it requires more iterations. This channel estimation is used for the AMP and the Joint-EM-AMP schemes in the data decoding part. Since BiG-AMP do not have previous knowledge of the sparsity of the scenario as HyGAMP and it jointly performs the AUD, CE and data decoding (using the proposed scheme), it clearly loses performance. Other hypothesis is that a different adaptive damping and/or an specific initialization should be considered to improve its efficiency for the mMTC scenario. Another hypothesis is that the mMTC system is not sparse enough for it, since the number of active devices vary from 1 to $10$\% of $N$.

    Using the channel estimation depicted in Fig.~\ref{fig:NMSE_channel}, the frame error rate (FER) performance of the schemes is shown in Fig.~\ref{fig:AER_FER}(a). Firstly, for the synchronous GFRA, we observe that for low SNR values ($< 10$ dB), we notice that the BiMSGAMP-type solutions outperform other approaches, getting even closer to the lower bound. For larger SNR values, BiMSGAMP-RBP and HyGAMP exhibit almost the same performance. The performance degradation of BiMSGAMP-RBP in SNRs between $10$ and $15$ dB, is due to the fact that the channel estimation was not efficient. On the other hand, the approach of the LLRs used in BiMSGAMP-AUD proved to be efficient for data decoding. Despite the fact that, as seen in Fig.~\ref{fig:AER_FER}(b), BiG-AMP and Joint-EM-AMP provide a satisfactory MDR and FAR values, their means and variances estimates that are the base of the LLRs for signal detection are not as accurate as those of HyGAMP and BiMSGAMP. Regarding the asynchronous scenario, besides the algorithms already discussed, we consider the Turbo-BiG-AMP~\cite{TDingTWC2019} that is designed for asynchronous mMTC scenarios. Unlike the scenario in our work, Turbo-BiG-AMP~\cite{TDingTWC2019} has the knowledge of the number of active devices. With this information, a subgraph is included in the original BiG-AMP with the aim of determining the location of the frames, and, consequently, the activity of devices. Including Turbo-BiG-AMP in the FER simulations, one can see that the hierarchy of performance of the algorithms remains the same as in the synchronous case. We remark that in the asynchronous scenario, the FER performances are in general improved. This can be explained by the fact that all considered algorithms are compressed sensing solutions, and the asynchronous scenario, which is more sparse than the synchronous one, favors them. It is important to remind that in the asynchronous scenario a frame is considered correct if and only if the whole frame is inside the observation window. If a false alarm occurs in the symbol interval immediately before or after a true frame, the frame is considered as wrong.

    As a crucial part of the study, the activity error rates for the asynchronous GFRA are shown in Figs~\ref{fig:AER_FER}(b). Evidently, there is a trade-off regarding the false alarm and missed detection rates. Since most of the approaches in the literature use the means to detect the activity, the threshold considered by all of them is $0.5$, while for BiMSGAMP-type schemes, that considers LLRs (for each symbol), is $0.95$. Thus, it is possible to see that for lower MDR values, the detector benefits in terms of FER since we focus on the active devices. Naturally, the activity threshold is a parameter that depends on the system designer.

\section{Conclusion} \label{sec:conc}

We have presented a joint channel estimation, activity detection and data decoding scheme for mMTC. By including the channel and the \textit{a priori} activity factor in the factor graph, we have devised BiMSGAMP, a message-passing solution that uses the channel decoder beliefs to refine the activity detection and data decoding. We have also included and presented two message-scheduling strategies based on RBP and AUD in which messages are evaluated and scheduled in every new iteration. Numerical results have shown that BiMSGAMP outperforms state-of-the-art algorithms, highlighting the gains achieved by using the dynamic scheduling strategies and the effects of the channel decoding part in the system and requiring much lower computational cost.

\appendices
\section{Derivation of messages of interest from variable to factor nodes}\label{app:B}

    We show the approximation of the messages from variable nodes to factor nodes. Since $\mathbf{Z}^\text{T} = \mathbf{X}^\text{T}\mathbf{H}^\text{T}$, the derivation of the approximation of $\Delta^{i+1}_{g_{mt}\rightarrow h_{nm}}\left(h_{nm}\right)$ is given by,
    \makeatletter%
    \if@twocolumn
        \begin{align} \label{eq:ApxA_approx_gtoh}
            &\hspace{-0.13cm}\Delta^{i+1}_{g_{mt}\rightarrow h_{nm}}\left(h_{nm}\right) \approx \left[\hat{s}^i_{mt}\, \hat{x}_{m,nt}^i + \nu^{si}_{mt}\,  \hat{x}^{i^2}_{nt}\, \hat{h}^{i}_{mn}\right]\, h_{mn} \\  \nonumber 
            &\hspace{2.5cm} - (\nicefrac{1}{2}) \left[\nu^{si}_{mt}\, \hat{x}_{nt}^{i^2} - \nu_{mn}^{xi}\,\left(\hat{s}_{mt}^{i^2} - \nu^{si}_{mt} \right)\right]\, h^2_{nt}.
        \end{align}  
    
    \else
        \begin{equation}\label{eq:ApxA_approx_gtoh}
            \Delta^{i+1}_{g_{mt}\rightarrow h_{nm}}\left(h_{nm}\right) \approx \left[\hat{s}^i_{mt}\, \hat{x}_{m,nt}^i + \nu^{si}_{mt}\,  \hat{x}^{i^2}_{nt}\, \hat{h}^{i}_{mn}\right]\, h_{mn} - (\nicefrac{1}{2}) \left[\nu^{si}_{mt}\, \hat{x}_{nt}^{i^2} - \nu_{mn}^{xi}\,\left(\hat{s}_{mt}^{i^2} - \nu^{si}_{mt} \right)\right]\, h^2_{nt}.
        \end{equation}   
    \fi
    \makeatother    
    
    Recalling (\ref{eq:htog}), converting the messages to the form of log-pdf and substituting (\ref{eq:ApxA_approx_gtoh}), we obtain
    \makeatletter%
    \if@twocolumn
        \begin{align} \nonumber
            & \Delta^{i+1}_{h_{mn}\rightarrow g_{mt}}\left(h_{mn}\right)\\ \nonumber
            &\hspace{0.3cm} \approx \Delta_{k_{mn}\rightarrow h_{mn}\left(h_{mn}\right)} \prod_{p \neq t} \Delta^i_{g_{mp}\rightarrow h_{mn}}\left(h_{mn}\right) \\ \nonumber
            &\hspace{0.3cm} = \log \left(\mathcal{P}\left(h_{mn}| \gamma_{nt}\right) \mathcal{P}\left(\gamma_{nt}\right)\right) + \sum_{p \neq t} \Delta^i_{g_{mp}\rightarrow h_{mn}}\left(h_{mn}\right) \\ \label{eq:apdxB_htog1}
            &\hspace{0.3cm} = \log \left(\mathcal{P}\left(h_{mn}| \gamma_{nt}\right) \mathcal{P}\left(\gamma_{nt}\right)\right) +\\ \nonumber
            &\hspace{0.9cm} \sum_{p \neq t} \left[\hat{s}^i_{mt}\, \hat{x}_{m,nt}^i + \nu^{si}_{mt}\,  \hat{x}^{i^2}_{nt}\, \hat{h}^{i}_{mn}\right]\, h_{mn} -\\ \nonumber
            &\hspace{0.9cm} (\nicefrac{1}{2}) \left[\nu^{si}_{mt}\, \hat{x}_{nt}^{i^2} - \nu_{mn}^{xi}\,\left(\hat{s}_{mt}^{i^2} - \nu^{si}_{mt} \right)\right]\, h^2_{nt}
        \end{align} 
    
    \else
        \begin{equation}\label{eq:apdxB_htog1}
            \begin{split}
                \Delta^{i+1}_{h_{mn}\rightarrow g_{mt}}\left(h_{mn}\right) \approx & \, \Delta_{k_{mn}\rightarrow h_{mn}\left(h_{mn}\right)} \prod_{p \neq t} \Delta^i_{g_{mp}\rightarrow h_{mn}}\left(h_{mn}\right) \\ 
                = & \, \log \left(\mathcal{P}\left(h_{mn}| \gamma_{nt}\right) \mathcal{P}\left(\gamma_{nt}\right)\right) + \sum_{p \neq t} \Delta^i_{g_{mp}\rightarrow h_{mn}}\left(h_{mn}\right) \\ 
                = & \, \log \left(\mathcal{P}\left(h_{mn}| \gamma_{nt}\right) \mathcal{P}\left(\gamma_{nt}\right)\right) + \sum_{p \neq t} \left[\hat{s}^i_{mt}\, \hat{x}_{m,nt}^i + \nu^{si}_{mt}\,  \hat{x}^{i^2}_{nt}\, \hat{h}^{i}_{mn}\right]\, h_{mn}\\
                & - (\nicefrac{1}{2}) \left[\nu^{si}_{mt}\, \hat{x}_{nt}^{i^2} - \nu_{mn}^{xi}\,\left(\hat{s}_{mt}^{i^2} - \nu^{si}_{mt} \right)\right]\, h^2_{nt}
            \end{split}        
        \end{equation}    
    \fi
    \makeatother        

    Since in our system model $x$ depends only on $n$ and $l$ and $h$ only on $m$ and $n$, we ignore the components of $x$ that mathematically depends on $m$ and the ones of $h$ that depends on $l$, following the same idea of~\cite{Rangan2012} and  ignoring the terms $<O(1/N)$. This approximation on (\ref{eq:apdxB_htog1}) leads to
    \makeatletter%
    \if@twocolumn
        \begin{align}\label{eq:apdxB_htog2}
            & \Delta^{i+1}_{h_{mn}\rightarrow g_{mt}}\left(h_{mn}\right)\\ \nonumber
            & \hspace{1cm} \approx \log \left(\mathcal{P}\left(h_{mn}| \gamma_{nt}\right) \mathcal{P}\left(\gamma_{nt}\right)\right)-\nicefrac{\left(h_{mn} - \hat{q}^i_{mn}\right)^2}{2\,\nu^{qi}_{mn}}\\ \nonumber
            & \hspace{1cm} = \log\bigg(\mathcal{P}\left(h_{mn}| \gamma_{nt}\right) \mathcal{P}\left(\gamma_{nt}\right) \mathcal{N}_c\left(h_{mn};\hat{q}^i_{mn},\nu^{qi}_{mn}\right) \bigg),
        \end{align}    
    
        \noindent for
        \vspace{-0.1cm}
        \begin{equation} \label{eq:apdxB_qvar}
            \hspace*{-1.5cm} \nu^{qi}_{mn} \approx \left(\sum_{t=1}^T \hat{x}^{i^2}_{nt} \nu^{si}_{mt}\right)^{-1} \hspace{-1cm}
        \end{equation}
        
        \noindent and 
        \vspace{-0.1cm}    
        \begin{equation} \label{eq:apdxB_qhat}
            \hat{q}^i_{mn} \approx \hat{h}^{i}_{mn} \left(\!1 - \nu^{qi}_{mn} \sum_{t=1}^{T} \nu^{xi}_{nt}\, \nu^{si}_{mt}\right)\! + \nu^{qi}_{mn} \sum_{t=1}^{T} \hat{x}^{i\ast}_{nt}\,\hat{s}^i_{mt}.
        \end{equation} 
    
        Therefore, the corresponding means and variances of interest are then further approximated as\\
            
            \begin{equation} \label{eq:apdxB_hhat_end}
                \hat{h}^{i+1}_{mn} \triangleq \underbrace{\frac{\int_h h\, \mathcal{P}\left(h| \gamma\right) \mathcal{P}\left(\gamma\right) \mathcal{N}_c\left(h;\hat{q},\nu^{q}\right)}{\int_h \mathcal{P}\left(h| \gamma\right) \mathcal{P}\left(\gamma\right) \mathcal{N}_c\left(h;\hat{q},\nu^{q}\right)}}_{g_{h_{mn}(\hat{q},\nu^q)}}
            \end{equation}
        
            \begin{equation} 
                \nu^{h\,i+1}_{mn} \triangleq \nu^{hi}_{mn}\, g'_{h_{mn}}\left(\hat{q}^i_{mn},\nu^{qi}_{mn}\right)\vphantom{\underbrace{\frac{\int_h h\, p_{\bsfh_{mn}| \bsfgamma_n}\left(h| \gamma\right) p_{\bsfgamma_n}\left(\gamma\right) \mathcal{N}_c\left(h;\hat{q},\nu^{q}\right)}{\int_h p_{\bsfh_{mn}| \bsfgamma_n}\left(h| \gamma\right) p_{\bsfgamma_n}\left(\gamma\right) \mathcal{N}_c\left(h;\hat{q},\nu^{q}\right)}}_{g_{h_{mn}(\hat{q},\nu^q)}}}
            \end{equation} 
            
            \vspace{-0.5cm}
            \noindent where $g'_{h_{mn}}$ is the first derivative of $g_{h_{mn}}$ in (\ref{eq:apdxB_hhat_end}). Similarly, following the same steps and using (\ref{eq:px_all}), $\Delta^{i+1}_{x_{nt}\rightarrow g_{mt}}\left(x_{nt}\right)$, the means and variances of interest for the problem are given by
            
            \begin{align} \label{eq:apdxB_xtog2}
                & \Delta^{i+1}_{x_{nt}\rightarrow g_{mt}}\left(x_{nt}\right)\\ \nonumber
                & \hspace{0.4cm} \approx \log\bigg( \sum_{s_{nt}}\! \sum_{c_{nt}} \mathcal{P}_{\bsfx_{d_{nt}}}\!\left(x_{nt},c_{nt},s_{nt}\right)  \mathcal{N}_c\left(x_{nt};\hat{r}^i_{nt},\nu^{ri}_{nt}\right) \bigg),
            \end{align}
            
            \noindent where the means and variances are
            
            \begin{equation} \label{eq:apdxB_rvar}
                \nu^{ri}_{nt} \approx \left(\sum_{m=1}^M \hat{h}^{i^2}_{mn} \nu^{si}_{mt}\right)^{-1}
            \end{equation}
            
            \noindent and
            \begin{equation} \label{eq:apdxB_rhat}
                \hat{r}^i_{nt} \approx \hat{x}^{i}_{nt} \left(\!1 - \nu^{ri}_{nt} \sum_{m=1}^{M} \nu^{hi}_{mn}\, \nu^{si}_{mt}\right)\! + \nu^{ri}_{nt} \sum_{m=1}^{M} \hat{h}^{i\ast}_{mn}\,\hat{s}^i_{mt}
            \end{equation} 
            
            \noindent with the mean and the variance to compute the data estimates given by
            
            \begin{equation} \label{eq:apdxB_xhat_end}
                \hat{x}^{i+1}_{mn} \triangleq \frac{\int_x x\, \sum_{s}\! \sum_{c} \mathcal{P}_{\bsfx_{d_{nt}}}\!\left(x,c,s\right) \mathcal{N}_c\left(x;\hat{r},\nu^{r}\right)}{\int_x \sum_{s}\! \sum_{c} \mathcal{P}_{\bsfx_{d_{nt}}}\!\left(x,c,s\right) \mathcal{N}_c\left(x;\hat{r},\nu^{r}\right)}        
            \end{equation}
            
            \begin{equation} \label{eq:apdxB_xvar_end}
                \nu^{x\,i+1}_{nt} \triangleq \nu^{xi}_{nt}\, g'_{x_{nt}}\left(\hat{x}^i_{nt},\nu^{ri}_{nt}\right)\hspace{-0.1cm}
            \end{equation} 
            
            \noindent as well as for the channel, the left part of (\ref{eq:apdxB_xhat_end}) is $g_{x_{nt}}\left(\hat{r},\nu^r\right)$ and $g'_{x_{nt}}$ in (\ref{eq:apdxB_xvar_end}) is its first derivative.
    
    \else
        \begin{equation}\label{eq:apdxB_htog2}
            \begin{split}
                \Delta^{i+1}_{h_{mn}\rightarrow g_{mt}}\left(h_{mn}\right) \approx & \log \left(\mathcal{P}\left(h_{mn}| \gamma_{nt}\right) \mathcal{P}\left(\gamma_{nt}\right)\right)-\nicefrac{\left(h_{mn} - \hat{q}^i_{mn}\right)^2}{2\,\nu^{qi}_{mn}}\\
                = &\log\bigg(\mathcal{P}\left(h_{mn}| \gamma_{nt}\right) \mathcal{P}\left(\gamma_{nt}\right) \mathcal{N}_c\left(h_{mn};\hat{q}^i_{mn},\nu^{qi}_{mn}\right) \bigg),
            \end{split}
        \end{equation}    
    
        \noindent for
    
        \begin{minipage}{0.3\textwidth}
            \begin{equation} \label{eq:apdxB_qvar}
                \hspace*{-1.5cm} \nu^{qi}_{mn} \approx \left(\sum_{t=1}^T \hat{x}^{i^2}_{nt} \nu^{si}_{mt}\right)^{-1} \hspace{-1cm}
            \end{equation}
        \end{minipage}
        \begin{minipage}{0.65\textwidth}
            \begin{equation} \label{eq:apdxB_qhat}
                \hat{q}^i_{mn} \approx \hat{h}^{i}_{mn} \left(\!1 - \nu^{qi}_{mn} \sum_{t=1}^{T} \nu^{xi}_{nt}\, \nu^{si}_{mt}\right)\! + \nu^{qi}_{mn} \sum_{t=1}^{T} \hat{x}^{i\ast}_{nt}\,\hat{s}^i_{mt}.
            \end{equation} 
        \end{minipage} 
        
        \vspace{0.5cm}   
         Therefore, the corresponding means and variances of interest are then further approximated as\\
            
            \begin{minipage}{0.5\textwidth}
                \begin{equation} \label{eq:apdxB_hhat_end}
                    \hat{h}^{i+1}_{mn} \triangleq \underbrace{\frac{\int_h h\, \mathcal{P}\left(h| \gamma\right) \mathcal{P}\left(\gamma\right) \mathcal{N}_c\left(h;\hat{q},\nu^{q}\right)}{\int_h \mathcal{P}\left(h| \gamma\right) \mathcal{P}\left(\gamma\right) \mathcal{N}_c\left(h;\hat{q},\nu^{q}\right)}}_{g_{h_{mn}(\hat{q},\nu^q)}}
                \end{equation}
            \end{minipage}
            \begin{minipage}{0.45\textwidth}
                \begin{equation} 
                    \nu^{h\,i+1}_{mn} \triangleq \nu^{hi}_{mn}\, g'_{h_{mn}}\left(\hat{q}^i_{mn},\nu^{qi}_{mn}\right)\vphantom{\underbrace{\frac{\int_h h\, p_{\bsfh_{mn}| \bsfgamma_n}\left(h| \gamma\right) p_{\bsfgamma_n}\left(\gamma\right) \mathcal{N}_c\left(h;\hat{q},\nu^{q}\right)}{\int_h p_{\bsfh_{mn}| \bsfgamma_n}\left(h| \gamma\right) p_{\bsfgamma_n}\left(\gamma\right) \mathcal{N}_c\left(h;\hat{q},\nu^{q}\right)}}_{g_{h_{mn}(\hat{q},\nu^q)}}}
                \end{equation} 
            \end{minipage}   
            
            \vspace{0.5cm}
            \noindent where $g'_{h_{mn}}$ is the first derivative of $g_{h_{mn}}$ in (\ref{eq:apdxB_hhat_end}). Similarly, following the same steps and using (\ref{eq:px_all}), $\Delta^{i+1}_{x_{nt}\rightarrow g_{mt}}\left(x_{nt}\right)$, the means and variances of interest for the problem are given by
            
            \begin{equation} \label{eq:apdxB_xtog2}
                \begin{split} 
                    \Delta^{i+1}_{x_{nt}\rightarrow g_{mt}}\left(x_{nt}\right) \approx & 
                    \log\bigg( \sum_{s_{nt}}\! \sum_{c_{nt}} \mathcal{P}_{\bsfx_{d_{nt}}}\!\left(x_{nt},c_{nt},s_{nt}\right)  \mathcal{N}_c\left(x_{nt};\hat{r}^i_{nt},\nu^{ri}_{nt}\right) \bigg),
                \end{split}
            \end{equation}
            
            \noindent where the means and variances are
            
            \vspace{0.1cm}
            \begin{minipage}{0.3\textwidth}
                \begin{equation} \label{eq:apdxB_rvar}
                    \nu^{ri}_{nt} \approx \left(\sum_{m=1}^M \hat{h}^{i^2}_{mn} \nu^{si}_{mt}\right)^{-1}
                \end{equation}
            \end{minipage}
            \begin{minipage}{0.65\textwidth}
                \begin{equation} \label{eq:apdxB_rhat}
                    \hat{r}^i_{nt} \approx \hat{x}^{i}_{nt} \left(\!1 - \nu^{ri}_{nt} \sum_{m=1}^{M} \nu^{hi}_{mn}\, \nu^{si}_{mt}\right)\! + \nu^{ri}_{nt} \sum_{m=1}^{M} \hat{h}^{i\ast}_{mn}\,\hat{s}^i_{mt}
                \end{equation} 
            \end{minipage}       
            
            \vspace{0.5cm}
            \noindent with the mean and the variance to compute the data estimates given by
            
            \vspace{0.5cm}
            \begin{minipage}{0.58\textwidth}
                \begin{equation} \label{eq:apdxB_xhat_end}
                    \hat{x}^{i+1}_{mn} \triangleq \frac{\int_x x\, \sum_{s}\! \sum_{c} \mathcal{P}_{\bsfx_{d_{nt}}}\!\left(x,c,s\right) \mathcal{N}_c\left(x;\hat{r},\nu^{r}\right)}{\int_x \sum_{s}\! \sum_{c} \mathcal{P}_{\bsfx_{d_{nt}}}\!\left(x,c,s\right) \mathcal{N}_c\left(x;\hat{r},\nu^{r}\right)}        
                \end{equation}
            \end{minipage}
            \begin{minipage}{0.38\textwidth}
                \begin{equation} \label{eq:apdxB_xvar_end}
                    \nu^{x\,i+1}_{nt} \triangleq \nu^{xi}_{nt}\, g'_{x_{nt}}\left(\hat{x}^i_{nt},\nu^{ri}_{nt}\right)\hspace{-0.1cm}
                \end{equation} 
            \end{minipage}    
        
            \vspace{0.5cm}
            \noindent as well as for the channel, the left part of (\ref{eq:apdxB_xhat_end}) is $g_{x_{nt}}\left(\hat{r},\nu^r\right)$ and $g'_{x_{nt}}$ in (\ref{eq:apdxB_xvar_end}) is its first derivative.        
    \fi
    \makeatother

\bibliographystyle{IEEEbib}
\bibliography{ref}

\end{document}